\documentclass[a4paper,11pt]{article}
\usepackage{jheppub} 
\usepackage{svg}
\usepackage{lineno}
\usepackage{isomath}
\usepackage{wasysym}
\usepackage{xcolor}
\usepackage{stmaryrd}
\usepackage{tikz}
\usetikzlibrary{patterns}
\usetikzlibrary{decorations.pathreplacing}
\usepackage{graphicx}
\usepackage{subcaption}
\usepackage[framemethod=tikz]{mdframed}

\pgfdeclarelayer{bg}    
\pgfsetlayers{bg,main}  

\def\propagatorPP{\tikz[baseline=.1ex]{
\fill (0,0.7ex) circle (2pt) coordinate (A);
\fill (10ex,0.7ex) circle (2pt) coordinate (B);
\begin{pgfonlayer}{bg}    
        \draw (A) -- (B);
\end{pgfonlayer}
\node at (0, 2ex) {$\eta_1$};
\node at (10ex, 2ex) {$\eta_2$};
}}
\def\propagatorPM{\tikz[baseline=.1ex]{
\fill (0,0.7ex) circle (2pt) coordinate (A);
\filldraw[color=black, fill=white] (10ex,0.7ex) circle (2pt) coordinate (B);
\begin{pgfonlayer}{bg}    
        \draw (A) -- (B);
\end{pgfonlayer}
\node at (0, 2ex) {$\eta_1$};
\node at (10ex, 2ex) {$\eta_2$};
}}
\def\propagatorMP{\tikz[baseline=.1ex]{
\filldraw[color=black, fill=white] (0,0.7ex) circle (2pt) coordinate (A);
\filldraw[color=black, fill=black] (10ex,0.7ex) circle (2pt) coordinate (B);
\begin{pgfonlayer}{bg}    
        \draw (A) -- (B);
\end{pgfonlayer}
\node at (0, 2ex) {$\eta_1$};
\node at (10ex, 2ex) {$\eta_2$};
}}
\def\propagatorMM{\tikz[baseline=.1ex]{
\filldraw[color=black, fill=white] (0,0.7ex) circle (2pt) coordinate (A);
\filldraw[color=black, fill=white] (10ex,0.7ex) circle (2pt) coordinate (B);
\begin{pgfonlayer}{bg}    
        \draw (A) -- (B);
\end{pgfonlayer}
\node at (0, 2ex) {$\eta_1$};
\node at (10ex, 2ex) {$\eta_2$};
}}

\def\blackdot{\tikz[baseline=.1ex]{
\filldraw[color=black, fill=black] (0,0.7ex) circle (2pt);
}}
\def\whitedot{\tikz[baseline=.1ex]{
\filldraw[color=black, fill=white] (0,0.7ex) circle (2pt);
}}

\def\propagatorP{\tikz[baseline=.1ex]{
\filldraw[color=black, fill=black] (4ex,-0.5ex) circle (2pt) coordinate (A);
\draw[color=black, thick] (0,2.5ex) -- (10ex,2.5ex);
\begin{pgfonlayer}{bg}    
        \draw (A) -- (6ex,2.5ex);
\end{pgfonlayer}
\node at (6ex, -0.5ex) {$\eta$};
}}

\def\propagatorM{\tikz[baseline=.1ex]{
\filldraw[color=black, fill=white] (4ex,-0.5ex) circle (2pt) coordinate (A);
\draw[color=black, thick] (0,2.5ex) -- (10ex,2.5ex);
\begin{pgfonlayer}{bg}    
        \draw (A) -- (6ex,2.5ex);
\end{pgfonlayer}
\node at (6ex, -0.5ex) {$\eta$};
}}

\def\vertexP{\tikz[baseline=.1ex]{
\filldraw[color=black, fill=black] (3ex,0.75ex) circle (2pt) coordinate (A);
\begin{pgfonlayer}{bg}    
        \draw (0,-2.25ex) -- (6ex, 3.75ex);
        \draw (0, 3.75ex) -- (6ex, -2.25ex);
\end{pgfonlayer}
}}
\def\vertexM{\tikz[baseline=.1ex]{
\filldraw[color=black, fill=white] (3ex,0.75ex) circle (2pt) coordinate (A);
\begin{pgfonlayer}{bg}    
        \draw (0,-2.25ex) -- (6ex, 3.75ex);
        \draw (0, 3.75ex) -- (6ex, -2.25ex);
\end{pgfonlayer}
}}

\newcommand{\halfblackvertex}[2]{
    \begin{scope}[shift={(#1,#2)}]
        \fill[black] (0,0) -- (0,2pt) arc (90:270:2pt) -- cycle;
        \fill[color=black, fill=white] (0,0) -- (0,2pt) arc (90:-90:2pt) -- cycle;
        \draw (0,0) circle (2pt);
    \end{scope}
}

\def\bubbleSimple{%
  \tikz[baseline={(-0.08,-0.08)}]{
    \draw[color=black, thick] (0,0) ellipse (0.9cm and 0.45cm);

      \fill[black] (-0.9,0) circle (2pt);
      \fill[black] (0.9,0) circle (2pt);

    \node at (0,0.65) {\scriptsize$\nu_1$};
    \node at (0,-0.65) {\scriptsize$\nu_2$};
  }%
}

\def\propagatorBW{%
\tikz[baseline={(-0.05,-0.05)}]{
  \draw[thick] (-0.9,0) -- (0.9,0);
  \fill[black] (-0.9,0) circle (2pt);

  \fill[black] (0.9,0) circle (2pt);

  \node at (0,0.35) {\scriptsize$\nu$};
}}

\def\ThreePointDiags{\tikz[baseline={(-0.05,-0.05)}]{
  \filldraw[color=black, fill=black] (-0.4,-0.2) circle (2pt) coordinate (A);
  \filldraw[color=black, fill=black] (0.4,-0.2) circle (2pt) coordinate (B);
\draw[color=black, thick] (-0.9,0.5) -- (0.9,0.5);
\begin{pgfonlayer}{bg}    
        \draw (A) -- (-0.8,0.5);
        \draw (A) -- (-0.4,0.5);
        \draw [dashed] (A) -- (0,0.5);
        \draw (B) -- (0.8,0.5);
        \draw (B) -- (0.4,0.5);
        \draw [dashed] (B) -- (0,0.5);
\end{pgfonlayer}
\node at (0,0) {\scriptsize$\nu$};
}}

\def\ExchangeDiagSigma{\tikz[baseline={(-0.05,-0.05)}]{
  \filldraw[color=lightgray] (0,-0.2) ellipse (0.5cm and 0.2cm);
  \filldraw[color=black, fill=black] (-0.5,-0.2) circle (2pt) coordinate (A);
  \filldraw[color=black, fill=black] (0.5,-0.2) circle (2pt) coordinate (B);
\draw[color=black, thick] (-1,0.5) -- (1,0.5);
\begin{pgfonlayer}{bg}    
        \draw (A) -- (-0.9,0.5);
        \draw (A) -- (-0.5,0.5);
        \draw (B) -- (0.9,0.5);
        \draw (B) -- (0.5,0.5);
\end{pgfonlayer}
\node at (0,-0.2) {\scriptsize$\Sigma$};
}}

\def\harmonicFunction{\tikz[baseline={(-0.05,-0.05)}]{
  \filldraw[color=black, fill=black] (-0.4,-0.2) circle (2pt) coordinate (A);
  \filldraw[color=black, fill=black] (0.4,-0.2) circle (2pt) coordinate (B);
\draw[color=black, thick] (-0.6,0.5) -- (0.6,0.5);
\begin{pgfonlayer}{bg}    
        \draw [dashed] (A) -- (0,0.5);
        \draw [dashed] (B) -- (0,0.5);
\end{pgfonlayer}
\node at (0,0) {\scriptsize$\nu$};
}}

\def\fourPointContact{\tikz[baseline={(-0.05,-0.05)}]{
  \filldraw[color=black, fill=black] (0,-0.2) circle (2pt) coordinate (A);
\draw[color=black, thick] (-0.7,0.5) -- (0.7,0.5);
\begin{pgfonlayer}{bg}    
        \draw (A) -- (-0.6,0.5);
        \draw (A) -- (-0.2,0.5);
        \draw (A) -- (0.6,0.5);
        \draw (A) -- (0.2,0.5);
\end{pgfonlayer}
}}

\def\fourPointContact{\tikz[baseline={(-0.05,-0.05)}]{
  \filldraw[color=black, fill=black] (0,-0.2) circle (2pt) coordinate (A);
\draw[color=black, thick] (-0.7,0.5) -- (0.7,0.5);
\begin{pgfonlayer}{bg}    
        \draw (A) -- (-0.6,0.5);
        \draw (A) -- (-0.2,0.5);
        \draw (A) -- (0.6,0.5);
        \draw (A) -- (0.2,0.5);
\end{pgfonlayer}
}}

\def\NLoopBubble{\tikz[baseline={(-0.05,-0.05)}]{
  \draw (-1.5,-0.2) ellipse (0.5cm and 0.3cm);
  \draw (-0.5,-0.2) ellipse (0.5cm and 0.3cm);
  \draw (1.5,-0.2) ellipse (0.5cm and 0.3cm);
  \halfblackvertex{-2.0}{-0.2};
  \halfblackvertex{-1.0}{-0.2};
  \halfblackvertex{0}{-0.2};
  \halfblackvertex{1.0}{-0.2};
  \halfblackvertex{2.0}{-0.2};
\draw[color=black, thick] (-2.5,0.5) -- (2.5,0.5);
\begin{pgfonlayer}{bg}    
        \draw (-2.0,-0.2) -- (-2.4,0.5);
        \draw (-2.0,-0.2) -- (-2.0,0.5);
        \draw (2.0,-0.2) -- (2.0,0.5);
        \draw (2.0,-0.2) -- (2.4,0.5);
\end{pgfonlayer}
\node at (0.5,-0.2) {$\dots$};
}}

\def\NChain{\tikz[baseline={(-0.05,-0.05)}]{
  \draw (-2.0,-0.2) --(-1.0, -0.2);
  \draw (-1.0,-0.2) -- (0, -0.2);
  \draw (1.0,-0.2) -- (2.0,-0.2);
  \halfblackvertex{-2.0}{-0.2};
  \halfblackvertex{-1.0}{-0.2};
  \halfblackvertex{0}{-0.2};
  \halfblackvertex{1.0}{-0.2};
  \halfblackvertex{2.0}{-0.2};
\draw[color=black, thick] (-2.5,0.5) -- (2.5,0.5);
\begin{pgfonlayer}{bg}    
        \draw (-2.0,-0.2) -- (-2.4,0.5);
        \draw (-2.0,-0.2) -- (-2.0,0.5);
        \draw (2.0,-0.2) -- (2.0,0.5);
        \draw (2.0,-0.2) -- (2.4,0.5);
\end{pgfonlayer}
\node at (0.5,-0.2) {$\dots$};
\node at (-1.5,0.1) {\scriptsize$\nu_1$};
\node at (-0.5,0.1) {\scriptsize$\nu_2$};
\node at (1.5,0.1) {\scriptsize$\nu_N$};
}}

\def\NLoopBubbleSplit{\tikz[baseline={(-0.05,-0.05)}]{
  \draw (-2.3,-0.2) ellipse (0.5cm and 0.3cm);
  \draw (-0.5,-0.2) ellipse (0.5cm and 0.3cm);
  \draw (1.5,-0.2) ellipse (0.5cm and 0.3cm);
  \halfblackvertex{-3.6}{-0.2};
  \halfblackvertex{-1.0}{-0.2};
  \halfblackvertex{0}{-0.2};
  \halfblackvertex{1.0}{-0.2};
  \halfblackvertex{2.8}{-0.2};
  \halfblackvertex{2.0}{-0.2};
  \halfblackvertex{-2.8}{-0.2};
  \halfblackvertex{-1.8}{-0.2};
\draw[color=black, thick] (-4.1,0.5) -- (3.3,0.5);
\begin{pgfonlayer}{bg}    
        \draw (-3.6,-0.2) -- (-4.0,0.5);
        \draw (-3.6,-0.2) -- (-3.6,0.5);
        \draw (2.8,-0.2) -- (2.8,0.5);
        \draw (2.8,-0.2) -- (3.2,0.5);
        \draw [dashed] (-3.6,-0.2) -- (-3.2,0.5);
        \draw [dashed] (-2.8,-0.2) -- (-3.2,0.5);
        \draw [dashed] (-1.8,-0.2) -- (-1.4,0.5);
        \draw [dashed] (-1.0,-0.2) -- (-1.4,0.5);
        \draw [dashed] (2.8,-0.2) -- (2.4,0.5);
        \draw [dashed] (2.0,-0.2) -- (2.4,0.5);
\end{pgfonlayer}
\node at (0.5,-0.2) {$\dots$};
\node at (-3.2,-0.4) {\scriptsize$\nu_1$};
\node at (-1.4,-0.4) {\scriptsize$\nu_2$};
\node at (2.5,-0.4) {\scriptsize$\nu_{N+1}$};
}}

\def\BubbleBuildingBlock{\tikz[baseline={(-0.05,-0.05)}]{
  \draw (0,-0.2) ellipse (0.5cm and 0.3cm);
  \halfblackvertex{-0.5}{-0.2}; 
  \halfblackvertex{0.5}{-0.2};
\draw[color=black, thick] (-1,0.5) -- (1,0.5);
\begin{pgfonlayer}{bg}    
        \draw[dashed] (-0.5,-0.2) -- (-0.9,0.5);
        \draw[dashed] (0.5,-0.2) -- (0.9,0.5);
\end{pgfonlayer}
\node at (-0.7,-0.5) {\scriptsize$a_j$};
\node at (0.8,-0.5) {\scriptsize$a_{j+1}$};
\node at (0,-0.2) {\scriptsize$\nu$};
\node at (-1.0,0.1) {\scriptsize$\nu_j$};
\node at (1.2,0.1) {\scriptsize$\nu_{j+1}$};
}}

\def\TreeBuildingBlock{\tikz[baseline={(-0.05,-0.05)}]{
  \draw (-0.5,-0.2) -- (0.5,-0.2);
  \halfblackvertex{-0.5}{-0.2}; 
  \halfblackvertex{0.5}{-0.2};
\draw[color=black, thick] (-1,0.5) -- (1,0.5);
\begin{pgfonlayer}{bg}    
        \draw[dashed] (-0.5,-0.2) -- (-0.9,0.5);
        \draw[dashed] (0.5,-0.2) -- (0.9,0.5);
\end{pgfonlayer}
\node at (-0.7,-0.5) {\scriptsize$a_j$};
\node at (0.8,-0.5) {\scriptsize$a_{j+1}$};
\node at (0,0.1) {\scriptsize$\nu'$};
\node at (-1.0,0.1) {\scriptsize$\nu_j$};
\node at (1.2,0.1) {\scriptsize$\nu_{j+1}$};
}}

\def\TreeBuildingBlockPP{\tikz[baseline={(-0.05,-0.05)}]{
  \draw (-0.5,-0.2) -- (0.5,-0.2);
  \filldraw[color=black, fill=black] (-0.5,-0.2) circle (2pt);
  \filldraw[color=black, fill=white] (0.5,-0.2) circle (2pt);
\draw[color=black, thick] (-1,0.5) -- (1,0.5);
\begin{pgfonlayer}{bg}    
        \draw[dashed] (-0.5,-0.2) -- (-0.9,0.5);
        \draw[dashed] (0.5,-0.2) -- (0.9,0.5);
\end{pgfonlayer}
\node at (0,0.1) {\scriptsize$\nu'$};
\node at (-1.0,0.1) {\scriptsize$\nu_j$};
\node at (1.2,0.1) {\scriptsize$\nu_{j+1}$};
}}

\def\TreeBuildingBlockFactorPP{\tikz[baseline={(-0.05,-0.05)}]{
  \filldraw[color=black, fill=black] (-0.4,-0.2) circle (2pt);
  \filldraw[color=black, fill=white] (0.4,-0.2) circle (2pt);
\draw[color=black, thick] (-0.9,0.5) -- (0.9,0.5);
\begin{pgfonlayer}{bg}    
        \draw[dashed] (-0.4,-0.2) -- (-0.8,0.5);
        \draw[dashed] (0.4,-0.2) -- (0.8,0.5);
        \draw[dashed] (0.4,-0.2) -- (0,0.5);
        \draw[dashed] (-0.4,-0.2) -- (0,0.5);
\end{pgfonlayer}
\node at (0,0) {\scriptsize$\nu'$};
\node at (-0.9,0.1) {\scriptsize$\nu_j$};
\node at (1.1,0.1) {\scriptsize$\nu_{j+1}$};
}}

\def\TreeBuildingBlockMM{\tikz[baseline={(-0.05,-0.05)}]{
  \draw (-0.5,-0.2) -- (0.5,-0.2);
  \filldraw[color=black, fill=white] (-0.5,-0.2) circle (2pt);
  \filldraw[color=black, fill=black] (0.5,-0.2) circle (2pt);
\draw[color=black, thick] (-1,0.5) -- (1,0.5);
\begin{pgfonlayer}{bg}    
        \draw[dashed] (-0.5,-0.2) -- (-0.9,0.5);
        \draw[dashed] (0.5,-0.2) -- (0.9,0.5);
\end{pgfonlayer}
\node at (0,0.1) {\scriptsize$\nu'$};
\node at (-1.0,0.1) {\scriptsize$\nu_j$};
\node at (1.2,0.1) {\scriptsize$\nu_{j+1}$};
}}

\def\TreeBuildingBlockFactorMM{\tikz[baseline={(-0.05,-0.05)}]{
  \filldraw[color=black, fill=white] (-0.4,-0.2) circle (2pt);
  \filldraw[color=black, fill=black] (0.4,-0.2) circle (2pt);
\draw[color=black, thick] (-0.9,0.5) -- (0.9,0.5);
\begin{pgfonlayer}{bg}    
        \draw[dashed] (-0.4,-0.2) -- (-0.8,0.5);
        \draw[dashed] (0.4,-0.2) -- (0.8,0.5);
        \draw[dashed] (0.4,-0.2) -- (0,0.5);
        \draw[dashed] (-0.4,-0.2) -- (0,0.5);
\end{pgfonlayer}
\node at (0,0) {\scriptsize$\nu'$};
\node at (-0.9,0.1) {\scriptsize$\nu_j$};
\node at (1.1,0.1) {\scriptsize$\nu_{j+1}$};
}}

\def\propagatorPPnu{\tikz[baseline={(-0.05,-0.05)}]{
  \draw (-0.4,0.0) -- (0.4,0.0);
  \filldraw[color=black, fill=black] (-0.4,0.0) circle (2pt);
  \filldraw[color=black, fill=black] (0.4,0.0) circle (2pt);
\node at (0,0.3) {\scriptsize$\nu'$};
}}

\def\propagatorPPnuFactor{\tikz[baseline={(-0.05,-0.05)}]{
  \filldraw[color=black, fill=black] (-0.4,-0.2) circle (2pt);
  \filldraw[color=black, fill=black] (0.4,-0.2) circle (2pt);
\draw[color=black, thick] (-0.5,0.5) -- (0.5,0.5);
\begin{pgfonlayer}{bg}    
        \draw[dashed] (0.4,-0.2) -- (0,0.5);
        \draw[dashed] (-0.4,-0.2) -- (0,0.5);
\end{pgfonlayer}
\node at (0,0) {\scriptsize$\mu$};
}}

\def\BubbleBuildingBlockNoText{\tikz[baseline={(-0.05,-0.05)}]{
  \draw (0,-0.2) ellipse (0.5cm and 0.3cm);
  \halfblackvertex{-0.5}{-0.2}; 
  \halfblackvertex{0.5}{-0.2};
\draw[color=black, thick] (-1,0.5) -- (1,0.5);
\begin{pgfonlayer}{bg}    
        \draw (-0.5,-0.2) -- (-0.9,0.5);
        \draw (0.5,-0.2) -- (0.9,0.5);
        \draw (-0.5,-0.2) -- (-0.5,0.5);
        \draw (0.5,-0.2) -- (0.5,0.5);
\end{pgfonlayer}
}}

\def\counterterm{\tikz[baseline={(-0.05,-0.05)}]{
  \filldraw[color=black, fill=white] (0,-0.2) circle (5pt);
  \draw[line width=0.02cm] (-0.13, -0.33) -- (0.13, -0.07);
  \draw[line width=0.02cm] (-0.13, -0.07) -- (0.13, -0.33);
\draw[color=black, thick] (-0.7,0.5) -- (0.7,0.5);
\begin{pgfonlayer}{bg}    
        \draw (-0,-0.2) -- (-0.2,0.5);
        \draw (0,-0.2) -- (0.2,0.5);
        \draw (0,-0.2) -- (-0.6,0.5);
        \draw (0,-0.2) -- (0.6,0.5);
\end{pgfonlayer}
}}

\def\RenormalizedBubbles{\tikz[baseline={(-0.05,-0.05)}]{
  \draw (0.7,-0.2) ellipse (0.5cm and 0.3cm);
  \draw (4.1,-0.2) ellipse (0.5cm and 0.3cm);
  \halfblackvertex{-0.2}{-0.2}; 
  \halfblackvertex{6.1}{-0.2};
  \node at (0.1,-0.2) {\scalebox{2}{(}};
  \node at (1.6, -0.2) {+};
  \filldraw[color=black, fill=white] (2.1,-0.2) circle (5pt);
  \draw[line width=0.02cm] (1.97, -0.33) -- (2.23, -0.07);
  \draw[line width=0.02cm] (1.97, -0.07) -- (2.23, -0.33);
  \node at (2.4,-0.2) {\scalebox{2}{)}};
  \node at (3.0, -0.2) {$\dots$};
  \node at (3.5,-0.2) {\scalebox{2}{(}};
  \node at (5.0, -0.2) {+};
  \filldraw[color=black, fill=white] (5.5,-0.2) circle (5pt);
  \draw[line width=0.02cm] (5.37, -0.33) -- (5.63, -0.07);
  \draw[line width=0.02cm] (5.37, -0.07) -- (5.63, -0.33);
  \node at (5.8,-0.2) {\scalebox{2}{)}};
\draw[color=black, thick] (-0.7,0.5) -- (6.6,0.5);
\begin{pgfonlayer}{bg}    
        \draw (-0.2,-0.2) -- (-0.2,0.5);
        \draw (-0.2,-0.2) -- (-0.6,0.5);
        \draw (6.1,-0.2) -- (6.1,0.5);
        \draw (6.1,-0.2) -- (6.5,0.5);
\end{pgfonlayer}
}}

\title{\boldmath Split Representations and Bubble Resummation for Massive de Sitter Correlators}

\author[a,b]{Jonathan Gräfe,}
\author[c]{Ivo Sachs}
\affiliation[a]{Max-Planck-Institut für Physik, Werner-Heisenberg-Institut,\\
Boltzmannstr. 8, 85748 Garching, Germany}
\affiliation[b]{Max Planck-IAS-NTU Center for Particle Physics, Cosmology and Geometry}
\affiliation[c]{Arnold-Sommerfeld-Center for Theoretical Physics, Ludwig-Maximilians-Universität München,\\
Theresienstr. 37, 80333 Munich, Germany}

\emailAdd{graefe@mpp.mpg.de}

\abstract{We combine spectral- and split representations to factorize multi-loop momentum space diagrams, in the Schwinger-Keldysh formulation for cosmological correlators, with massive scalars in the loop. This allows us to extend the resummation of loop contributions from flat to de Sitter space. Furthermore, in our split representation the signal part of the correlators can be identified directly on the integrand level from the spectral function. We apply this to  describe the non-perturbative flow of the EFT background and the cosmological collider signals in a large-$N$ model. }

\begin{document}
\maketitle
\flushbottom
\section{Introduction}
Measurements of cosmological correlation functions imprinted into the inhomogeneities and anisotropies of, for example, the Cosmic Microwave Background \cite{plancks:2018in, plancks:2018ng,Hinshaw_2013} or large-scale structures \cite{ferraro2022snowmass2021cosmicfrontierwhite} open a window to the physics of the very early universe. It is widely believed that the origin of the underlying density perturbations stems from a period of inflation at primordial times even before the hot big bang \cite{Starobinsky:1980te,Guth:1980zm,Linde:1981mu}. Although the details of this inflationary expansion are still uncertain, it provides an elegant mechanism to create late-time inhomogeneities from quantum vacuum fluctuations \cite{Mukhanov:1981xt,Achucarro:2022qrl}. \\
In these models, the observable late-time correlation functions can be directly traced back to correlation functions of some quantum fields on the inflationary background \cite{Maldacena_2003}. These could be scalar isocurvature modes or even tensor modes of metric fluctuations (measurable, among other things, by B-mode polarizations of the CMB \cite{PhysRevLett.112.241101, 1997ApJ...482....6S}). To bridge the gap between the observable data and the theoretical models, it is therefore crucial to understand quantum field theories on an expanding background. In most models that can be aligned with the cosmological data the space-time of inflation is very close to the Poincaré patch of a de Sitter space-time. De Sitter space is a maximally symmetric Lorentzian manifold, meaning it has a constant Ricci scalar. While quantum field theories in Minkowski or anti-de Sitter space have been well-studied for decades and an extensive literature of perturbative and non-perturbative results for correlators are available, quantum field theories in de Sitter space are still very much at the beginning. Nonetheless, it is of high interest with regard to its cosmological applications. 

One major discovery that sparked this interest was the realization that cosmological correlators can not only be used to study properties of inflaton fields and metric perturbations but that they offer a way to study particle phenomenologies at energy scales much higher than the available scales for experiments on earth \cite{PhysRevD.81.063511,XingangChen_2010,PhysRevD.85.103520,XingangChen_2012,ShiPi_2012, Noumi_2013,Jinn-OukGong_2013,arkanihamed2015cosmologicalcolliderphysics}. In particular, it was found that heavy particles that were present during the period of inflation leave imprints on cosmological correlators in the form of oscillatory signals. This observation gave rise to the term \textit{cosmological collider physics} \cite{arkanihamed2015cosmologicalcolliderphysics}. Therefore, the study of cosmological correlators could potentially unravel the particle contents of the primordial universe as well as several properties of interest like masses, spins and interaction types. 

The perturbative computation of correlators on a de Sitter background itself turns out to be quite challenging. The arising Feynman integrals have a much more complicated structure than their flat-space analogs due to the lack of time-translation symmetry (introducing time-ordering Heaviside functions in the integrands and leading to nested time integrals) as well as the rather complex structure of the mode functions themselves (consisting of Hankel functions). Only in a few special cases the computation of cosmological correlators has advanced to the multi-loop level. For conformally coupled scalars (for which the mass is $m^2=2H^2$, $H$ being the Hubble rate) the mode functions reduce to flat-space mode functions up to an overall scaling with the scale factor of the background metric. This leads to integrals very similar to flat space integrals that have been studied extensively in the literature and have led to the development of a large variety of techniques\cite{Arkani-Hamed:CosmoPolytopes,Arkani-Hamed:2018bjr,Benincasa:2019vqr,Benincasa:2022gtd,Benincasa:2018ssx,Hillman:2019wgh, Goodhew:2020hob, Melville:2021lst, Goodhew:2021oqg,Meltzer:2020qbr, Arkani-Hamed:2023kig,Baumann:2024mvm, Arkani-Hamed:2024jbp}. Another special case that has been broadly studied is the case of massless scalar fields, and in particular the arising infrared divergences of the correlators \cite{PhysRevD.32.3136,PhysRevD.82.123522,Beneke:2012kn,Hollands2011,beneke2023cosmological}. One of the main motivations for this is the fact, that only fluctuations of massless scalars and massless spin-2 fields are not getting suppressed during the expansion of the background and survive the late-time limit. Also, the mode functions of massless scalars are not too different from the conformally coupled case. When it comes to massive fields (especially massive fields in the bulk are of interest to phenomenologists) however, only very few full analytic results have been obtained yet (mostly at tree-level) \cite{Sleight:2019mgd,Pacifico:2024dyo,Sleight:2019hfp,Qin:mellinBarnesBootstrapEq, Bootstrap:fromSymSing,Bootstrap:weightShift,Bootstrap:spinningCorr,Bootstrap:snowmass,Bootstrap:nonpert,Bootstrap:boostless,Bootstrap:localityUnitarity,Bootstrap:linkingSing, Ema:2024hkj,Qin:2023bjk, Liu:2024xyi, Xianyu:2025lbk, Fan:2025scu, Liu:2024str}.  

Concerning massive bulk fields at loop order, there is currently only one full analytic result for a one-loop diagram available \cite{Xianyu:2022jwk,Liu:2024xyi,Qin:2024gtr}. While many cosmological collider signals are expected to appear only at loop order \cite{Chen:2016uwp,Chen:2016hrz,Chen:2016nrs,Chen:2018xck,Hook:2019vcn}, for example, due to pair production effects or charge conservation, the calculation of these diagrams is even more difficult than the tree-level diagrams due to the additional loop integral. It is also not clear at the moment, how some of the developed bootstrap methods can be modified to be applicable in this case as well. Recently, progress was made by using results for bubble loops in Euclidean de Sitter space (which is basically a sphere) in order to compute the one-loop bubble diagram in de Sitter \cite{Xianyu:2022jwk}. The key idea was to utilize a spectral representation of the bubble loop integral \cite{Marolf:2010zp} and to rewrite the integral over the loop momentum as a spectral integral. In this way, the one-loop result could be related to the result for a simpler tree-level diagram via a spectral integral. The final integration was then carried out with the help of the residue theorem, which enabled the derivation of full analytic expressions for the bubble diagram. Additionally, the expected UV divergence of the bubble loop was discovered and with the help of the modified minimal subtraction scheme, the one-loop bubble could be regulated. 

While perturbation theory in de Sitter space is already challenging, we should keep non-perturbative effects in mind as well since they play an important role for questions including vacuum stability \cite{DiPietro:2023inn,Sachs:2023eph} or infrared singularities, see for instance \cite{Starobinsky:1982ee,Beneke:2012kn,Gorbenko:2019rza,Cespedes:2023aal}. An important class of non-perturbative contributions comes from the resummation of perturbative loop diagrams that give rise to resummed propagators of fundamental (e.g. \cite{Itzykson:1980rh} ) and composite fields in large-$N$ models \cite{Moshe:2003xn}. To date, very little is known about resummed loops in curved space time beyond three dimensions \cite{DiPietro:2023inn} and conformally coupled fields \cite{Sachs:2023eph, Nowinski:2025cvw}. One reason for this scarcity is that the usual flat space methods for resummation do not apply in curved space. For one thing, energy is not conserved in an expanding universe, and for another the integration over the different Schwinger-Keldysh contours leads to combinatorial artifacts that hide expected factorization properties from flat space analogues. 

In this paper we develop a spectral method to compute \emph{and} resum scalar loops of massive fields in de Sitter momentum space. A familiar spectral method in flat space is the Källén-Lehman representation which expresses loop diagrams in terms of a spectral integral over a simpler tree-level diagram. Schematically, one can write 
\begin{equation}
    \bubbleSimple = \int_{-\infty}^{+\infty}\mathrm{d}\tilde{\nu}\,\rho_{\nu_1\nu_2}(\tilde{\nu})\,\,\, \propagatorBW.
\end{equation}
For massive scalar one-loop diagrams, this has been adapted to de Sitter space in \cite{Xianyu:2022jwk,Zhang:2025nzd}. However, it turns out that the (although straightforward) generalization to higher loops involves tedious prior computations of tree-level diagrams that we would like to circumvent. Instead, we will make use of a different spectral representation where the tree-level diagram is further reduced to a product over disconnected diagrams using a \emph{split} representation \cite{Penedones:2010ue,Sleight:2019hfp,Sleight:2020obc,Loparco:2023rug}.
\begin{equation}
    \ExchangeDiagSigma = \int_{-\infty}^{+\infty}\mathrm{d}\tilde{\nu}\,\Sigma(\tilde{\nu})\,\,\,\ThreePointDiags
\end{equation}

We will show that with this representation scalar loops in de Sitter can be resummed much like in flat space. As a consequence, we can then read off the  background and cosmological signal from this non-perturbative contribution. In fact, as we will see our (non-perturbative) spectral function provides directly the frequencies (and amplitudes) of the cosmological signals without having to perform the tedious integral over the spectral parameter. In other words, our spectral function directly provides the frequency analysis of the signal, which is the object of interest in the cosmological collider program. \\
Concretely we obtain the resummed spectral function for a chain of massive scalar loops interacting with conformally coupled bulk-to-boundary propagators. Replacing the external legs by massive scalars does not pose a conceptual problem but we do not see a strong cosmological motivation for it (the transition to massless external legs can be achieved with the help of weight-shifting operators \cite{Bootstrap:weightShift, Benincasa:2019vqr}). For the  $d=2$ case, we can explicitly describe the non-perturbative flow of the cosmological signals. In the process, we infer the compactness of the coupling space and establish the completeness of the principal series for the corresponding spectral decomposition. For a four-dimensional de Sitter space ($d=3$) we similarly obtain the non-perturbatively renormalized spectral function. Unlike for $d=2$ we have not found a closed expression for the latter but we are still able to discuss the corresponding non-perturbative flow and some features of the signal. \\ 

The outline of the paper is as follows:  We begin in section \ref{ref:preliminaries} with a brief review of the Schwinger-Keldysh formalism to compute in-in correlators on a cosmological background, introducing necessary notations and diagrammatics. There we also briefly review a class of large-$N$ models and motivate the non-perturbative computations we are about to carry out. In section \ref{sec:split} we derive the split representation starting from an integral identity for Hankel functions and apply it as a test to a simple 4-point contact diagram, recovering a set of intriguing identities for sums of hypergeometric functions. In section \ref{sec:bubble-chains} we use the split representation in combination with the Källén-Lehmann spectral representation for bubble loops in order to find a factorized expression for the spectral function of a chain of bubble loops. We then  discuss renormalization in dimensional regularization and in the special case $d=2$, we find a closed-form expression for the spectral function. We first rederive the one-loop result using our method and then discuss the procedure of resummation and its effect on the pole structure of the spectral function. In $d=2$ we give explicit numerical results for this pole flow. Some mathematical details are relegated to the appendix.

\paragraph{Notations and conventions} We work in a $d+1$ dimensional de Sitter space dS$_{d+1}$ with metric $\mathrm{d}s^2 = (\eta H)^{-2}(-\mathrm{d}\eta^2 +\mathrm{d}\mathbfit{x}^2)$ in conformal coordinates $(\eta,\mathbfit{x})$ where $\eta\in(-\infty,0)$, $\mathbfit{x}\in\mathbb{R}^d$ and the signature is $(-+,\dots,+)$. The Hubble parameter $H$ is set to $H\equiv1$ throughout this paper. To characterize the kinematic dependence of a four-point exchange correlator, we define the energies $k_i \equiv |\mathbfit{k}_i|$, where $\mathbfit{k}_i$ are the external momenta. We will further use the shorthand notations $k_{ij} \equiv k_i + k_j$ and $k_{1234} = \sum_{i=1}^4 k_i$ for sums of energies. The magnitude of the exchanged momentum in the s-channel is defined as $k_s \equiv |\mathbfit{k}_1+\mathbfit{k}_2| = |\mathbfit{k}_3+\mathbfit{k}_4|$. The results will depend only on the conformal momentum ratios $r_1\equiv k_s/k_{12}$ and $r_2 \equiv k_s/ k_{34}$. 

Throughout this paper we focus on principal series scalars with spin $l=0$ and scaling dimension $\Delta = \frac{d}{2}+\nu$ where $\nu=i \tilde{\nu}$ is imaginary, i.e. $\tilde{\nu}\in\mathbb{R}$. 

\section{Preliminaries}\label{ref:preliminaries}

\subsection{Schwinger-Keldysh formalism}

In contrast to transition amplitudes in particle physics, in cosmology we are usually interested in equal-time in-in correlation functions, i.e. expectation values
\begin{equation}
    \langle \Omega\vert \phi^{a_1}(\eta,\mathbfit{x}_1)\dots \phi^{a_N}(\eta,\mathbfit{x}_N)\vert\Omega\rangle
\end{equation}
of certain field operators $\phi^{a}$ (where $a=\pm$ is some index characterizing the type of field) evaluated on an initial vacuum state $\Omega$ (we choose the standard Bunch-Davies vacuum conditions). The Schwinger-Keldysh formalism allows us to express this correlation function in terms of a path integral. To this end, one has to introduce two copies $\phi_\pm$ for each field in the theory which are integrated over the two branches of the in-in contour and sewn together at the final time-slice $\eta_*$. The (+) branch therefore captures the forward time evolution from some initial time $\eta_0$ to the final time $\eta_*$, while the ($-$) contour represents the backward time evolution. The sewing condition then reads $\phi_+(\eta_*,\mathbfit{x}) =\phi_-(\eta_*,\mathbfit{x})$. Hence, if we evaluate the correlator on the final time slice $\eta_*$, it should not matter how we choose the indices of the fields in the correlator.

For a given theory with classical Lagrangian $\mathcal{L}$ one defines the generating functional
\begin{equation}
    Z[J_+,J_-] = \int\mathcal{D}[\phi_+, \phi_-] e^{i\int_{\eta_0}^{\eta_*} \mathrm{d}\eta \int \mathrm{d}^3\mathbfit{x}\, (\mathcal{L}[\phi_+] -\mathcal{L}[\phi_-] + J_+ \phi_+ - J_- \phi_-)}
\end{equation}
and by taking functional derivatives, one recovers the desired correlator:
\begin{align} \label{eq:expectation-from-genfunc}
    \langle \phi^{a_1}(\eta,\mathbfit{x}_1)\dots\phi^{a_N}(\eta,\mathbfit{x}_N)\rangle = (-i a_1)\frac{\delta}{\delta J_{a_1}(\eta,\mathbfit{x}_1)}\cdots (-i a_N)\frac{\delta}{\delta J_{a_N}(\eta,\mathbfit{x}_N)} Z[J_+, J_-] \Big\vert_{J_\pm = 0}.
\end{align}
If we are dealing with an interacting theory, we usually treat the interactions as perturbations. So, we start by splitting the Lagrangian $\mathcal{L} = \mathcal{L}_0 + \mathcal{L}_{\text{int}}$ into the free (quadratic) part $\mathcal{L}_0$ and the interacting part $\mathcal{L}_{\text{int}}$. By introducing the generating functional of the free theory
\begin{equation}
    Z_0[J_+,J_-] = \int\mathcal{D}[\phi_+, \phi_-] e^{i\int_{\eta_0}^{\eta_*} \mathrm{d}\eta \int \mathrm{d}^3\mathbfit{x}\, (\mathcal{L}_0 [\phi_+] -\mathcal{L}_0[\phi_-] + J_+ \phi_+ - J_- \phi_-)},
\end{equation}
we can rewrite $Z[J_+,J_-]$ as
\begin{equation}\label{eq:interacting-genfunc}
    Z[J_+, J_-] = \exp{\left\{i\int_{\eta_0}^{\eta_*}\mathrm{d}\eta \int \mathrm{d}^3 \mathbfit{x} \,\left(\mathcal{L}_{\text{int}}\left[\frac{1}{i}\frac{\delta}{\delta J_+}\right] - \mathcal{L}_{\text{int}}\left[\frac{1}{i}\frac{\delta}{\delta J_-}\right]\right) \right\}}Z_0[J_+, J_-]. 
\end{equation}
From the free generating functional $Z_0$, which is just a Gaussian integral and can be solved explicitly, one obtains the tree-level propagators of the theory. Then, by plugging \eqref{eq:interacting-genfunc} into \eqref{eq:expectation-from-genfunc} one can calculate correlation functions to arbitrary order in the perturbation expansion. \\
\\
\noindent
We start the discussion of diagrammatic rules by reviewing the tree-level propagators in this theory. Since there are two types of fields in the path integral, there are in total four different propagators defined as the two-point functions
\begin{equation}
    -i\Delta_{ab} (\eta_1,\mathbfit{x}_1;\eta_2,\mathbfit{x}_2) \equiv \frac{1}{ia}\frac{\delta}{\delta J_a (\eta_1,\mathbfit{x}_1)} \frac{1}{ib}\frac{\delta}{\delta J_b(\eta_2,\mathbfit{x}_2)} Z_0[J_+,J_-]\Big\vert_{J_\pm = 0}
\end{equation}
where $a,b=\pm$. Using the underlying path integral formalism, we can rewrite these propagators in terms of Green's and Wightman functions:
\begin{align}
    -i\Delta_{++}(\eta_1,\mathbfit{x}_1;\eta_2,\mathbfit{x}_2) &= \langle 0 |T\{\phi(\eta_1,\mathbfit{x}_1)\phi(\eta_2,\mathbfit{x}_2)\}|0\rangle, \\
    -i\Delta_{--}(\eta_1,\mathbfit{x}_1;\eta_2,\mathbfit{x}_2) 
    &= \langle 0 |\bar{T}\{\phi(\eta_1,\mathbfit{x}_1)\phi(\eta_2,\mathbfit{x}_2)\}|0\rangle, \\
    -i\Delta_{+-}(\eta_1,\mathbfit{x}_1;\eta_2,\mathbfit{x}_2) 
    &= \langle 0 |\phi(\eta_2,\mathbfit{x}_2)\phi(\eta_1,\mathbfit{x}_1)|0\rangle, \\
    -i\Delta_{-+}(\eta_1,\mathbfit{x}_1;\eta_2,\mathbfit{x}_2) 
    &= \langle 0 |\phi(\eta_1,\mathbfit{x}_1)\phi(\eta_2,\mathbfit{x}_2)|0\rangle .
\end{align}
Here, $\vert 0\rangle$ is the vacuum state of the free theory.

In cosmology, calculations are typically carried out in momentum space, thereby  providing the input data for the subsequent non-linear evolution leading to the relevant observables for the cosmic microwave background  and large scale structure \cite{Achucarro:2022qrl}. Therefore, we use the mode expansion of the field operators,
\begin{equation}
    \phi(\eta,\mathbfit{x}) = \int\frac{\mathrm{d}^3\mathbfit{k}}{(2\pi)^3} \left\{u_\nu(k;\eta) \hat{b}_{\mathbfit{k}} + u_\nu^*(k;\eta) \hat{b}_{-\mathbfit{k}}^\dagger\right\}e^{i\mathbfit{k}\cdot\mathbfit{x}}
\end{equation}
to rewrite all the propagators in momentum space. Here, the functions $u_\nu(\eta;k)$ are the mode functions of the free field, i.e. solutions to the Klein-Gordon equation in de Sitter space,
\begin{equation}
    \left[\eta^2\partial_\eta^2 -(d-1)\eta\partial_\eta +k^2\eta^2+\frac{d^2}{4} -\nu^2\right]u_\nu(\eta;k) =0,
\end{equation}
where $\nu = \sqrt{d^2/4 -m^2/H^2}$.
Expressing the field operators in terms of these mode functions and introducing the Fourier transformed propagators of the free theory,
\begin{equation}
    G_{ab}(k;\eta_1,\eta_2) \equiv -i \int \mathrm{d}^3\mathbfit{x}\, e^{-i\mathbfit{k}\cdot\mathbfit{x}} \Delta_{ab}(\eta_1,\mathbfit{x};\eta_2,0),
\end{equation}
we find
\begin{align}
    G_{-+}(k;\eta_1,\eta_2) &= u_\nu(\eta_1,k)u_\nu^*(\eta_2,k),\\
    G_{+-}(k;\eta_1,\eta_2)
    &= u_\nu^*(\eta_1,k)u_\nu(\eta_2,k).
\end{align}
Here we stress that the propagators as well as the mode functions themselves only depend on the absolute value $k = |\mathbfit{k}|$ of the momentum, due to rotational symmetry. 
By taking the time-ordering into account, we get the remaining two propagators:
\begin{align}
    G_{++}(k;\eta_1,\eta_2) &= G_{-+}(k;\eta_1,\eta_2)\theta(\eta_1-\eta_2) + G_{+-}(k;\eta_1,\eta_2)\theta(\eta_2-\eta_1), \\
    G_{--}(k;\eta_1,\eta_2) &= G_{+-}(k;\eta_1,\eta_2)\theta(\eta_1-\eta_2) + G_{-+}(k;\eta_1,\eta_2)\theta(\eta_2-\eta_1).
\end{align}
Only three of the four propagators are linearly independent and they satisfy the complex conjugate relations $G_{+-}=G_{-+}^*$ as well as $G_{--}=G_{++}^*$.

Due to the possible sign choices, whenever we draw Feynman diagrams we need to assign a sign $a=\pm$ to each vertex and then draw the respective propagator $G_{ab}$:
\begin{align}
    \propagatorPP\,&= \,G_{++}(k;\eta_1,\eta_2) \nonumber\\
    \propagatorPM\,&= \,G_{+-}(k;\eta_1,\eta_2) \nonumber\\
    \propagatorMP\,&= \,G_{-+}(k;\eta_1,\eta_2) \nonumber\\
    \propagatorMM\,&= \,G_{--}(k;\eta_1,\eta_2)\,. 
\end{align}
A black dot ($\blackdot$) in a diagram corresponds to a $+$ vertex, while a white dot ($\whitedot$) corresponds to a $-$ vertex. \\
Ultimately, we are interested in late-time correlators, i.e. we want to consider diagrams for which the external legs are attached to the boundary $\eta \to \eta_*$. We therefore define a set of \textit{bulk-to-boundary propagators} in addition to the above \textit{bulk-to-bulk propagators} by taking the late-time limit of one of the coordinates. On the boundary itself, we can make use of the boundary condition $\phi_+(\eta_*) = \phi_-(\eta_*)$. Hence the bulk-to-boundary propagators only depend on the sign of the bulk vertex and therefore there are only two different such propagators:
\begin{align}
    \propagatorP\,&= \, G_+(k;\eta) \equiv G_{++}(k;\eta,\eta_f),\nonumber\\
    \propagatorM\,&= \, G_-(k;\eta) \equiv G_{-+}(k;\eta,\eta_f).
\end{align}
In the chosen basis, writing down Feynman rules for vertices is particularly easy: The two path integrals contribute vertices (\blackdot) and (\whitedot) that come with a different sign due to the sign difference in the action for the two types of fields. Deriving the vertex rules is straightforward and pretty much the same as in the usual flat-space context. The only difference is that we Fourier transform only the position coordinates but not the conformal time. Thus, the vertex rules look like a mix of position and momentum space Feynman rules. \\
As an example, let us consider a $\phi^4$-theory with interaction Lagrangian $\mathcal{L}_{\text{int}} = \frac{\lambda}{4!} a(\eta)^4 \phi^4$. The time-dependent scale factor stems from the metric determinant $\sqrt{-g}$ in the volume element of the action. Then we get the following vertex rules:
\begin{align}
    \vertexP\,&= +i \lambda \int_{\eta_0}^{\eta_f} \mathrm{d}\eta\, a(\eta)^4 \dots, \nonumber \\
    \vertexM\,&= -i \lambda \int_{\eta_0}^{\eta_f} \mathrm{d}\eta\, a(\eta)^4 \dots.
\end{align}
The dots $\dots$ contain all the $\eta$-dependent terms from the propagators connected to the vertex. In that sense, at each vertex one has to integrate over all the possible times at which the interaction could have taken place. \\
In this work, we only consider couplings of the bulk fields without derivatives. However, dealing with derivative interactions is also straightforward. To find a more detailed derivation of diagrammatic rules in the Schwinger-Keldysh formalism, see \cite{Chen:2017ryl}.

\subsection{Resummation and large $N$ models}
There are at least two instances when resummation provides a powerful tool to analyze quantum field theories. One concerns the resummed propagators which are a very common tool in perturbation theory e.g. \cite{Itzykson:1980rh}. Another application concerns large-$N$ models. They have played an important role in exploring non-perturbative effects in quantum field theories, such as stability \cite{Coleman:1974jh} and  phase transitions \cite{Moshe:2003xn}. What makes this calculation possible is the factorization of higher loop  momentum space diagrams into products of 1-loop diagrams, as a consequence of energy-momentum conservation. In time dependent backgrounds we do not have energy conservation. This is the case, in particular for de Sitter (and anti-de Sitter) space-time, in spite of being maximally symmetric. On the other hand, the spectral parameter in harmonic analysis is an invariant quantity and the split representation reviewed in section \ref{sec:split} shows that it is conserved. This paves the way to carry over large-$N$ analysis to (anti-) de Sitter space \cite{Carmi:2018qzm,DiPietro:2023inn,Sachs:2023eph,Cacciatori:2024zbe,Cacciatori:2024zrv}. In  \cite{Carmi:2018qzm,Sachs:2023eph} this was worked out explicitly for AdS$_3$ (see also \cite{DiPietro:2023inn} for an analysis in dS$_3$). The exact large-$N$ anomalous dimensions in AdS$_4$ were obtained using the \emph{split} representation, reviewed below, together with a \emph{renormalized} spectral density. Among other insights, this provided a new interpretation of the Landau pole as well as the large-$N$ tachyonic instability \cite{Coleman:1974jh} in this theory. The renormalization and resummation of conformally coupled scalar loops in dS$_4$ was found in \cite{Nowinski:2025cvw} using the  shadow action formalism of \cite{Sleight:2020obc,Sleight:2021plv,DiPietro:2021sjt,Chowdhury:2023arc}. The resummation of massive loops, presented below, was so far elusive. 

For concreteness, in this paper, we are interested in an $O(N)$-invariant scalar theory on a de Sitter background that interacts with a conformally coupled spectator scalar field $\varphi$ whose external legs probe the $O(N)$ dynamics through a bi-quadratic interaction. A suitable action is given by
\begin{equation}
    S = S_\phi + S_\varphi + S_\text{int},
\end{equation}
where
\begin{equation}
    S_\phi = \int \mathrm{d}^{d+1}x \,\sqrt{-g} \left\{\tfrac12 g^{\mu\nu} \partial_\mu\phi_a \partial_\nu\phi_a -\tfrac{1}{2}m_\phi^2 \phi_a\phi_a -\tfrac{\lambda}{4N}(\phi_a\phi_a)^2\right\}, \,\,\,\,\,\,\, a = 1,\dots, N
\end{equation}
is the action of the $O(N)$ theory, 
\begin{equation}
    S_\varphi = \int \mathrm{d}^{d+1}x\,\sqrt{-g}\left\{\tfrac{1}{2} g^{\mu\nu}\partial_\mu\varphi\partial_\nu\varphi - \tfrac{1}{2}\xi_c \varphi^2\right\}
\end{equation}
is the action of the conformally coupled scalar and 
\begin{equation}
    S_\text{int} = -\int \mathrm{d}^{d+1}x \sqrt{-g} \tfrac{\lambda}{2N} \varphi^2 \phi_a\phi_a
\end{equation}
describes the coupling between the bulk scalars and the spectator field. Here, $R = d(d+1)H^2$ is the Ricci scalar of the $(d+1)$-dimensional de Sitter space and the condition that $\varphi$ is conformally coupled is encoded by
\begin{equation}
    \xi_c = \frac{d-1}{4d}.
\end{equation}
For simplicity of all our further computations, we set the two coupling constants in this theory to be equal. At the end, one can easily introduce a different coupling constant for the bi-quadratic interaction to study the more general case. 

In section \ref{sec:bubble-chains} we want to apply our split representation formalism to chains of bubble loop diagrams. Here, we will argue why this would provide the leading contribution to a four-point function for such an $O(N)$ model with large $N$. For this, we can do some counting for the $1/N$ expansion for any given Schwinger-Keldysh diagram:
\begin{itemize}
    \item Each quartic $(\phi_a\phi_a)^2$ vertex contributes a factor $\lambda/N$.
    \item Each mixed $\varphi^2(\phi_a\phi_a)$ vertex also contributes a factor $\lambda/N$.
    \item Each \textit{closed index} $\phi_a$ loop contributes a factor $N$. 
\end{itemize}
A diagram with $V_4$ pure $\phi^4$ vertices, $V_m$ mixed vertices and $L_\phi$ closed index loops scales with $\sim \lambda^{V_4+V_m} N^{L_\phi-V_4-V_m}$. For diagrams with conformally coupled external legs, the leading contribution stems from diagrams in which the $\phi$ propagators form bubble loops because then the number of closed index loops equals the number of quartic insertions along the chain, i.e. $L_\phi = V_4+V_m-1$ so that these diagrams scale as $\sim N^0$. 

\section{Split representations}\label{sec:split}
The split representation \cite{Costa:2014kfa,Sleight:2019mgd,Sleight:2019hfp,Loparco:2023rug} has proved to be a very useful tool to analyze the structure of correlators in de Sitter and anti-de Sitter space. It is based on harmonic analysis \cite{Bros:1995js,Bros:1998ik,Costa:2014kfa} and allows to write bulk-to-bulk propagators as a spectral integral over a composition of bulk-to-boundary propagators (harmonic functions) with a suitable spectral function, which amounts to factorization. Multiloop bulk correlators can be similarly factorized  by inserting a decomposition of the identity at the interaction vertices. This has been instrumental for the resummation of loops in \cite{Carmi:2018qzm,Sachs:2023eph,DiPietro:2023inn}. A thorough discussion of this with more references can be found in \cite{Loparco:2023rug}. Most  of the literature in de Sitter on this is formulated in position space. Below we will give the relevant formulae in momentum space, appropriate for our analysis.  

\subsection{De Sitter harmonic function in momentum space}
In the following we will construct a de Sitter harmonic function in momentum space, starting from a simple observation about the behavior of a certain integral of Hankel functions. In appendix \ref{sec:App:Hankel} we explicitly show the following orthogonality relation for Hankel functions:
\begin{equation}\label{eq:orthoHankel}
    \int_{-\infty}^0 \frac{\mathrm{d}\eta}{-\eta} H_{\nu}^{(1)}(-k\eta) H_{\nu'}^{(1)}(-k\eta) = \tfrac{2}{\pi} e^{-i\frac{\pi}{2}(\nu+\nu')}\mathrm{\Gamma}(\nu)\mathrm{\Gamma}(-\nu)\left[\delta(\tilde{\nu}-\tilde{\nu}')+\delta(\tilde{\nu}+\tilde{\nu}')\right].
\end{equation}
We want to use this relation in order to construct a function $\Omega_\nu(k;\eta_1,\eta_2)$ that possesses the property
\begin{equation}\label{eq:harmonicFunctionOrtho}
    \int_{-\infty}^0 \frac{\mathrm{d}\eta}{(-\eta)^{d+1}} \Omega_{\nu'}(k;\eta_1,\eta)\Omega_\nu(k;\eta,\eta_2) = \frac12\left[\delta(\tilde{\nu}-\tilde{\nu}')+\delta(\tilde{\nu}+\tilde{\nu}')\right] \Omega_\nu(k;\eta_1,\eta_2). 
\end{equation}
By comparing this integral to the orthogonality relation \eqref{eq:orthoHankel} and remembering that the mode function for a scalar field in de Sitter is given by
\begin{equation}
    u_\nu(k;\eta) = \frac{\sqrt{\pi}}{2} e^{i\frac{\pi}{2}(\nu+\frac12)} (-\eta)^{d/2} H_\nu^{(1)}(-k\eta),
\end{equation}
we can make the ansatz 
\begin{equation}\label{eq:harmonicAnsatz}
    \Omega_\nu(k;\eta_1,\eta_2) = -i\mathcal{N}_\nu u_\nu(k;\eta_1)u_\nu(k;\eta_2)
\end{equation}
with some yet to be determined coefficient $\mathcal{N}_\nu$. The r.h.s. of (\ref{eq:harmonicFunctionOrtho}) clearly possesses a shadow symmetry $\nu' \leftrightarrow -\nu'$. Also, the mode functions themselves fulfill $u_\nu(k;\eta)=u_{-\nu}(k;\eta)$ due to the property $H_{-i\tilde{\nu}}^{(1)}(z)=e^{-\pi\tilde{\nu}}H_{i\tilde{\nu}}^{(1)}(z)$ of the Hankel functions. Hence we must require $\mathcal{N}_{-\nu} = \mathcal{N}_\nu$. With the help of these relations, plugging the ansatz  (\ref{eq:harmonicAnsatz}) into (\ref{eq:harmonicFunctionOrtho}) leads to 
\begin{equation}\label{eq:Nn}
    \mathcal{N}_\nu = -\frac{1}{\mathrm{\Gamma}(\nu)\mathrm{\Gamma}(-\nu)} = -\frac{\nu}{\pi}\sin(\pi\nu) = \frac{\tilde{\nu}}{\pi}\sinh(\pi\tilde{\nu}).
\end{equation}
This is nothing else but the usual de Sitter density of states that also appeared in the expression of the spectral function for the time-ordered bulk-to-bulk propagator in \cite{Werth:2024mjg} (this is, of course, not a coincidence). 

Having found $\Omega_\nu$, we can integrate over the variable $\nu$ in (\ref{eq:harmonicFunctionOrtho}) and use the fact that $\Omega_{-\nu}(k;\eta_1,\eta_2)=\Omega_\nu(k;\eta_1,\eta_2)$ so that the r.h.s. simplifies to
\begin{equation}
    \int_{-\infty}^0\frac{\mathrm{d}\eta}{(-\eta)^{d+1}}\Omega_{\nu'}(k;\eta_1,\eta)\int_{-\infty}^{+\infty} \mathrm{d}\tilde{\nu}\, \Omega_\nu(k;\eta,\eta_2) = \Omega_{\nu'}(k;\eta_1,\eta_2).
\end{equation}
Thus, if we compare the two sides, the spectral integral simply amounts to a  $\delta$-function on the conformal time domain
\begin{equation}\label{eq:harmonicDeltaIdentity}
    \int_{-\infty}^{+\infty}\mathrm{d}\tilde{\nu}\,\Omega_\nu(k;\eta_1,\eta_2) = (-\eta_1)^{d+1}\delta(\eta_1-\eta_2).
\end{equation}
This identity also has a representation in terms of the Schwinger-Keldysh diagrammatics. Basically, what is stated above translates to the possibility to split a vertex by inserting a pair of bulk-to-boundary propagators. Remember that bulk-to-boundary propagators are simply the mode functions of the free fields (up to a constant prefactor that amounts to the evaluation of one of the propagator vertices on the late-time boundary). Graphically:
\begin{equation}
    \blackdot = \int_{-\infty}^{+\infty}\mathrm{d}\tilde{\nu}\, \mathcal{N}_\nu\,\harmonicFunction .
\end{equation}
The integration contour chosen above amounts to principal series contributions to the decomposition of the Hilbert space. It has been argued in \cite{Loparco:2023rug} that for correlation functions of scalar operators considered in this paper, the other representations do not contribute. We will confirm this below by showing that several 4-point functions previously computed in the literature (like contact diagrams or the 1-loop bubble) are correctly reproduced with this decomposition.

\subsection{A warm up example: 4-point contact diagram}

To see how this harmonic function can be applied to computations of cosmological correlators, we first take a look at a well known toy example: We consider the cross diagram, i.e. a 4-point contact diagram with conformally coupled external legs in $d=3$ dimensions. For conformally coupled scalars, the bulk-to-boundary propagators read
\begin{equation}
    G_\pm(k;\eta) = \frac{\eta\eta_*}{2k} e^{\pm i k\eta}\,,
\end{equation}
where $\eta_*$ is an IR cutoff (the final time-slice on which we evaluate the correlation function). For simplicity, we have set the Hubble rate to $H\equiv 1$. Then, according to the diagrammatic rules, the cross diagram can be written as
\begin{align}
    \mathcal{I}_\times^\pm(r_1,r_2) &= \pm i\lambda\int_{-\infty}^0\frac{\mathrm{d}\eta}{(-\eta)^4} G_\pm(k_1;\eta)G_\pm(k_2;\eta)G_\pm(k_3;\eta)G_\pm(k_4;\eta) \nonumber \\
    &=\pm\frac{i\lambda \eta_*^4}{16k_1k_2k_3k_4} \int_{-\infty}^0\mathrm{d}\eta\,e^{\pm i k_{1234}\eta}  = \frac{\lambda \eta_*^4}{16 k_1k_2k_3k_4k_s}\frac{r_1r_2}{r_1+r_2}.
\end{align}
Here, we expressed the result in terms of the s-channel exchanged momentum $k_s\equiv|\mathbfit{k}_1+\mathbfit{k}_2|$ and the conformal momentum ratios $r_1\equiv k_s/k_{12}$ and $r_2\equiv k_s/k_{34}$. We also introduced the short-hand notation $k_{ij} = k_i+k_j$. 

Instead of the straightforward solution via direct integration we will now describe an alternative approach: We  insert the identity (\ref{eq:harmonicDeltaIdentity}) into the 4-point function which is equivalent to \textit{splitting} the diagram into two 3-point functions. There is however a small subtlety when inserting this $\delta$-function. If we would like to split the diagram into consistent sub-diagrams, we must ensure that we attach the correct bulk-to-boundary propagators to a vertex of a given sign. This means, for a ($-$)-vertex, we can use the relation as stated in (\ref{eq:harmonicDeltaIdentity}), but for a ($+$)-vertex we will use the complex conjugate of this equation which contains the conjugated mode functions to build up $\Omega_\nu^*(k;\eta_1,\eta_2)$. For the diagram with ($+$)-vertex we find
\begin{align}
    \mathcal{I}_\times^+(r_1,r_2) &= i \lambda \int_{-\infty}^0\frac{\mathrm{d}\eta_1 \mathrm{d}\eta_2}{(\eta_1\eta_2)^4} G_+(k_1;\eta_1)G_+(k_2;\eta_1)G_+(k_3;\eta_2)G_+(k_4;\eta_2)\int_{-\infty}^{+\infty}\mathrm{d}\tilde{\nu}\,\Omega_\nu^*(k_s;\eta_1,\eta_2) \nonumber \\
    &= -\lambda \int_{-\infty}^{+\infty}\mathrm{d}\tilde{\nu}\,\mathcal{N}_\nu \,\mathcal{I}_{\Yright}^+(r_1;\nu)\mathcal{I}_{\Yright}^+(r_2;\nu).
\end{align}
Here, $\mathcal{I}_{\Yright}^+(r_1;\nu)$ is the 3-point function 
\begin{equation}\label{eq:3pt}
    \mathcal{I}_{\Yright}^+(r_1;\nu) = \int_{-\infty}^0\frac{\mathrm{d}\eta}{(-\eta)^4} G_+(k_1;\eta)G_+(k_2;\eta) u_\nu^*(k_s;\eta)\,,
\end{equation}
which we compute explicitly in appendix \ref{sec:buildingBlocks}, resulting in
\begin{equation}
    \mathcal{I}_{\Yright}^+(r_1;\nu) = \frac{\eta_*^2}{8\sqrt{\pi k_{12}}k_1k_2}\left\{F_+^{\nu,\frac12}(r_1) + F_-^{\nu,\frac12}(r_1)\right\}.
\end{equation}
For convenience we introduced the functions
\begin{equation}\label{eq:defF}
    F_\pm^{\nu,p}(r) \equiv \mathrm{\Gamma}(\mp\nu)\mathrm{\Gamma}(p\pm\nu) \left(\frac{r}{2}\right)^{\pm\nu} {}_2F_1\left[\begin{matrix}\tfrac{p\pm\nu}{2},\tfrac{p+1\pm\nu}{2} \\ 1\pm\nu\end{matrix}\Big\vert r^2\right]
\end{equation}
that will appear frequently in discussions of 4-point functions.
From the diagrammatic point of view, we have rewritten the 4-point contact diagram in terms of a product of two 3-point diagrams:
\begin{equation}
    \fourPointContact = \int_{-\infty}^{+\infty}\mathrm{d}\tilde{\nu}\,\mathcal{N}_\nu\,\,\ThreePointDiags
\end{equation}
To proceed, we need to evaluate the spectral integrals
\begin{equation}
    i_{\pm\pm} = \int_{-\infty}^{+\infty}\mathrm{d}\tilde{\nu}\,\mathcal{N}_\nu F_\pm^{\nu,p}(r_1)F_\pm^{\nu,p}(r_2).
\end{equation}
We are going to evaluate many similar integrals later on, so we will go through the computation once in detail. Since we integrate $\tilde{\nu}$ over the whole real axis and $\mathcal{N}_\nu$ is symmetric in $\tilde{\nu}$, we can immediately see that $i_{++} = i_{--}$ and $i_{+-}=i_{-+}$ must hold. Therefore, we only have to compute two integrals. We start by evaluating the same-sign part
\begin{align}
    i_{++} &= -\int_{-\infty}^{+\infty} \mathrm{d}\tilde{\nu}
    \frac{\nu}{\pi}\sin(\pi\nu)F_+^{\nu,p} (r_1) F_+^{\nu,p}(r_2) \nonumber\\
    &= -\int_{-\infty}^{+\infty} \mathrm{d}\tilde{\nu} \frac{\nu\pi}{\sin(\pi\nu)}\left(\frac{r_1r_2}{4}\right)^\nu \mathrm{\Gamma}(p +\nu)^2 {}_2\tilde{\mathcal{F}}_1\left[\begin{matrix}\tfrac{p+\nu}{2},\tfrac{p+1+\nu}{2} \\ 1+\nu\end{matrix}\Big\vert r_1^2\right]{}_2\tilde{\mathcal{F}}_1\left[\begin{matrix}\tfrac{p+\nu}{2},\tfrac{p+1+\nu}{2} \\ 1+\nu\end{matrix}\Big\vert r_2^2\right]\,,
\end{align}
where we introduced the regularized hypergeometric function 
\begin{equation}
    {}_2F_1\left[\begin{matrix}a,b \\ c\end{matrix}\Big\vert z\right] = \mathrm{\Gamma}(c){}_2\tilde{\mathcal{F}}_1\left[\begin{matrix}a,b \\ c\end{matrix}\Big\vert z\right]\,,
\end{equation}
which is an entire function of the parameters $a,b,c$. We have simplified the pre-factor $\mathrm{\Gamma}(1+\nu)^2\mathrm{\Gamma}(-\nu)^2=\pi^2/\sin^2(\pi\nu)$ and find that the integrand has simple poles on the imaginary axis at $\tilde{\nu} = in,\,n\in\mathbb{Z}$ stemming from the $\csc(\pi\nu)$ term (with residue $(-1)^n/\pi i$) as well as second-order poles at $\tilde{\nu} = i(n +p),\,n\in\mathbb{N}$ stemming from the factor $\mathrm{\Gamma}(p+\nu)^2$. The asymptotic behavior of the integrand is dominated by the term $(r_1r_2/4)^\nu$ and hence for $0<r_1,r_2<1$, if we close the contour in the lower half-plane, the integrand vanishes along the additional semi-circle and the integral only picks up the poles with negative imaginary part. The pole at $\nu=0$ does not contribute due to the additional factor $\nu$ in the integrand. We end up with
\begin{align}
    i_{++} 
    &= 2\pi \sum\limits_{n=0}^\infty (-1)^n n\frac{\mathrm{\Gamma}(p+n)^2}{(n!)^2} \left(\frac{r_1r_2}{4}\right)^n {}_2F_1\left[\begin{matrix}\tfrac{p+n}{2},\tfrac{p+1+n}{2} \\ 1+n\end{matrix}\Big\vert r_1^2\right]{}_2F_1\left[\begin{matrix}\tfrac{p+n}{2},\tfrac{p+1+n}{2} \\ 1+n\end{matrix}\Big\vert r_2^2\right]\,.
\end{align}
Similarly, we can compute the mixed integrals.
\begin{align}
    i_{+-} &= \int_{-\infty}^{+\infty}\mathrm{d}\tilde{\nu} \frac{\nu\pi}{\sin(\pi\nu)}\left(\frac{r_1}{r_2}\right)^\nu \mathrm{\Gamma}(p+\nu)\mathrm{\Gamma}(p-\nu) {}_2\tilde{\mathcal{F}}_1\left[\begin{matrix}\tfrac{p+\nu}{2},\tfrac{p+1+\nu}{2} \\ 1+\nu\end{matrix}\Big\vert r_1^2\right]{}_2\tilde{\mathcal{F}}_1\left[\begin{matrix}\tfrac{p-\nu}{2},\tfrac{p+1-\nu}{2} \\ 1-\nu\end{matrix}\Big\vert r_2^2\right] \,.
\end{align}
This time, the integrand has poles on the imaginary axis at $\tilde{\nu}=in,\,n\in\mathbb{Z}$ (with residue $(-1)^n/\pi i$) from the $\csc(\pi\nu)$ term as well as poles at $\tilde{\nu}=\pm i(n+p),\,n\in\mathbb{N}$ (with residue $(-1)^n/n! i $) from The $\mathrm{\Gamma}$-factors. The asymptotic behavior is dominated by the term $(r_1/r_2)^\nu$. Contrary to the same-sign case, here \emph{we need to specify a kinematic region} in which our result shall hold, i.e. we have to set either $r_1<r_2$ or vice versa. In the following we stick to the case $r_1<r_2$ so that we can close the contour in the lower half-plane again and pick up the poles with negative imaginary part. We call the two contributions from the different sets of poles the even and the odd part (the even part coming from the $\csc$ function). 
\begin{align}
    i_{+-}^{\text{even}} 
    &= -2\pi \sum\limits_{n=0}^\infty (-1)^n n\frac{ \mathrm{\Gamma}(p+n)\mathrm{\Gamma}(p-n)}{n!\mathrm{\Gamma}(1-n)} \left(\frac{r_1}{r_2}\right)^{n}{}_2F_1\left[\begin{matrix}\tfrac{p+n}{2},\tfrac{p+1+n}{2} \\ 1+n\end{matrix}\Big\vert r_1^2\right] {}_2F_1\left[\begin{matrix}\tfrac{p-n}{2},\tfrac{p+1-n}{2} \\ 1-n\end{matrix}\Big\vert r_2^2\right] .
\end{align}
Here we need to pay attention since $\mathrm{\Gamma}(1-n)$ diverges but the hypergeometric function in $r_2^2$ also diverges due to the lower parameter $1-n$. The quotient of both is finite however:
\begin{align}
    \lim\limits_{c\to1-n} \frac{1}{\mathrm{\Gamma}(c)} {}_2F_1\left[\begin{matrix}\tfrac{p-n}{2},\tfrac{p+1-n}{2} \\ c\end{matrix}\Big\vert r_2^2\right] = \frac{\mathrm{\Gamma}(\tfrac{p+n}{2})\mathrm{\Gamma}(\tfrac{p+1+n}{2})}{\mathrm{\Gamma}(\tfrac{p-n}{2})\mathrm{\Gamma}(\tfrac{p+1-n}{2}) n!} r_2^{2n} {}_2F_1\left[\begin{matrix}\tfrac{p+n}{2},\tfrac{p+1+n}{2} \\ 1+n\end{matrix}\Big\vert r_2^2\right].
\end{align}
Using $\mathrm{\Gamma}(\tfrac{p\pm n}{2})\mathrm{\Gamma}(\tfrac{p+1\pm n}{2}) = 2^{1-p\mp n}\sqrt{\pi}\mathrm{\Gamma}(p\pm n)$, we can simplify the even part to
\begin{align}
    i_{+-}^{\text{even}} &= - 2\pi \sum\limits_{n=0}^\infty (-1)^n n\frac{\mathrm{\Gamma}(p+n)^2}{(n!)^2} \left(\frac{r_1r_2}{4}\right)^n {}_2F_1\left[\begin{matrix}\tfrac{p+n}{2},\tfrac{p+1+n}{2} \\ 1+n\end{matrix}\Big\vert r_1^2\right]{}_2F_1\left[\begin{matrix}\tfrac{p+n}{2},\tfrac{p+1+n}{2} \\ 1+n\end{matrix}\Big\vert r_2^2\right].
\end{align}
Therefore, we have
\begin{equation}
    i_{++}+i_{--}+i_{+-}^{\text{even}}+i_{-+}^{\text{even}} = 0
\end{equation}
and only the odd part of the mixed terms will contribute to the final result. The odd part on the other hand is given by
\begin{align}
    i_{+-}^{\text{odd}} &= -2\pi^2 \sum\limits_{n=0}^\infty \frac{(-1)^n}{n!}\frac{(n+p)\mathrm{\Gamma}(2p+n)}{\sin(\pi(n+p))} \left(\frac{r_1}{r_2}\right)^{n+p}{}_2\tilde{\mathcal{F}}_1\left[\begin{matrix}\tfrac{2p+n}{2},\tfrac{2p+1+n}{2} \\ p+1+n\end{matrix}\Big\vert r_1^2\right] {}_2\tilde{\mathcal{F}}_1\left[\begin{matrix}-\tfrac{n}{2},\tfrac{1-n}{2} \\ \tfrac12-n\end{matrix}\Big\vert r_2^2\right].
\end{align}
Since we assume the external legs to be conformally coupled, we have $p=\frac12$, so that
\begin{equation}
    i_{+-}^{\text{odd}} = -2\pi  \sum\limits_{n=0}^\infty (-1)^n \left(\frac{r_1}{r_2}\right)^{n+\frac12} {}_2F_1\left[\begin{matrix}\tfrac12+\tfrac{n}{2},1+\tfrac{n}{2} \\ \tfrac32+n\end{matrix}\Big\vert r_1^2\right] {}_2F_1\left[\begin{matrix}-\tfrac{n}{2},\tfrac12-\tfrac{n}{2} \\ \tfrac12-n\end{matrix}\Big\vert r_2^2\right].
\end{equation}
Therefore, the full result is obtained by adding up the contributions from $i_{+-}$ and $i_{-+}$:
\begin{equation}
    \mathcal{I}_\times^+(r_1,r_2) = \frac{\lambda \eta_*^4}{16k_1k_2k_3k_4k_s} r_1 \sum\limits_{n=0}^\infty (-1)^n \left(\frac{r_1}{r_2}\right)^{n} {}_2F_1\left[\begin{matrix}\tfrac12+\tfrac{n}{2},1+\tfrac{n}{2} \\ \tfrac32+n\end{matrix}\Big\vert r_1^2\right] {}_2F_1\left[\begin{matrix}-\tfrac{n}{2},\tfrac12-\tfrac{n}{2} \\ \tfrac12-n\end{matrix}\Big\vert r_2^2\right].
\end{equation}
The integral $\mathcal{I}_\times^-(r_1,r_2)$ is obtained by taking the complex conjugate of $\mathcal{I}_\times^+(r_1,r_2)$. Since the result was real, the full answer is therefore just twice the contribution of a single Schwinger-Keldysh diagram.
By comparing to the previous result, we also find an intriguing summation formula for hypergeometric functions,
\begin{equation}\label{eq:hyperIdentity}
    \frac{1}{1+\frac{r_1}{r_2}} = \sum\limits_{n=0}^\infty (-1)^n \left(\frac{r_1}{r_2}\right)^{n} {}_2F_1\left[\begin{matrix}\tfrac12+\tfrac{n}{2},1+\tfrac{n}{2} \\ \tfrac32+n\end{matrix}\Big\vert r_1^2\right] {}_2F_1\left[\begin{matrix}-\tfrac{n}{2},\tfrac12-\tfrac{n}{2} \\ \tfrac12-n\end{matrix}\Big\vert r_2^2\right].
\end{equation}
We checked numerically that this holds for all $r_1 < r_2$.\footnote{ The numerical convergence is quite slow, however, when approaching the diagonal $r_1=r_2$.} It is interesting to observe that the coefficient in front of the hypergeometric functions is exactly the coefficient in the Taylor expansion of $(1+r_1/r_2)^{-1}$ (which in this case is just a geometric series)\footnote{There are some arguments why these coefficients must be the same: The given integral must be a solution of the conformal Ward identity $(\Delta_{r_1}-\Delta_{r_2})\mathcal{I}_\times =0$ with some suitable boundary conditions. In particular, the series representation must yield the correct squeezed limit $r_1,r_2\to 0$. Taking these limits while fixing $r_1/r_2$ requires the coefficient to match the Taylor coefficient since the hypergeometric functions behave as ${}_2F_1[\dots\vert 0] =1$.}. One may wonder whether our expression involves only some special values of hypergeometric functions that are equal to 1, but this is actually not the case. Instead there is some fascinating, non-trivial resummation at work. 

\begin{figure}[t]
  \centering
  \includegraphics[width=0.95\linewidth]{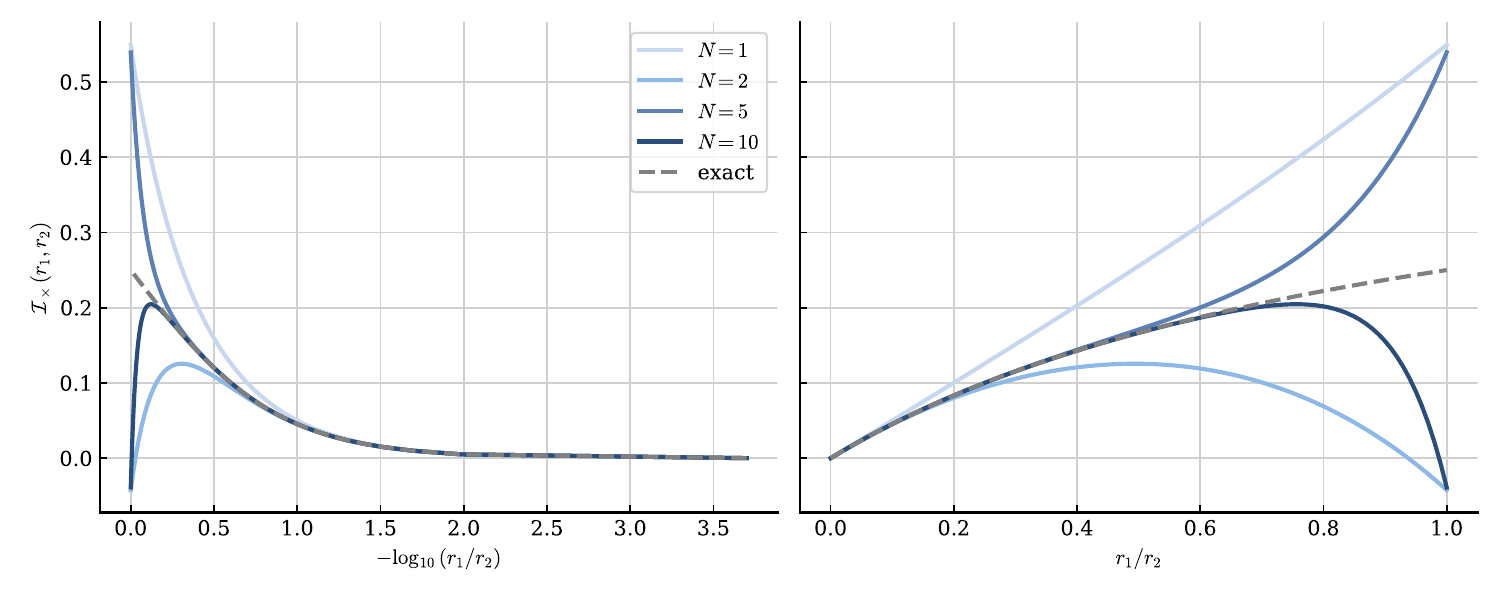}
  \caption{Numerical comparison of the contact diagram via split representations where we truncate the sum over hypergeometric functions after $N=1,2,5,10$ terms, to the exact result from direct integration. For the evaluation, we fixed the conformal momentum ratio $r_2= 0.5$ and varied $r_1$. The two plots show the same curve with different scaling on the horizontal axis. We will use the logarithmic scaling later for displays of cosmological collider signals. The linear scaling in the right plot highlights the fact that the series representation approximates the exact solution very well even for a small number of terms until it gets close to the diagonal where $r_1=r_2$. The results are plotted in units of $\lambda \eta_*^4/(16k_1k_2k_3k_4k_s)$.}
\end{figure}

\subsection{A second example: boost-breaking contact diagram}
Again, we consider a 4-point contact diagram, but this time we allow for a boost-breaking time-dependence of the interaction. To avoid clutter from irrelevant pre-factors, we will just focus on the following dimensionless seed integral
\begin{equation}
    \mathcal{I}_\times^{+,p}(r_1,r_2) = ik_s^{1+p} \int_{-\infty}^0\mathrm{d}\eta\,(-\eta)^p e^{ik_{1234}\eta} = e^{-i\frac{\pi}{2}p} \mathrm{\Gamma}(1+p)\left(\frac{r_1r_2}{r_1+r_2}\right)^{1+p}.
\end{equation}
We can play the same game as before: We introduce a second time integral and connect the two by a $\delta$-function which we then rewrite in terms of a spectral integral over the harmonic function. In doing so, we actually have an interesting degree of freedom - we can split the time-dependence in an arbitrary way:
\begin{align}
    \mathcal{I}_\times^{+,p}(r_1,r_2) = ik_s^{1+p} \int_{-\infty}^0\mathrm{d}\eta_1\,\mathrm{d}\eta_2\,(-\eta_1)^{p-\alpha}(-\eta_2)^{\alpha-4} e^{ik_{12}\eta_1}e^{ik_{34}\eta_2} (-\eta_2)^4\delta(\eta_1-\eta_2).
\end{align}
This freedom was of course present in the previous calculation too, but here we will try to capture the consequences of it. The computation itself is similar to the one above, the same cancellations appear and at the end we arrive at a single series expression for the final result, 
The final result is then
\begin{equation}
    \mathcal{I}_\times^{+,p} = e^{-i\frac{\pi}{2}p} r_1^{1+p}\sum\limits_{n=0}^\infty \frac{(-1)^n}{n!}\mathrm{\Gamma}(1+p+n)\left(\frac{r_1}{r_2}\right)^n {}_2F_1\left[\begin{matrix}\tfrac{1+p+n}{2},\tfrac{2+p+n}{2} \\ \alpha-\tfrac12+n\end{matrix}\Big\vert r_1^2\right] {}_2F_1\left[\begin{matrix}-\tfrac{n}{2},\tfrac12-\tfrac{n}{2} \\ \tfrac52-\alpha-n\end{matrix}\Big\vert r_2^2\right].
\end{equation}
There are two key observations here: First, the coefficient in front of the hypergeometric functions is again just the respective term of the Taylor expansion of $(1+r_1/r_2)^{-1-p}$ and we found another 'magical' summation formula for hypergeometric functions. Second, the freedom of choosing the parameter $\alpha$ is also visible in the final result. We checked numerically that the result is indeed independent of the choice of $\alpha$ as long as $\alpha\neq \frac{1}{2}+n, \,n\in\mathbb{Z}$ (in this case the second hypergeometric function would diverge).

A valid question is why we should solve a trivial integral that can be written down as a one-liner via a method that requires three pages of advanced calculus just to get a solution that is not a closed-form expression but instead a rather complicated infinite sum over hypergeometric functions. However, we will see below that we have already uncovered a rather natural structure for describing \emph{generic} 4-point functions. For related discussions see \cite{Melville:2024ove,Lee:2025kgs}.

\section{Bubble chains}\label{sec:bubble-chains}
In this section we consider multi-loop diagrams of massive scalars interacting with conformally coupled external legs. We will see that our spectral representation allows us to resum such diagrams much as in flat space and thus uncover some non-perturbative features in cosmological correlation functions. 

Suppose we want to compute a chain of $N$ bubble loops. Such a diagram can be described with the help of the following seed integral 
\begin{align}\label{eq:bubble-seed}
\mathcal{I}_{\rangle\!\bigcirc^N\!\langle}^\nu(r_1,r_2) &= \NLoopBubble\nonumber \\
&= \sum\limits_{a_l=\pm}  \int_{-\infty}^0 \left[\prod\limits_{l=1}^{N+1}\mathrm{d}\eta_l \,ia_l(-\eta_l)^{p_l} \right] e^{ia_1 k_{12}\eta_1+ia_{N+1}k_{34}\eta_{N+1}} \prod\limits_{j=1}^N\mathcal{Q}_{a_j a_{j+1}}^\nu (k_s;\eta_j,\eta_{j+1}).
\end{align}
Here, the loop integrals are given by
\begin{align}
    \mathcal{Q}_{a_1 a_2}^\nu (k_s;\eta_1,\eta_2) = \int \frac{\mathrm{d}^d\mathbfit{q}}{(2\pi)^d} G_{a_1a_2}^\nu(q;\eta_1,\eta_2)G_{a_1a_2}^\nu(|\mathbfit{k_s}-\mathbfit{q}|;\eta_1,\eta_2),
\end{align}
and we stripped off additional pre-factors from the bulk-to-boundary propagators $G_a(k_i;\eta)$ as well as the coupling constant $\lambda$,  that we will re-insert later. In that sense, our notation is a bit sloppy, because later we will refer to the same integral with the additional pre-factors as $\mathcal{I}_{\rangle\!\bigcirc^N\!\langle}^\nu(r_1,r_2)$ too. 

The term seed integrals refers to the fact that we could use the above integral to describe several different $N$-loop correlators in different physical theories. First, we can adjust the time-dependence at the vertices by the parameters $p_l$. Second, the above integral represents conformally coupled scalars as external legs because there mode functions are simple exponential functions (up to scaling). If we were interested in massless bulk-to-boundary propagators instead (typical in inflationary cosmology), we could act with differential operators on our seed integral or the final result in order to obtain the massless correlator. This procedure is called weight-shifting \cite{Bootstrap:weightShift,Benincasa:2019vqr}. 

In the successful computation of the 1-loop bubble diagram with the help of the spectral function \cite{,Marolf:2010zp,Xianyu:2022jwk}, the key idea was to write down a spectral integral over a tree-level exchange diagram. The generalization of this approach to a chain of $N$ bubble loops, is straightforward: One would first compute a tree-level diagram with a chain of $N$ massive propagators of mass $\nu_j$,
\begin{align}
    \mathcal{I}_{\rangle\!-^N\!\langle}^{\nu_1,\dots,\nu_N}(r_1,r_2) &= \NChain \nonumber \\
    &=\sum\limits_{a_l=\pm}  \int_{-\infty}^0 \left[\prod\limits_{l=1}^{N+1}\mathrm{d}\eta_l \,ia_l(-\eta_l)^{p_l} \right] e^{ia_1 k_{12}\eta_1+ia_{N+1}k_{34}\eta_{N+1}} \prod\limits_{j=1}^N G_{a_j a_{j+1}}^{\nu_j} (k_s;\eta_j,\eta_{j+1}),
\end{align}
and then perform $N$ spectral integrals,
\begin{align}
    \mathcal{I}_{\rangle\!\bigcirc^N\!\langle}^\nu(r_1,r_2) = \int_{-\infty}^{+\infty}\left[\prod\limits_{j=1}^N \mathrm{d}\tilde{\nu}_j\,\rho_\nu(\tilde{\nu}_j)\right] \mathcal{I}_{\rangle\!-^N\!\langle}^{\nu_1,\dots,\nu_N}(r_1,r_2)\,,
\end{align}
with the spectral function obtained in \cite{,Marolf:2010zp,Xianyu:2022jwk}. The computation of the tree-level diagram is inherently complex (although straightforward). Specifically, the signals arise from a nontrivial summation over mass parameters associated with the intermediate states. While the individual terms do not exhibit a level of complexity substantially beyond that encountered in the one-loop case, the combinatorial nature of the problem demands considerable analytical effort.

\subsection{Bubble building blocks}
We will now argue that even for the 1-loop case there is a convenient choice of spectral integrals which, moreover, easily generalizes to the computation of $N$-loop diagrams in a way that actually factorizes for de Sitter invariant interactions. The key idea to achieve this factorization is to use our split representation and to split the chain of bubbles into individual \textit{bubble building blocks}. To be precise, we double the number of time integrals in \eqref{eq:bubble-seed}, with extra integrations of $\mathrm{d}\eta'$ at each vertex, insert $\delta$-functions and then rewrite those $\delta$-functions with the help of our harmonic function as in \eqref{eq:harmonicDeltaIdentity}. With this, \eqref{eq:bubble-seed} becomes:
\begin{align}
    \mathcal{I}_{\rangle\!\bigcirc^N\!\langle}^\nu(r_1,r_2) = \int_{-\infty}^{+\infty}\prod\limits_{i=1}^{N+1}\mathrm{d}\tilde{\nu}_i\sum\limits_{a_l=\pm} \int_{-\infty}^{0} \left[\prod\limits_{l=1}^{N+1}\mathrm{d}\eta_l \,\mathrm{d}\eta_l'\,ia_l(-\eta_l)^{p_l}(-\eta_l')^{-d-1} \Omega_{\nu_l}^{a_l}(k_s;\eta_l,\eta_l')  \right]\nonumber\\
    e^{ia_1 k_{12}\eta_1+ia_{N+1}k_{34}\eta_{N+1}} \prod\limits_{j=1}^N\mathcal{Q}_{a_j a_{j+1}}^\nu (k_s;\eta_j',\eta_{j+1}).
\end{align}
Here, we introduced the notation $\Omega_\nu^- = \Omega_\nu$ and $\Omega_\nu^+ = \Omega_\nu^*$. Diagrammatically this corresponds to
\begin{align}
    \mathcal{I}_{\rangle\!\bigcirc^N\!\langle}^\nu(r_1,r_2) = \int_{-\infty}^{+\infty}\prod\limits_{i=1}^{N+1}\mathrm{d}\tilde{\nu}_i \mathcal{N}_{\nu_i}\,\,\NLoopBubbleSplit\,.
\end{align}

At present, the situation appears even more challenging than the case of the $N$ spectral integrals over the $N$-site chain described previously, as the current formulation involves $N+1$ spectral integrals - prior to any consideration of the loop integrals themselves. Nevertheless, it is possible to mitigate this complexity by reorganizing the terms and identifying a set of simpler, foundational building blocks. To proceed, we recall the definition
\begin{equation}
    \Omega_\nu^a(k;\eta_1,\eta_2) = ia\mathcal{N}_\nu u_\nu^a(k;\eta_1) u_\nu^a(k;\eta_2)
\end{equation}
where $u_\nu^-=u_\nu$ and $u_\nu^+=u_\nu^*$, and, with the help of a generalization of the 3-point functions in \eqref{eq:3pt} for arbitrary powers of the conformal time in the integrand,
\begin{align}
    \mathcal{I}_{\Yright,p}^+(r_1;\nu) &= \int_{-\infty}^0\frac{\mathrm{d}\eta}{(-\eta)^{p+7/2}} G_+(k_1;\eta)G_+(k_2;\eta) u_\nu^*(k_s;\eta) \nonumber \\
    &= \frac{\sqrt{\pi}\eta_*^2}{8k_1k_2} e^{-i\frac{\pi}{2}(\nu^*+\frac12)} \int_{-\infty}^0\frac{\mathrm{d}\eta}{(-\eta)^{p}} e^{ik_{12}\eta} H_{\nu^*}^{(2)}(-k_s\eta)
\end{align}
and $\mathcal{I}_{\Yright, p}^- =(\mathcal{I}_{\Yright,p}^+)^*$
(note that we reintroduced the pre-factors from the bulk-to-boundary propagators at this point), we write (\ref{eq:bubble-seed}) as 
\begin{align}
\mathcal{I}_{\rangle\!\bigcirc^N\!\langle}^\nu(r_1,r_2) = \int_{-\infty}^{+\infty}\prod\limits_{i=1}^{N+1}\left[-\mathrm{d}\tilde{\nu}_i \,\mathcal{N}_{\nu_i}\right]\sum\limits_{a_l=\pm} \mathcal{I}_{\Yright,p_1}^{a_1}(r_1;\nu_1) \mathcal{I}_{\Yright,p_{N+1}}^{a_{N+1}}(r_2;\nu_{N+1})\prod\limits_{j=1}^N\tilde{Q}_{a_j a_{j+1}}^\nu (j).
\end{align}
Here, we introduced the bubble building blocks
\begin{align}
    \tilde{Q}_{a_j a_{j+1}}^\nu (j) &= \BubbleBuildingBlock \nonumber \\
    &=\int_{-\infty}^0 \frac{\mathrm{d}\eta_j'}{(-\eta_j')^{d+1}}\frac{\mathrm{d}\eta_{j+1}}{(-\eta_{j+1})^{d+1}} u_{\nu_j}^{a_j}(k_s;\eta_j') Q_{a_j a_{j+1}}^\nu (k_s;\eta_j',\eta_{j+1}) u_{\nu_{j+1}}^{a_{j+1}}(k_s;\eta_{j+1})
\end{align}
and we assume de Sitter invariant self-couplings of the massive scalars in the loops so that $p_l = -d-1$ at all of the internal vertices. As an intermediate step our goal now will be to compute this building block explicitly and to solve the two time integrals. For this purpose, the next step is to rewrite the bubble loop integrals with the help of a spectral representation, i.e. we use the identity
\begin{equation}\label{eq:Q+}
    \mathcal{Q}_{a b}^\nu(k;\eta_1,\eta_2) = \int_{-\infty}^{+\infty} \mathrm{d}\tilde{\nu}'\, \rho_\nu^+(\tilde{\nu}') G_{ab}^{\nu'}(k;\eta_1,\eta_2).
\end{equation}
Throughout this paper, we will use a symmetrized version of the spectral function, 
\begin{equation}\label{eq:symr}
    \rho_\nu^+(\tilde{\nu}') = \frac{\mathrm{\Gamma}(\tfrac{d+1}{2})\tilde{\nu}'\sinh(\pi\tilde{\nu}')}{2^{6-d}\pi^{\frac{d+7}{2}}\mathrm{\Gamma}(d)\mathrm{\Gamma}(\tfrac{d}{2}+i\tilde{\nu}')\mathrm{\Gamma}(\tfrac{d}{2}-i\tilde{\nu}')} \prod\limits_{\pm,\pm,\pm} \mathrm{\Gamma}(\tfrac{d}{4}\pm \tfrac{i\tilde{\nu}}{2} \pm  \tfrac{i\tilde{\nu}}{2}\pm  \tfrac{i\tilde{\nu}'}{2}).
\end{equation}
In \cite{Xianyu:2022jwk}, a different spectral function was used (which can be expressed in terms of a generalized hypergeometric function ${}_7F_6$ with argument 1 and that has parameters depending on the spectral parameters $\nu$ and $\nu'$) which was inferred from calculations in EdS / the sphere \cite{Marolf:2010zp}. In \cite{Zhang:2025nzd} it was shown that the symmetrization of that spectral function is exactly the Källén-Lehmann spectral function \eqref{eq:symr} that was computed in \cite{Loparco:2023rug}, relying on previous works in \cite{Bros:2009bz, Penedones:2010ue}. Both spectral functions lead to the same contour integral in \eqref{eq:Q+} and hence we could use either one of them, but due to the symmetry properties of \eqref{eq:symr} we will stick to $\rho_\nu^+$.

Diagrammatically, what we have achieved now is the following simplification:
\begin{equation}
    \BubbleBuildingBlock = \int_{-\infty}^{+\infty} \mathrm{d}\tilde{\nu}'\, \rho_\nu^+(\tilde{\nu}') \,\TreeBuildingBlock
\end{equation}
For opposite signs, i.e. $G_{+-}^{\nu'}$ or $G_{-+}^{\nu'}$, we immediately obtain a factorization of the remaining time integrals due to $G_{-+}(k;\eta_1,\eta_2) = u_{\nu'}^-(k;\eta_1) u_{\nu'}^+(k;\eta_2)$ with a similar expression  for $G_{+-}$:
\begin{align}
    \TreeBuildingBlockPP &= \TreeBuildingBlockFactorPP \nonumber \\
    \nonumber \\
    \TreeBuildingBlockMM &= \TreeBuildingBlockFactorMM
\end{align}
The equal-sign propagators contain time-ordering Heaviside functions and hence time integrals do not factorize automatically. To bypass this nested structure, we can employ the spectral representation
\begin{equation}\label{eq:tospec}
    G_{aa}^{\nu'}(k;\eta_1,\eta_2) = ia\int_{-\infty}^{+\infty}\mathrm{d}\tilde{\mu}\,\mathcal{N}_\mu \frac{u_{\mu}^a(k;\eta_1)u_{\mu}^a(k;\eta_2)}{(\tilde{\mu}^2-\tilde{\nu}'^2)_{ia\epsilon}},
\end{equation}
of the time-ordered propagators with spectral density
\begin{equation}
    \frac{1}{(\tilde{\mu}^2-\tilde{\nu}'^2)_{ia\epsilon}} \equiv \frac{1}{2\sinh(\pi\tilde{\nu}')}\left[\frac{e^{+\pi\tilde{\nu}'}}{\tilde{\mu}^2-\tilde{\nu}'^2 +ia\epsilon}-\frac{e^{-\pi\tilde{\nu}'}}{\tilde{\mu}^2-\tilde{\nu}'^2 -ia\epsilon}\right]
\end{equation}
that was found in \cite{Melville:2024ove} and further discussed in \cite{Werth:2024mjg}. Again, diagrammatically this corresponds to
\begin{equation}
    \propagatorPPnu = i\int_{-\infty}^{+\infty}\mathrm{d}\tilde{\mu}\,\frac{\mathcal{N}_\mu}{(\tilde{\mu}^2-\tilde{\nu}'^2)_{i\epsilon}} \,\,\,\,\propagatorPPnuFactor
\end{equation}
The advantage of using these factorized representations of the Schwinger-Keldysh diagrams is that they break up the nested structure of the time integrals which then allows us to employ the orthogonality property \eqref{eq:harmonicFunctionOrtho} of the harmonic function (i.e. products of mode functions) to solve the time integrals.
For opposite signs we can thus write
\begin{align}
    \tilde{Q}_{-+}^\nu (j) &= \int_{-\infty}^{+\infty}\mathrm{d}\tilde{\nu}'_j \,\rho_\nu^+(\tilde\nu'_j)\int_{-\infty}^0 \frac{\mathrm{d}\eta_j'}{(-\eta_j')^{d+1}} u_{\nu_j}^{-}(k_s;\eta_j')u_{\nu_j'}^{-}(k_s;\eta_j') \nonumber \\ 
    &\,\,\,\,\,\,\,\,\cdot\int_{-\infty}^0  \frac{\mathrm{d}\eta_{j+1}}{(-\eta_{j+1})^{d+1}}u_{\nu_j'}^{+}(k_s;\eta_{j+1})u_{\nu_{j+1}}^{+}(k_s;\eta_{j+1}) \nonumber \\
    &= \int_{-\infty}^{+\infty}\mathrm{d}\tilde{\nu}'_j \,\rho_\nu^+(\tilde\nu'_j) \mathcal{N}_{\nu_j}^{-1} \mathcal{N}_{\nu_{j+1}}^{-1}\frac12\left[\delta(\tilde{\nu}_j-\tilde{\nu}_j')+\delta(\tilde{\nu}_j+\tilde{\nu}_j')\right]\frac12\left[\delta(\tilde{\nu}_{j+1}-\tilde{\nu}_j')+\delta(\tilde{\nu}_{j+1}+\tilde{\nu}_j')\right] \nonumber \\
    &= \rho_\nu^+(\tilde\nu_j) \mathcal{N}_{\nu_j}^{-2} \frac12 \left[\delta(\tilde{\nu}_{j}-\tilde{\nu}_{j+1})+\delta(\tilde{\nu}_{j}+\tilde{\nu}_{j+1})\right].
\end{align}
In the second equality we evaluated the factorized time integrals using our identity for integrals over Hankel functions \eqref{eq:orthoHankel}. Throughout this calculation we have used the symmetry $\rho_\nu^+(\tilde{\nu}_j) = \rho_\nu^+(-\tilde{\nu}_j)$ of the Källén-Lehmann spectral function. Since the result is real, we have $\tilde{Q}_{-+}^\nu (j)=\tilde{Q}_{+-}^\nu (j)$. 

On the other hand, the equal-sign loops contain one additional layer of spectral integrals stemming from the time-ordered propagators \eqref{eq:tospec}. We can rewrite them as 
\begin{align}
    \tilde{Q}_{++}^\nu (j) =& i\int_{-\infty}^{+\infty}\mathrm{d}\tilde{\nu}'_j \,\rho_\nu(\tilde\nu'_j)\int_{-\infty}^{+\infty}\mathrm{d}\tilde{\mu}\frac{\mathcal{N}_\mu}{(\tilde{\mu}^2-\tilde{\nu}_j'^2)_{i\epsilon}}\int_{-\infty}^0 \frac{\mathrm{d}\eta_j'}{(-\eta_j')^{d+1}} u_{\nu_j}^{+}(k_s;\eta_j')u_{\mu}^{+}(k_s;\eta_j')\\
    &\,\,\,\,\,\,\,\,\cdot\int_{-\infty}^0  \frac{\mathrm{d}\eta_{j+1}}{(-\eta_{j+1})^{d+1}}u_{\mu}^{+}(k_s;\eta_{j+1})u_{\nu_{j+1}}^{+}(k_s;\eta_{j+1})\nonumber \\
    =& -i\int_{-\infty}^{+\infty} \mathrm{d}\tilde{\nu}_j'\,\mathrm{d}\tilde{\mu}\,\frac{\rho_\nu(\tilde\nu'_j)\mathcal{N}_{\nu_j}^{-1}}{(\tilde{\mu}^2-\tilde{\nu}_j'^2)_{i\epsilon}}\, \frac12\left[\delta(\tilde{\nu}_j-\tilde{\mu})+\delta(\tilde{\nu}_j+\tilde{\mu})\right]\frac12\left[\delta(\tilde{\nu}_{j+1}-\tilde{\mu})+\delta(\tilde{\nu}_{j+1}+\tilde{\mu})\right] \nonumber \\
    =& -i\mathcal{N}_{\nu_j}^{-1}\frac12\left[\delta(\tilde{\nu}_j-\tilde{\nu}_{j+1})+\delta(\tilde{\nu}_j+\tilde{\nu}_{j+1})\right]\int_{-\infty}^{+\infty}\mathrm{d}\tilde{\nu}_j' \frac{\rho_\nu(\tilde\nu_j')}{(\tilde{\nu}_j^2-\tilde{\nu}_j'^2)_{i\epsilon}}.
\end{align}
It remains to evaluate the spectral integral over the two spectral densities for the bubble and the time-ordered propagator \eqref{eq:tospec}. To this end, it is helpful to rewrite the $i\epsilon$-prescription in a distributional sense utilizing the Sokhotski–Plemelj identity
\begin{equation}
    \frac{1}{x\pm i \epsilon} = \mp i\pi \delta(x) + \mathcal{P}\left(\frac{1}{x}\right)\,,
\end{equation}
in terms of a $\delta$-function and a principal value. This leads to
\begin{equation}
    \frac{1}{(\tilde{\nu}_j^2 - \tilde{\nu}_j'^2)_{i\epsilon}} = -i\pi\coth(\pi\tilde{\nu}_j')\delta(\tilde{\nu}_j^2 - \tilde{\nu}_j'^2) + \mathcal{P}\left(\frac{1}{\tilde{\nu}_j^2-\tilde{\nu}_j'^2}\right).
\end{equation}
Using the symmetry of the Källén-Lehmann spectral function again,  we deduce that the first term from the $i\epsilon$-prescription cannot contribute to the spectral integral because it is odd in $\tilde{\nu}_j' \leftrightarrow -\tilde{\nu}_j'$ due to the $\coth(\pi\tilde{\nu}_j')$ factor. We then define
\begin{equation}\label{eq:Pdef}
    P_\nu(\tilde\nu_j) = -i \int_{-\infty}^{+\infty}\mathrm{d}\tilde{\nu}_j'\, \rho_\nu^+(\tilde\nu_j') \mathcal{P}\left(\frac{1}{\tilde{\nu}_j^2-\tilde{\nu}_j'^2}\right)
\end{equation}
and end up with
\begin{equation}
    \tilde{Q}_{++}^\nu (j) = \tilde{Q}_{--}^\nu (j) = P_\nu(\tilde\nu_j)\mathcal{N}_{\nu_j}^{-1}\frac12\left[\delta(\tilde{\nu}_j-\tilde{\nu}_{j+1})+\delta(\tilde{\nu}_j+\tilde{\nu}_{j+1})\right]\,.
\end{equation}
The Källén-Lehmann spectral function \eqref{eq:symr} has the following two sets of simple poles stemming from the $\mathrm{\Gamma}$-factors:
\begin{align}
    \text{Set A:}\,\,\,\,\, &\tilde{\nu}' = -i\left(\frac{d}{2} + 2m\right),\, m\in\mathbb{N} \nonumber \\
    \text{Set B:}\,\,\,\,\, &\tilde{\nu}' = -i\left(\frac{d}{2} + 2m \pm 2i\tilde{\nu}\right),\, m\in\mathbb{N}
\end{align}
By explicitly carrying out the integral in \eqref{eq:Pdef} (closing the contour in the complex plane and summing over the residues at the poles of $\rho_\nu^+$) we find the following expression for $P_\nu$ in terms of an infinite series:
\begin{align}\label{eq:P-nu}
    P_\nu(\tilde\nu') =& \left.\sum\limits_{m=0}^\infty  \frac{(-1)^m}{16 \pi^{\frac{d+3}{2}}} \mathrm{\Gamma}\left[\begin{matrix}\frac{d+1}{2}, \frac{d}{2}+m\\ d,m+1\end{matrix}\right] \right\{ \\
    &\frac{(d+4m)\sin(\pi\frac{d}{2})}{\left(\frac{d}{2}+2m\right)^2+\tilde{\nu}'^2} \mathrm{\Gamma} \left[\begin{matrix}-m- i \tilde{\nu},-m + i \tilde{\nu},\frac{d}{2}+m - i\tilde{\nu},\frac{d}{2}+m + i\tilde{\nu}\\ \frac12-m, \frac{d+1}{2}+m \end{matrix}\right]  \nonumber \\
    +&   \frac{(d+4m+4i\tilde{\nu})\sin(\frac{\pi}{2}(d+4i\tilde{\nu}))}{\left(\frac{d}{2}+2m+2i\tilde{\nu}\right)^2+\tilde{\nu}'^2}\mathrm{\Gamma} \left[\begin{matrix}-m- i \tilde{\nu},-m-2 i \tilde{\nu},\frac{d}{2}+m + i\tilde{\nu},\frac{d}{2}+m +2 i\tilde{\nu}\\ \frac12-m-i\tilde{\nu},\frac{d+1}{2}+m+i\tilde{\nu}\end{matrix}\right]\nonumber \\
    +&  \left.\frac{(d+4m-4i\tilde{\nu})\sin(\frac{\pi}{2}(d-4i\tilde{\nu}))}{\left(\frac{d}{2}+2m-2i\tilde{\nu}\right)^2+\tilde{\nu}'^2}\mathrm{\Gamma} \left[\begin{matrix}-m+i \tilde{\nu},-m+2 i \tilde{\nu},\frac{d}{2}+m - i\tilde{\nu},\frac{d}{2}+m -2 i\tilde{\nu}\\ \frac12-m+i\tilde{\nu},\frac{d+1}{2}+m-i\tilde{\nu}\end{matrix}\right]\right\} \nonumber \\
\end{align}
This series is divergent for $d=3$. Although this might seem like an issue for our goal of finding a spectral representation in the first place, this is actually expected and we can directly relate the divergence of $P_\nu$ to the UV divergence of the 1-loop bubble diagram. We will discuss this topic in section \ref{sec:renormalization}.\\

\subsection{Factorization of the spectral function}

If we assume de Sitter invariant couplings for the interactions involving the external legs, then $\mathcal{I}_{\Yright,p}^{+} = \mathcal{I}_{\Yright,p}^{-} = \mathcal{I}_{\Yright}$ and we can pull these factors out of the sum over the vertex indices. Thus,
\begin{equation}
    \mathcal{I}_{\rangle\!\bigcirc^N\!\langle}^\nu(r_1,r_2) = \int_{-\infty}^{+\infty}\prod\limits_{i=1}^{N+1}\left[-\mathrm{d}\tilde{\nu}_i\,\mathcal{N}_{\nu_i}\right]\mathcal{I}_{\Yright}(r_1;\nu_1)\mathcal{I}_{\Yright}(r_2;\nu_{N+1}) \sum\limits_{aj=\pm}\prod\limits_{j=1}^N \tilde{Q}_{a_j a_{j+1}}^\nu(j)\,.
\end{equation}
We can always write 
\begin{equation}
    \tilde{Q}_{ab}^\nu(j) = \tilde{q}_{ab}^\nu \mathcal{N}_{\nu_j}^{-1} \frac12 \left[\delta(\tilde{\nu}_j-\tilde{\nu}_{j+1})+\delta(\tilde{\nu}_j+\tilde{\nu}_{j+1})\right]\,,
\end{equation}
with $\tilde{q}_{+-}^\nu = \tilde{q}_{-+}^\nu = \mathcal{N}_{\nu_j}^{-1} \rho_\nu^+(\tilde\nu_j)$ and $\tilde{q}_{++}^\nu= \tilde{q}_{--}^\nu = P_\nu(\tilde\nu_j)$. Therefore, the $N+1$ spectral integrals actually collapse to a single spectral integral
\begin{equation}
    \mathcal{I}_{\rangle\!\bigcirc^N\!\langle}^\nu(r_1,r_2) = (-1)^{N+1}\int_{-\infty}^{+\infty}\mathrm{d}\tilde{\nu}_1\,\mathcal{N}_{\nu_1}\mathcal{I}_{\Yright}(r_1;\nu_1)\mathcal{I}_{\Yright}(r_2;\nu_1) \sum\limits_{aj=\pm}\prod\limits_{j=1}^N \tilde{q}_{a_j a_{j+1}}^\nu(1).
\end{equation}
To continue we want to use induction to show that we can write
\begin{equation}
    \sum\limits_{a_j=\pm}\prod\limits_{j=1}^N \tilde{q}_{a_j a_{j+1}}^\nu (1)= 2 \left\{\mathcal{N}_{\nu_1}^{-1}\rho^+_\nu(\tilde\nu_1) + P_\nu(\tilde\nu_1)\right\}^N \equiv 2 \,\Sigma_\nu(\tilde\nu_1)^N.
\end{equation}
For $N=1$ the statement is obviously true. Suppose it holds for $N$. Then, via induction (we omit the argument 1 for the bubble functions $\tilde{q}_{ab}^\nu$ to avoid clutter)
\begin{align}
    \sum\limits_{\substack{a_l=\pm \\l\in\{1,\dots,N+1\}}}\prod\limits_{j=1}^{N+1} \tilde{q}_{a_j a_{j+1}}^\nu (1) &= \sum\limits_{\substack{a_l=\pm\\l\in\{1,\dots,N-1\}}}\left[\prod\limits_{j=1}^{N-1} \tilde{q}_{a_j a_{j+1}}^\nu \right] \left[\tilde{q}_{a_{j-1}+}^\nu\tilde{q}_{++}^\nu + \tilde{q}_{a_{j-1}+}^\nu\tilde{q}_{+-}^\nu \right.\nonumber \\
    &\left.\,\,\,\,\,\,\,\,\,\,\,\,\,\,\,\,\,\,\,\,\,\,\,\,\,\,\,\,\,\,\,\,\,\,\,\,\,\,\,\,\,\,\,\,\,\,\,\,\,\,\,\,\,\,\,\,\,\,\,\,\,\,\,\,\,\,\,\,\,\,\,
    + \tilde{q}_{a_{j-1}-}^\nu\tilde{q}_{-+}^\nu+\tilde{q}_{a_{j-1}-}^\nu\tilde{q}_{--}^\nu\right] \nonumber \\
    &= \sum\limits_{\substack{a_l=\pm\\l\in\{1,\dots,N-1\}}}\left[\prod\limits_{j=1}^{N-1} \tilde{q}_{a_j a_{j+1}}^\nu \right] \left[\tilde{q}_{a_{j-1}+}^\nu+\tilde{q}_{a_{j-1}-}^\nu\right]\Sigma_\nu(\tilde\nu_1) \nonumber \\
    &= \sum\limits_{\substack{a_l=\pm\\l\in\{1,\dots,N\}}}\left[\prod\limits_{j=1}^{N} \tilde{q}_{a_j a_{j+1}}^\nu \right]\Sigma_\nu(\tilde\nu_1) \nonumber \\
    &= 2\left\{\mathcal{N}_{\nu_1}^{-1}\rho^+_\nu(\tilde\nu_1) + P_\nu(\tilde\nu_1)\right\}^{N+1}\,,
\end{align}
which proves the result. 
Thus, we can express the $N$-loop bubble chain as a single spectral integral
\begin{equation}
    \mathcal{I}_{\rangle\!\bigcirc^N\!\langle}^\nu(r_1,r_2) = 2(-1)^{N+1}\int_{-\infty}^{+\infty}\mathrm{d}\tilde{\nu}_1\,\mathcal{N}_{\nu_1}\mathcal{I}_{\Yright}(r_1;\nu_1)\mathcal{I}_{\Yright}(r_2;\nu_1) \Sigma_\nu(\tilde\nu_1)^N
\end{equation}
where the spectral density is given by the $N$th power of the 1-loop spectral density $\Sigma_\nu$. 

The factorization structure of the series of $N$-loop diagrams allows us to resum them in terms of a geometric series. For a given interaction scale $\lambda$, the resummed (non-perturbative) correlator is given by
\begin{align}
    \mathcal{I}_\text{np}^\nu(r_1,r_2) = \sum\limits_{N=0}^\infty \lambda^{N+1} \mathcal{I}_{\rangle\!\bigcirc^N\!\langle}^\nu = -2\lambda \int_{-\infty}^{+\infty}\mathrm{d}\tilde{\nu}'\,\mathcal{N}_{\nu'}\mathcal{I}_{\Yright}(r_1;\nu')\mathcal{I}_{\Yright}(r_2;\nu') \frac{1}{1+\lambda \Sigma_\nu(\tilde\nu')} .
\end{align}
We would like to stress that this factorization and resummation works for de Sitter invariant couplings. In more general cases,  $\mathcal{I}_{\Yright, p}^+ \neq \mathcal{I}_{\Yright,p}^-$ spoils the factorization. This is expected since in these boost-breaking cases we lack some of the symmetries of the spacetime that would restore the correct factorization in the flat-space limit. However, assuming de Sitter invariant internal vertices still allows the factorization of individual diagrams (just not the full correlator). 

\subsection{Renormalization}\label{sec:renormalization}

The series in $P_\nu$ contains the UV divergence of the 1-loop bubble. One can compare this to previous computations in \cite{Xianyu:2022jwk} which used a spectral integral over a tree-level correlator to obtain the 1-loop result. There, the authors used a spectral function that was bootstrapped from computations in EdS (i.e. the sphere) and can be expressed in terms of a hypergeometric function ${}_7F_6$. In the limit $d\to 3 $ this spectral function is divergent. It is possible to remove this divergence with a counterterm as usual. However, the Källén-Lehmann spectral function being the symmetrized version of that function is itself finite in the limit $d\to 3$. When using the Källén-Lehmann spectral function directly for the computation \cite{Zhang:2025nzd}, after integration one obtains a divergent series similar to our expression for $P_\nu$. One can regularize the result by subtracting the asymptotic part of the series (which does not depend on the mass of the field and again amounts to a counterterm). Here, we will follow the same strategy but apply it already to the spectral function in the integrand, making the UV divergence manifest at the integrand level and outlining a strategy to remove the divergence for higher-loop diagrams by local counterterms as well. 

We find that asymptotically, for large $m$, the series expansion in $P_\nu$ behaves like
\begin{equation}
    P_\nu(\tilde\nu') \sim \frac{\mathrm{\Gamma}(\frac{d+1}{2})\sin(\pi\frac{d}{2})}{8\pi^{\frac{d+1}{2}}\mathrm{\Gamma}(d)} \sum\limits_{m=0}^\infty (m+1)^{d-4} = \frac{\mathrm{\Gamma}(\frac{d+1}{2})\sin(\pi\frac{d}{2})}{8\pi^{\frac{d+1}{2}}\mathrm{\Gamma}(d)}\zeta(4-d) \equiv P_\nu^{\text{asy}}(\tilde\nu').
\end{equation}
Here, $\zeta(4-d)$ refers to the Riemann zeta function and diverges for $d\to 3$. We can expand $P_\nu^\text{asy}$ around $d=3$ by setting $d=3-\varepsilon$. This leads to
\begin{equation}
    P_\nu^\text{asy}(\tilde\nu') = -\frac{1}{(4\pi)^2 \varepsilon} -\frac{2+\gamma_E +\log(\pi)}{32\pi^2} + \mathcal{O}(\varepsilon). 
\end{equation}
Hence, we can introduce a renormalized spectral function $\Sigma_\nu^\text{ren}$ by subtracting the UV divergent asymptotic term $P_\nu^\text{asy}$:
\begin{equation}\label{eq:sigmaren}
    \Sigma_\nu^\text{ren} (\tilde\nu') = \mathcal{N}_{\nu'}^{-1}\rho^+_\nu(\tilde\nu') + P_\nu(\tilde\nu') - P_\nu^\text{asy}(\tilde\nu'). 
\end{equation}
Since the zeroth order in $\varepsilon$ of $P_\nu^\text{asy}$ is non-zero, our renormalization scheme is a form of a modified minimal subtraction scheme. 

Although the renormalized spectral function is finite and in principle can be used to evaluate the spectral integral analytically or numerically, let us quickly comment on the feasibility of numerical integration. It turns out that the series expansion \eqref{eq:P-nu} of $P_\nu$ converges extremely slowly. The quality of convergence depends on the choice of the masses $\nu$ and $\nu'$, but for several examples that we checked we still observed about 25\% deviations when truncating the sum after $10^5$ terms compared to $10^6$ terms. Even with acceleration techniques like Wynn's epsilon algorithm \cite{Wynn}, this series representation of the spectral function is therefore not well-suited for attempts of efficient numerical integration. \\

If we take a look at the effect on the spectral integral, we see that for the 1-loop case the subtraction of $P_\nu^\text{asy}$ amounts to adding a counterterm to the Lagrangian. For a moment, let us focus purely on the UV pole. The subtraction of $-1/(4\pi)^2\varepsilon$ from the spectral function means adding the integral 
\begin{equation}
    \mathcal{I}_C(r_1,r_2) = \frac{\lambda^2_\text{ren}}{8\pi^2 \varepsilon} \int_{-\infty}^{+\infty} \mathrm{d}\tilde{\nu}'\,\mathcal{N}_{\nu'}\mathcal{I}_{\Yright}(r_1;\nu')\mathcal{I}_{\Yright}(r_2;\nu').
\end{equation}
But this is nothing else than the split representation of the 4-point contact diagram that we discussed in the beginning. Hence,
\begin{equation}
    \mathcal{I}_C(r_1,r_2) = \frac{\delta_\lambda \eta_*^4}{8 k_1k_2k_3k_4k_s}\frac{r_1r_2}{r_1+r_2},
\end{equation}
where $\delta_\lambda = -3 \lambda_{\text{ren}}^2/(4\pi)^2 \varepsilon$
belongs to the local counterterm $\mathcal{L} \supset -\frac{\delta_\lambda}{4!} \varphi^4$ that is added to the Lagrangian. Of course the argument also holds for the other terms in case of a modified minimal subtraction, and diagramatically we remove the UV divergence by 
\begin{equation}
    \BubbleBuildingBlockNoText + \counterterm\,.
\end{equation}
Now that we understand the renormalization of the 1-loop diagram, let us come back to the chain of $N$ bubble loops. In this case, the spectral function is given by $\Sigma_\nu(\tilde\nu')^N$ and we already know that we can render $\Sigma_\nu$ finite by subtracting $P_\nu^\text{asy}$. The natural extension would therefore be to set $(\Sigma_\nu^N)^\text{ren} = (\Sigma_\nu^\text{ren})^N$. And indeed, this immediately provides the correct combinatorial structure, because this is equivalent to summing over all possible diagrams where bubble loops can be replaced by a bulk contact interaction 
\begin{equation}
    \RenormalizedBubbles\,.
\end{equation}
Of course there is one small subtlety here: We actually need three different counterterms, one for the interaction of the boundary fields $\mathcal{L} \supset -\frac{\delta_{\lambda_1}}{4!} \varphi^4$, the interaction of the bulk fields $\mathcal{L} \supset -\frac{\delta_{\lambda_2}}{4} (\phi^2)^2$ as well as the interaction of the bulk with the boundary fields $\mathcal{L} \supset -\frac{\delta_{\lambda_3}}{4} \varphi^2\phi^2$.
Hence, the renormalized $N$-loop integral is given by 
\begin{equation}
    \mathcal{I}_{\rangle\!\bigcirc^N\!\langle}(r_1,r_2) = 2(-1)^{N+1}\int_{-\infty}^{+\infty}\mathrm{d}\tilde{\nu}'\,\mathcal{N}_{\nu'}\mathcal{I}_{\Yright}(r_1;\nu')\mathcal{I}_{\Yright}(r_2;\nu') \Sigma_\nu^\text{ren}(\tilde\nu')^N
\end{equation}
and for the resummation of the bubble diagrams we can simply use the renormalized spectral function in the geometric series: 
\begin{equation}\label{eq:Inp}
\mathcal{I}_\text{np}^\nu(r_1,r_2) = -2\int_{-\infty}^{+\infty}\mathrm{d}\tilde{\nu}'\,\mathcal{N}_{\nu'}\mathcal{I}_{\Yright}(r_1;\nu')\mathcal{I}_{\Yright}(r_2;\nu') \frac{\lambda_{\text{ren}}}{1+\lambda_{\text{ren}}\Sigma_\nu^\text{ren}(\tilde\nu')}.
\end{equation}

\subsection{Spectral function in $d=2$ dimensions}
As previously pointed out in the literature \cite{DiPietro:2023inn, Loparco:2023rug}, major simplifications of the spectral function occur in $d=2$ dimensions. This is due to the fact that the $d=2$ case hits certain special cases of $\mathrm{\Gamma}$-functions which themselves can be simplified further using reflection and related identities for $\mathrm{\Gamma}$-functions. First, we take a look at the rather complicated looking expression \eqref{eq:P-nu} for $P_\nu(\tilde\nu')$. In $d=2$ we do not need to bother with dimensional regularization because the expression is already convergent. Plugging in $d=2$ and using
\begin{equation}
    \sum\limits_{m=0}^\infty \frac{1}{\left(m+\frac{d}{4}\pm i\tilde{\nu}\right)^2+\frac{\tilde{\nu}'^2}{4}} = \frac{1}{i\tilde{\nu}'}\left[\psi\left(\tfrac{d}{4}\pm i\tilde{\nu}+\tfrac{i\tilde{\nu}'}{2}\right)-\psi\left(\tfrac{d}{4}\pm i\tilde{\nu}-\tfrac{i\tilde{\nu}'}{2}\right)\right]
\end{equation}
with the digamma function $\psi$, we can simplify the sums to 
\begin{equation}
    P_\nu(\tilde\nu',d=2) = \frac{i\coth(\pi\tilde{\nu})}{32\pi\tilde{\nu}'} \sum\limits_{a,b=\pm}ab\,\psi\left(\tfrac{d}{4}+ i a\,\tilde{\nu}+ib\,\tfrac{\tilde{\nu}'}{2}\right).
\end{equation}
The first of three parts of the series representation drops completely due to the factor $\sin(\pi d/2)$.
Likewise, the Källén-Lehmann spectral function \eqref{eq:symr} can be simplified using $\mathrm{\Gamma}$-function identities. Using the reflection formula $\mathrm{\Gamma}(z)\mathrm{\Gamma}(1-z)= \pi/\sin(\pi z)$, we first find
\begin{equation}
    \rho_\nu^+(\tilde\nu') = \frac{\sinh^2(\pi\tilde{\nu}')}{32\pi \cosh\left(\pi(\tilde{\nu}+\frac{\tilde{\nu}'}{2})\right)\cosh\left(\pi(\tilde{\nu}-\frac{\tilde{\nu}'}{2})\right) \cosh^2\left(\pi \frac{\tilde{\nu}'}{2}\right)}.
\end{equation}
On the other hand, we can utilize the digamma reflection formula $\psi(\frac12 +z)-\psi(\frac12-z)= \pi\tan(\pi z)$ in order to rewrite the Källén-Lehmann spectral function in terms of digamma functions too. Multiplying by the pre-factor $\mathcal{N}_{\nu'}^{-1}$ from \eqref{eq:Nn} leads to the following expression for the Källén-Lehmann part in $d=2$:
\begin{equation}\label{eq:Sd=2}
    \mathcal{N}_{\nu}^{-1}\rho_\nu^+(\tilde{\nu}', d=2) = -\frac{i}{32\pi \tilde{\nu}'\cosh^2\left(\pi\frac{\tilde{\nu}'}{2}\right)}\sum\limits_{a,b=\pm}b\,\psi\left(\tfrac{d}{4}+ i a\,\tilde{\nu}+ib\,\tfrac{\tilde{\nu}'}{2}\right).
\end{equation}
Be aware, that the relative sign in the sum of digamma functions is different for $P_\nu$ and $\mathcal{N}_{\nu}^{-1}\rho_\nu^+$ (one is kind of the real and one the imaginary part of a difference between two digamma functions). The expression itself agrees quite well with previous expressions for the spectral function in terms of digamma functions \cite{DiPietro:2023inn, Sachs:2023eph}. However, here (for the first time) we can clarify and distinguish contributions from Wightman and time-ordered propagators, which was somewhat vaguely defined in previous literature on bubble resummations in de Sitter. For example, \cite{Loparco:2023rug} provides only the Källén-Lehmann part, while \cite{DiPietro:2023inn} only used the time-ordered part. 

 A natural follow-up question is whether there exists a similar (simple) expression for the spectral function in $d=3$. More precisely, it would be interesting whether we can reverse-engineer the previous argument and also write $P_\nu \propto \rho_\nu^+$ with some pre-factor that might be dimension-dependent and that we have not found so far. There are actually some arguments against such a general relation in arbitrary $d$:  In $d=3$ the Källén-Lehmann spectral function $\rho_\nu^+$ is a mereomorphic function that falls off exponentially in $\tilde\nu'$, while $P_\nu$ is given by a divergent sum. The renormalized $P_\nu-P_\nu^\text{asy}$ has a slower fall off at infinity. Furtermore, it is not necessarily positive since this property depends on the magnitude of the finite subtraction. Nonetheless, in appendix \ref{sec:3d-spectral} we were still able to fit a sum of digamma functions to the spectral function in $d=3$. 

\begin{figure}[t]
  \centering
  \begin{subfigure}{0.48\textwidth}
        \centering
        \includegraphics[width=\linewidth]{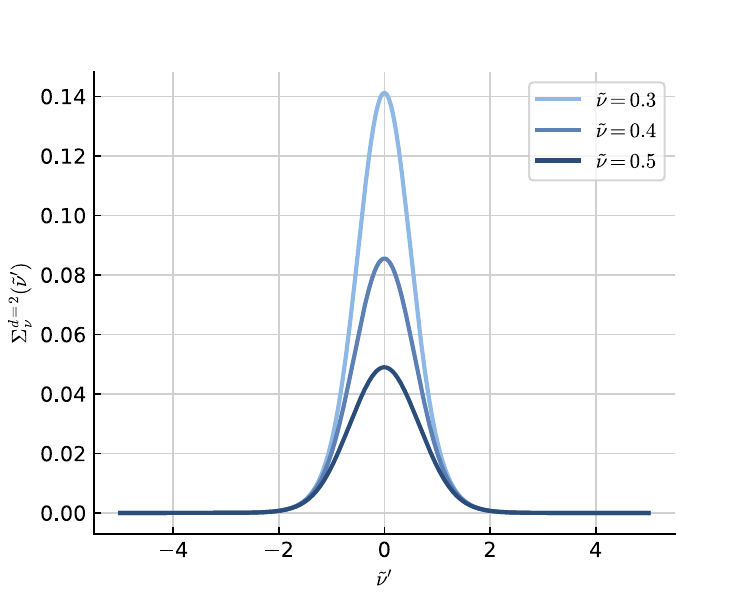}
    \end{subfigure}
    \hfill
    \begin{subfigure}{0.48\textwidth}
        \centering
        \includegraphics[width=\linewidth]{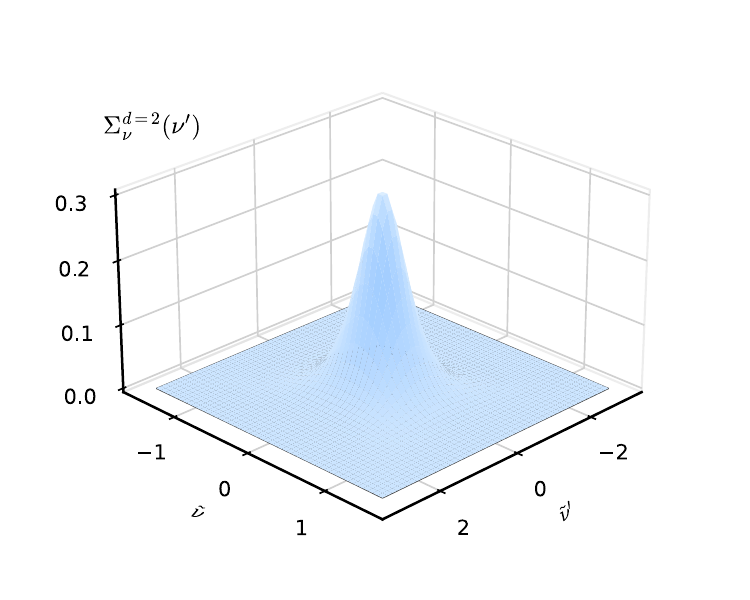}
    \end{subfigure}
    \label{fig:spectral2D}
  \caption{Spectral function in $d=2$ dimensions for different masses $\nu$ and $\nu'$. From these numerical studies we can see that the spectral function is indeed positive along the real axis. }
\end{figure}

\subsection{One-loop bubble}
Returning to $d=3$, the massive one-loop bubble diagram on a de Sitter background has been computed in the literature with different approaches. First, via a spectral integral over a tree-level diagram - using the spectral function bootstrapped from EdS / the sphere in \cite{Xianyu:2022jwk} and later with the Källen-Lehmann spectral function in \cite{Zhang:2025nzd}. Second, with the help of dispersion relations \cite{Liu:2024xyi}, together with the Partial Mellin-Barnes representations in \cite{Qin:2024gtr}. Here we will repeat the one-loop computation using our new spectral representation. The one-loop spectral integral is then given by
\begin{equation}\label{eq:Jn1}
    \mathcal{I}_{\rangle\!\bigcirc\!\langle}^\nu(r_1,r_2) = 2\int_{-\infty}^{+\infty}\mathrm{d}\tilde{\nu}'\, \mathcal{N}_{\nu'}\mathcal{I}_{\Yright}(r_1;\nu')\mathcal{I}_{\Yright}(r_2;\nu')\Sigma_\nu^\text{ren}(\tilde\nu')\,.
\end{equation}
To compute this spectral integral we close the contour appropriately and use the residue theorem. To simplify further computations, we use the following result that can be proven with a calculation closely following the derivation of (\ref{eq:hyperIdentity}):\\
\paragraph{Lemma:} Let a  $f(\tilde{\nu}') = f(-\tilde{\nu}')$ be a mereomorphic function and $\mathcal{P}(f)$ its the principal value in the sense of neglecting the poles from $f$ in the residue theorem. Then
\begin{align}\label{eq:basicResidueFormula}
    \int_{-\infty}^{+\infty}\mathrm{d}\tilde{\nu}'\, \mathcal{N}_{\nu'}\mathcal{I}_{\Yright}(r_1;\nu')\mathcal{I}_{\Yright}(r_2;\nu') \mathcal{P}\left(f(\tilde\nu')\right) = \frac{\eta_*^4 r_1}{32k_1k_2k_3k_4k_s}  \sum\limits_{n=0}^\infty \left\{(-1)^n f\left(-i(n+\tfrac12)\right)\left(\frac{r_1}{r_2}\right)^n\right.\nonumber \\
    \left. {}_2F_1\left[\begin{matrix}\tfrac12+\tfrac{n}{2},1+\tfrac{n}{2} \\ \tfrac32+n\end{matrix}\Big\vert r_1^2\right] {}_2F_1\left[\begin{matrix}-\tfrac{n}{2},\tfrac12-\tfrac{n}{2} \\ \tfrac12-n\end{matrix}\Big\vert r_2^2\right]\right\}.
\end{align}
With the help of this we can immediately read off one of the \emph{background terms} of the 1-loop result
\begin{align}
\mathcal{I}_{\rangle\!\bigcirc\!\langle,BG}^\nu(r_1,r_2) = -\frac{\eta_*^4 r_1}{16k_1k_2k_3k_4k_s}  \sum\limits_{n=0}^\infty \left\{(-1)^n \Sigma_\nu^\text{ren}\left(-i(n+\tfrac12)\right)\left(\frac{r_1}{r_2}\right)^n\right.\nonumber \\
    \left. {}_2F_1\left[\begin{matrix}\tfrac12+\tfrac{n}{2},1+\tfrac{n}{2} \\ \tfrac32+n\end{matrix}\Big\vert r_1^2\right] {}_2F_1\left[\begin{matrix}-\tfrac{n}{2},\tfrac12-\tfrac{n}{2} \\ \tfrac12-n\end{matrix}\Big\vert r_2^2\right]\right\}
\end{align}
which is an analytic function in $r_1/r_2$. This is exactly the EFT part of the correlator and in \cite{Lee:2025kgs} it was also argued how the above structure of the integral of an exchange diagram can be used to infer this EFT background. If we had different masses $\nu_1$ and $\nu_2$ running in the bubble loop, the Källén-Lehmann spectral function would contain the product $\prod_{\pm,\pm,\pm} \mathrm{\Gamma}\left(\tfrac12(\frac{d}{4}\pm i\tilde{\nu}_1\pm i\tilde{\nu}_2\pm i\tilde{\nu}')\right)$ which leads to four sets of poles corresponding to $\pm\nu_1\pm \nu_2$ and hence four contributions to the signal part of the correlator. For equal masses however, the combinations $\pm\nu_1\mp\nu_2$ would vanish and lead to poles contributing to the analytic background. This argument of course also translates to the poles of $P_\nu$. Hence, in the equal mass case it is not a priori obvious that the term above captures the full EFT background. However, it was already discovered in \cite{Xianyu:2022jwk} that the additional contribution will vanish in the de Sitter invariant case because the pre-factor has zeros at the locations of the poles at $\pm i (\frac{d}{2} + 2m),\, m\in\mathbb{N}$. \\ 

The remaining contributions to \eqref{eq:Jn1} stem from the poles of the spectral functions  $\rho_\nu^+(\tilde\nu')$ and $P_\nu(\tilde\nu')$ which have poles at the same locations, with all of them being  simple poles. \\

When performing the spectral integrals, we have to close the contour in an appropriate way. The functions appearing in the integrand have a sufficient fall-off behavior at infinity, so the only relevant terms that we have to focus on are the factors $(r_1r_2)^{\pm i\tilde{\nu}'}$ stemming from $F_\pm^{\nu',\frac12}(r_1)F_\pm^{\nu',\frac12}(r_2)$ and $(r_1/r_2)^{\pm i \tilde{\nu}'}$ stemming from $F_\pm^{\nu',\frac12}(r_1)F_\mp^{\nu',\frac12}(r_2)$. We have to split the integral into two separate pieces and close the contour in the contour in the lower half plane for a positive power $+i\tilde{\nu}'$ of the momentum ratios (remember that $0<r_1,r_2<1$ and we look at the kinematic region where $r_1<r_2$), and close in the upper half plane for negative powers $-i\tilde{\nu}'$. The contributions from $F_\pm^{\nu',\frac12}(r_1)F_\pm^{\nu',\frac12}(r_2)$ will then generate the non-local signal $\mathcal{I}_{\rangle\!\bigcirc\!\langle,NS}^\nu$ and the terms from $F_\pm^{\nu',\frac12}(r_1)F_\mp^{\nu',\frac12}(r_2)$ the local signal $\mathcal{I}_{\rangle\!\bigcirc\!\langle,LS}^\nu$. These two differ in their analytic properties in the squeezed limit $r_{1,2} \to 0$.

Explicitly, we find:
\begin{align}
    \mathcal{I}_{\rangle\!\bigcirc\!\langle,NS}^\nu = -\frac{\eta_*^4 \left[1+\sin\left(\frac{\pi}{2}(d+4\nu)\right)\right]}{64 \pi^{\frac{d+3}{2}}\sqrt{k_{12}k_{34}}k_1k_2k_3k_4} \sum\limits_{m=0}^\infty (-1)^m \mathcal{N}_{\frac{d}{2}+2m+2\nu} F_+^{\frac{d}{2}+2m+2\nu,\frac12}(r_1)F_+^{\frac{d}{2}+2m+2\nu,\frac12}(r_2) \nonumber\\\mathrm{\Gamma}\left[\begin{matrix}\frac{d+1}{2}, \frac{d}{2}+m, -m-\nu, -m-2\nu, \frac{d}{2}+m+\nu,\frac{d}{2}+m+2\nu \\ d,m+1,\frac12 -m -\nu, \frac{d+1}{2}+m+\nu\end{matrix}\right] \nonumber \\
    +(\nu \leftrightarrow -\nu),
\end{align}
\begin{align}
    \mathcal{I}_{\rangle\!\bigcirc\!\langle,LS}^\nu = -\frac{\eta_*^4 \left[1+\sin\left(\frac{\pi}{2}(d+4\nu)\right)\right]}{64 \pi^{\frac{d+3}{2}}\sqrt{k_{12}k_{34}}k_1k_2k_3k_4} \sum\limits_{m=0}^\infty (-1)^m \mathcal{N}_{\frac{d}{2}+2m+2\nu} F_+^{\frac{d}{2}+2m+2\nu,\frac12}(r_1)F_-^{\frac{d}{2}+2m+2\nu,\frac12}(r_2) \nonumber\\\mathrm{\Gamma}\left[\begin{matrix}\frac{d+1}{2}, \frac{d}{2}+m, -m-\nu, -m-2\nu, \frac{d}{2}+m+\nu,\frac{d}{2}+m+2\nu \\ d,m+1,\frac12 -m -\nu, \frac{d+1}{2}+m+\nu\end{matrix}\right] \nonumber \\
    +(\nu \leftrightarrow -\nu).
\end{align}

\begin{figure}[t]
  \centering
  \begin{subfigure}{0.49\textwidth}
    \centering
    \includegraphics[width=\linewidth]{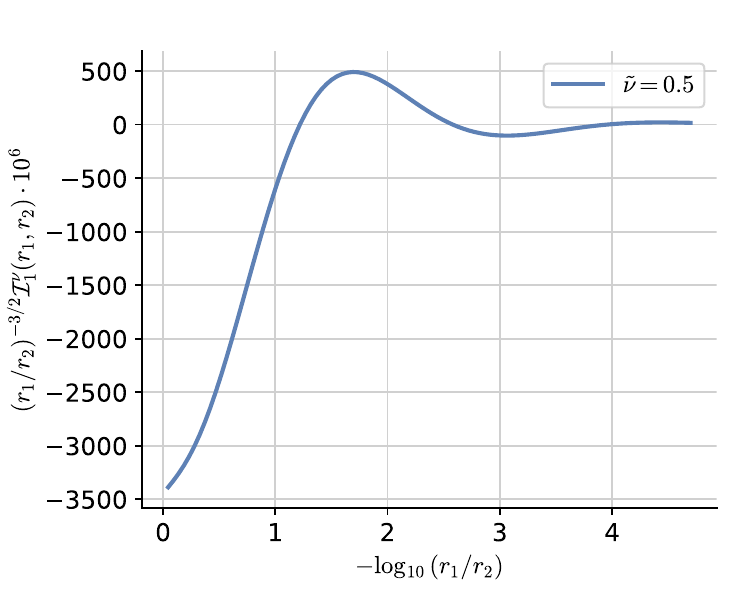}
  \end{subfigure}
  \hfill
  \begin{subfigure}{0.49\textwidth}
    \centering
    \includegraphics[width=\linewidth]{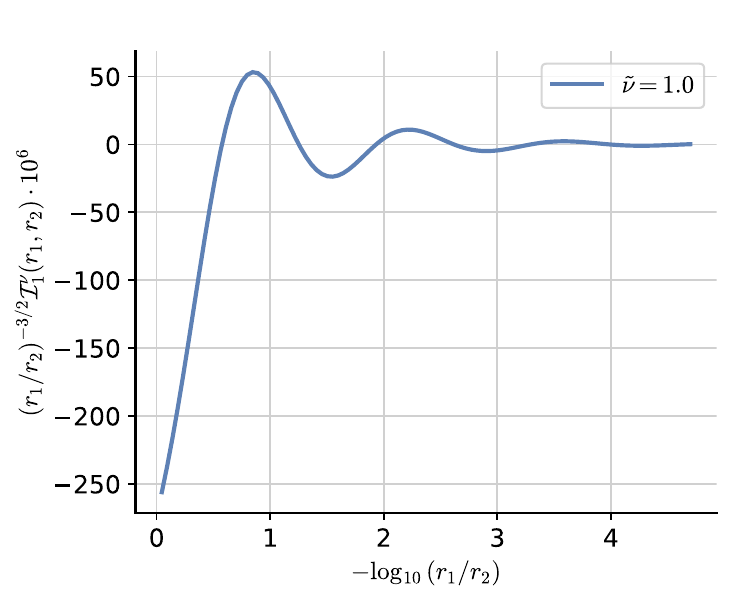}
  \end{subfigure}

  \medskip

  \begin{subfigure}{0.49\textwidth}
    \centering
    \includegraphics[width=\linewidth]{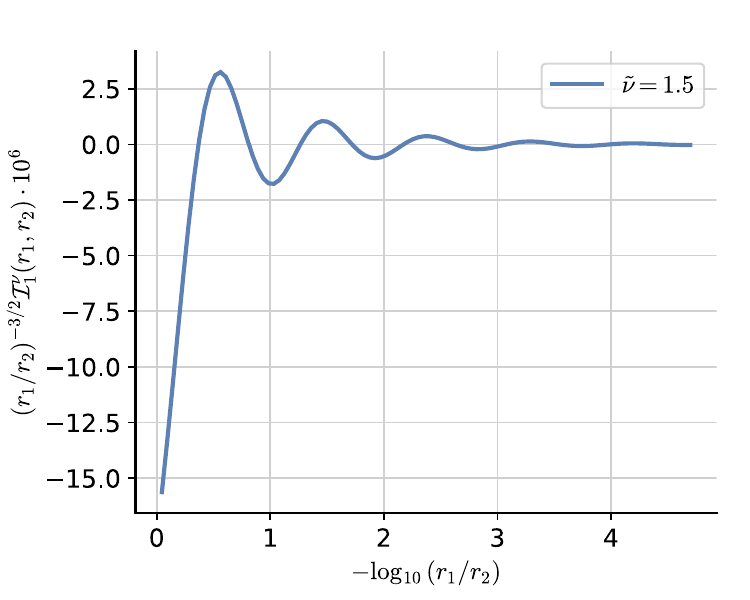}
  \end{subfigure}
  \hfill
  \begin{subfigure}{0.49\textwidth}
    \centering
    \includegraphics[width=\linewidth]{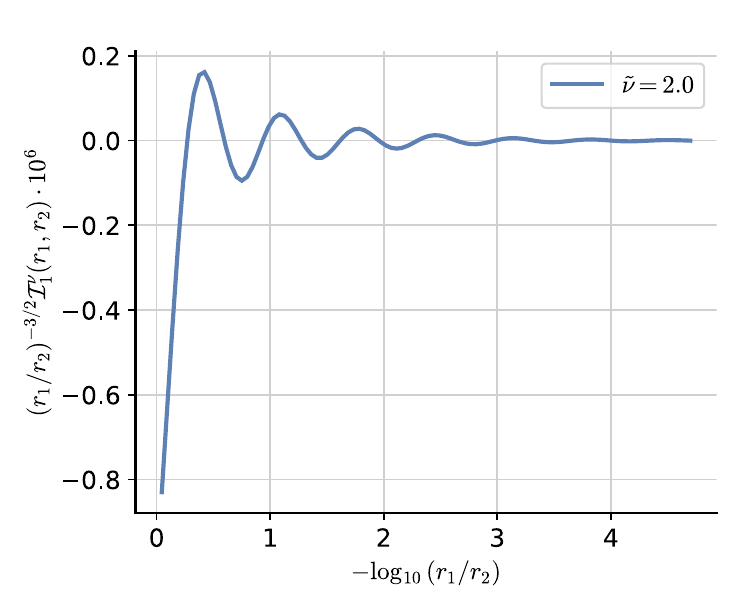}
  \end{subfigure}

  \caption{Plot of the signal part of the 1-loop correction to the four-point function, i.e. the sum of $\mathcal{I}_{\rangle\!\bigcirc\!\langle,NS}^\nu$ and $\mathcal{I}_{\rangle\!\bigcirc\!\langle,LS}^\nu$, in $d=3$ dimensions for four different masses $\nu$. We fixed $r_2 = 0.5$ and varied $r_1\in(10^{-5}, r_2)$. The sums in the expressions of the signals were truncated after 30 terms. For better visibility of the oscillations we scaled the result by  $(r_1/r_2)^{-3/2}$ so that only the $r_1$-dependence with power $\pm 2\nu$ survives in the squeezed limit, as well as a numerical factor of $10^6$. }
  \label{fig:signals-1L}
\end{figure}
We plotted the signal parts of the 1-loop correlator in Figure \ref{fig:signals-1L} where one can clearly see the oscillations $\propto \cos\left[2\tilde{\nu}\log r_1+ \theta\right]$ which coined the term cosmological collider signals. The amplitude of oscillations is still decaying in the scaling chosen in these plots because there is another dependence on $\sqrt{r_1r_2}$ in the prefactor stemming from $(k_{12}k_{34})^{-1/2}$. So, if we would scale by $(r_1/r_2)^{-2}$ instead of $(r_1/r_2)^{-3/2}$, we would get oscillations with constant amplitude. 
We did not plot the background (EFT) part of the correlator because the signals are suppressed compared to the background by several orders of magnitude for our example. Furthermore, when evaluating the background one needs an explicit expression for $\Sigma_\nu^\text{ren}$ in $d=3$ and as we mentioned earlier, the respective series expression for the time-ordered part is extremely slowly converging and might screw up the numerical evaluation. 

From these results we can also read off the squeezed limit, which is interesting for typical cosmological collider applications. For $r_1 \ll r_2\ll 1$ we find the squeezed signals
\begin{align}
    \lim\limits_{r_1\ll r_2\ll1 } \mathcal{I}_{\rangle\!\bigcirc\!\langle,NS}^\nu =\frac{\eta_*^4 \left[1+\sin\left(\frac{\pi}{2}(d+4\nu)\right)\right]}{2^{5-d} \pi^{\frac{d+2}{2}}\sqrt{k_{12}k_{34}}k_1k_2k_3k_4}  \mathrm{\Gamma}\left[\begin{matrix}- \nu, -2\nu, \frac{d}{2}+\nu,-\frac{d}{2}-2\nu\\ \frac12 -\nu, \frac{d+1}{2}+\nu\end{matrix}\right]  \nonumber \\
    \mathrm{\Gamma}(\tfrac{d+1}{2}+2\nu)^2\left(\frac{r_1r_2}{4}\right)^{\frac{d}{2}+2\nu}+(\nu \leftrightarrow -\nu) \\
    \lim\limits_{r_1\ll r_2\ll1 } \mathcal{I}_{\rangle\!\bigcirc\!\langle,LS}^\nu =\frac{\eta_*^4 \left[1+\sin\left(\frac{\pi}{2}(d+4\nu)\right)\right]}{2^{5-d} \pi^{\frac{d+2}{2}}\sqrt{k_{12}k_{34}}k_1k_2k_3k_4}  \mathrm{\Gamma}\left[\begin{matrix}- \nu, -2\nu, \frac{d}{2}+\nu,+\frac{d}{2}+2\nu\\ \frac12 -\nu, \frac{d+1}{2}+\nu\end{matrix}\right]  \nonumber \\
    \frac{\pi}{\cos\left(\pi(\tfrac{d}{2}+2\nu)\right)}\left(\frac{r_1}{r_2}\right)^{\frac{d}{2}+2\nu}+(\nu \leftrightarrow -\nu)
\end{align}
that provide the typical oscillations in the momentum ratios $\propto (r_1r_2)^{\pm 2\nu}$ and $\propto (r_1/r_2)^{\pm 2\nu}$. The background term in the squeezed limit behaves as 
\begin{align}
    \lim\limits_{r_1\ll r_2\ll1 } \mathcal{I}_{\rangle\!\bigcirc\!\langle,BG}^\nu =-\frac{\eta_*^4 \,\Sigma_\nu(-\tfrac{i}{2})}{16 \sqrt{k_{12}k_{34}}k_1k_2k_3k_4} \left(\frac{r_1}{r_2}\right)^{1/2}.
\end{align}
Interestingly, for our choice of time-dependence at the vertices connected to the boundary (de Sitter invariant couplings with $p=-2$) the background decays slower in the squeezed limit $\propto (r_1/r_2)^{1/2}$ compared to the signals which decay as $(r_1r_2)^{3/2}$ and $(r_1/r_2)^{3/2}$ respectively. Hence the signals will not be very well visible in cosmological applications. This is different from the time-dependence that was investigated in \cite{Xianyu:2022jwk}, where the trispectrum for the case $p=0$ was explored and in which the background decays with $(r_1/r_2)^{5/2}$, so that the signal was enhanced in the squeezed limit. Therefore, claims that the signal parts of cosmological correlators dominate over the background in the squeezed limit are in general not true and require certain bounds on the time dependence of vertices. 

\subsection{Resummed bubble chain}

In this section we will discuss some aspects of the resummed correlator. Using our Lemma \eqref{eq:basicResidueFormula}, we can immediately write down the analytic background of any $N$-loop diagram, which is given by
\begin{align}
    \mathcal{I}_{\rangle\!\bigcirc^N\!\langle,BG}^\nu (r_1,r_2) = (-1)^N\frac{\eta_*^4 r_1}{16 k_1k_2k_3k_4k_s}\sum\limits_{n=0}^\infty \left\{(-1)^n \Sigma_\nu^\text{ren}\left(-i(n+\tfrac12)\right)^N \left(\frac{r_1}{r_2}\right)^n\right. \nonumber \\
    \left. {}_2F_1\left[\begin{matrix}\tfrac12+\tfrac{n}{2},1+\tfrac{n}{2} \\ \tfrac32+n\end{matrix}\Big\vert r_1^2\right] {}_2F_1\left[\begin{matrix}-\tfrac{n}{2},\tfrac12-\tfrac{n}{2} \\ \tfrac12-n\end{matrix}\Big\vert r_2^2\right]\right\}.
\end{align}
Likewise, the resummed analytic background is simply given by
\begin{align}\label{eq:resummed-bg}
    \mathcal{I}_{\text{np},BG}^\nu (r_1,r_2) = \frac{\lambda_\text{ren}\eta_*^4 r_1}{16 k_1k_2k_3k_4k_s}\sum\limits_{n=0}^\infty \left\{ \frac{(-1)^n}{1+\lambda_\text{ren} \Sigma_\nu^\text{ren}\left(-i(n+\tfrac12)\right)} \left(\frac{r_1}{r_2}\right)^n\right. \nonumber \\
    \left. {}_2F_1\left[\begin{matrix}\tfrac12+\tfrac{n}{2},1+\tfrac{n}{2} \\ \tfrac32+n\end{matrix}\Big\vert r_1^2\right] {}_2F_1\left[\begin{matrix}-\tfrac{n}{2},\tfrac12-\tfrac{n}{2} \\ \tfrac12-n\end{matrix}\Big\vert r_2^2\right]\right\}.
\end{align}

As we have seen for the 1-loop case, the background dominates in the squeezed limit since the signals have a suppression with a factor $r_{1,2}^{d/2}$. We would expect a very similar behavior from the resummed correlator, hence in our particular choice of de Sitter invariant vertices we expect the leading contribution to stem from the background.

The trickier part of the analysis is finding the signal part of the resummed correlator. For this, we would need the poles of the resummed spectral function $\left(1+\lambda_{\text{ren}} \Sigma_\nu^\text{ren}\right)^{-1}$ in \eqref{eq:Inp}, which is equivalent to finding the zeros of $1+\lambda_{\text{ren}} \Sigma_\nu^\text{ren}$. Due to the complicated form of our spectral function \eqref{eq:sigmaren}, it is not obvious how to find the exact solutions to this equation. However, we can discuss the general structure of the solution as well as some limiting cases. \\

Suppose that $\mathcal{L}$ is an index set describing the set of poles of $ \Sigma_\nu$ and let $p_l,\,l\in\mathcal{L}$ be one such pole. For notational convenience we drop the sub-/ superscript 'ren' in this discusssion. On the real axis, the spectral function $\Sigma_\nu$ is positive. Extended to the full complex plane, the spectral function has predominantly positive real part except for local sign reversals at the poles. Hence, for positive $\lambda > 0$, the resummed spectral function $\left(1+\lambda  \Sigma_\nu\right)^{-1}$ can only have poles in a neighborhood of the poles $p_l$. That means, there exists a bijective mapping $p_l \mapsto \tilde{p}_l(\lambda)$ where $\tilde{p}_l$ is a pole of the resummed spectral function. To see why this must hold, we can formalize our argument:\\
\begin{figure}[t]
  \centering
  \begin{subfigure}{0.48\textwidth}
        \centering
        \includegraphics[width=\linewidth]{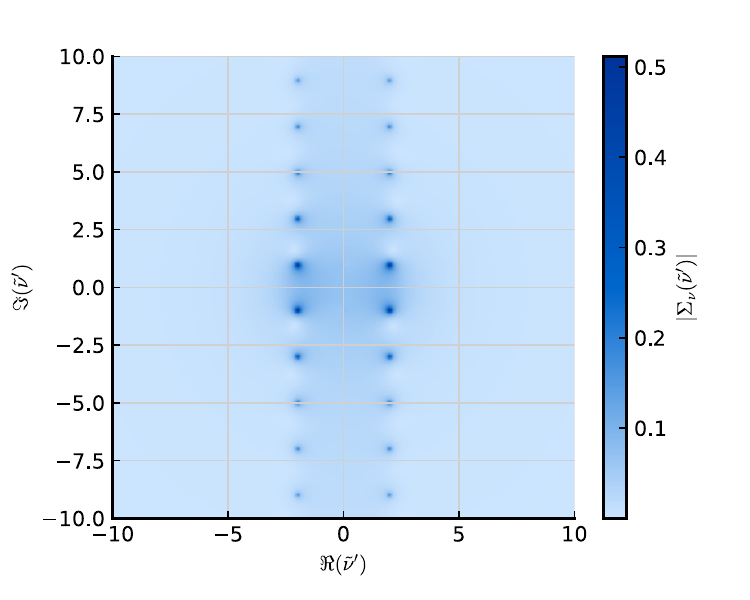}
    \end{subfigure}
    \hfill
    \begin{subfigure}{0.48\textwidth}
        \centering
        \includegraphics[width=\linewidth]{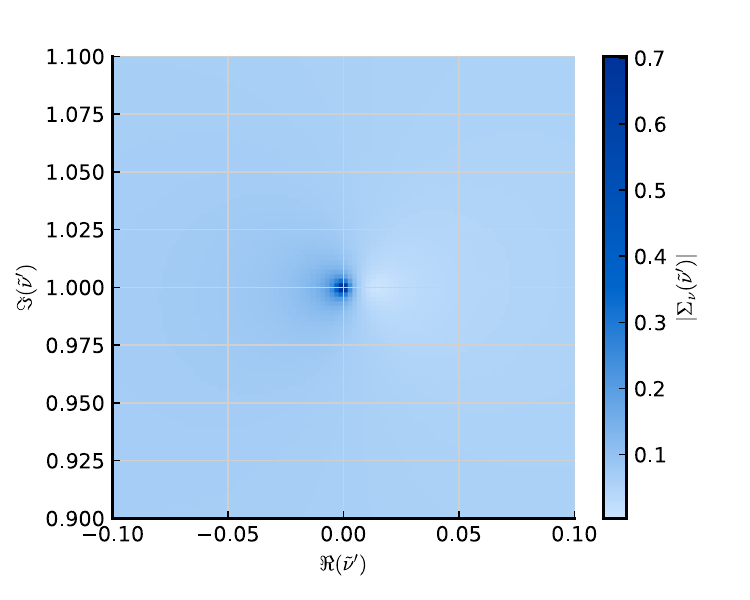}
    \end{subfigure}
    \label{fig:cspectral2D}
  \caption{Pole structure of the 1-loop spectral function $\Sigma_\nu(\tilde{\nu}')$ for $\tilde{\nu}=2.0$ in the complex plane which has poles at $\pm i(\frac{d}{2}+2m)$ and $\pm i(\frac{d}{2}+2m\pm \nu),\,m\in\mathbb{N}$. The poles at $\pm i(\frac{d}{2}+2m)$ are not visible in the left plot, so we zoomed in on one of these poles on the right.
  For simplicity, we plotted the spectral function in $d=2$. The $d=2$ spectral function is not UV divergent and hence we do not need to renormalize it. It is clearly visible that $\Sigma_\nu$ has positive real part almost everywhere in the complex plane except for small neighborhoods around the poles. }
\end{figure}\\
Let $\Sigma_\nu(z)$ be a meromorphic function such that $\Sigma_\nu(z) >0$ for $z\in\mathbb{R}$ and so that $\Sigma_\nu$ has simple poles $p_l,\,l\in\mathcal{L}$ in $\mathbb{C}\setminus\mathbb{R}$. Furthermore, there exists $c>0$ such that $\Re[\Sigma_\nu(z)] \geq c$ for all $z \in\mathbb{C}$ outside a sufficiently small neighborhood of its poles. 
For $\lambda>0$, we define a function
\begin{equation}
    \tilde{\Sigma}_\nu^\lambda(z) = \frac{1}{1+\lambda\Sigma_\nu(z)}.
\end{equation}
The poles of this function $\tilde{\Sigma}_\nu^\lambda(z)$ are the zeros of the function $g_\lambda(z) = 1+\lambda \Sigma_\nu(z)$. We choose disjoint disks $D_\varepsilon(p_l)$ around the poles $p_l$ of $\Sigma_\nu$ so that $D_\varepsilon(p_l) \cap D_\varepsilon(p_j) = \varnothing$ for $l\neq j$ and on the complement
\begin{equation}
    \mathcal{C} = \mathbb{C}\setminus \bigcup\limits_{l\in\mathcal{L}}D_\varepsilon(p_l)
\end{equation}
we have $\Re[\Sigma_\nu(z)] \geq c >0 $ for all $z\in\mathcal{C}$. Then, $g_\lambda (z) = 1+\lambda\Sigma_\nu(z) \geq 1+\lambda c >0$ for all $z\in\mathcal{C}$ and therefore $g_\lambda \neq 0$ on $\mathcal{C}$. Thus, all poles of $\tilde{\Sigma}_\nu^\lambda$ must lie in close neighborhoods of the poles of $\Sigma_\nu$.

Locally around a pole $p_l$ of $\Sigma_\nu$ we can write
\begin{equation}
    \Sigma_\nu(z) = \frac{r_l}{z-p_l} + h_l(z),
\end{equation}
where $r_l$ is the residue of $p_l$ and $h_l(z)$ is a holomorphic function near $p_l$. Therefore we can locally write
\begin{equation}
    (z-p_l)g_\lambda(z) = \lambda r_l + (z-p_l)\left[1+\lambda h_l(z)\right].
\end{equation}
The right hand side is holomorphic near $p_l$ and for sufficiently small $z-p_l$ the term $\lambda r_l$ dominates. By Rouché's theorem (or by the holomorphic implicit function theorem), there exists exactly one zero $\tilde{p}_l^\lambda$ of $g_\lambda$ in a sufficiently small disk $D_\varepsilon(p_l)$. To leading order, we find 
\begin{equation}
    \tilde{p}_l^\lambda = p_l -\lambda r_l + \mathcal{O}(\lambda^2).
\end{equation}
Hence, the poles are shifted (to first order) in the direction opposite to the sign of its residues. Moreover, we can also show that the pole of $\tilde{\Sigma}_\nu^\lambda$ is a simple pole too. By differentiation, we have $g_\lambda' = \lambda \Sigma_\nu'$ and near $p_l$ we have $\Sigma_\nu' \sim -r_l/(z-p_l)^2$ so that $g_\lambda ' \neq 0$. Since the zero of $g_\lambda$ is simple, also the pole of $\tilde{\Sigma}_\nu^\lambda$ must be simple. 

To summarize: As $\lambda \in\mathbb{R}_+$ varies, the poles of $(1+\lambda \Sigma_\nu)^{-1}$ move continuously and in fact analytically in the complex plane. Each simple pole $p_l$ of $\Sigma_\nu$ generates a unique analytic trajectory of simple poles $\tilde{p}_l(\lambda)$ of $(1+\lambda \Sigma_\nu)^{-1}$, yielding a one-to-one correspondence between the two pole sets. We refer to this one-parameter family as the \textit{pole flow} associated with $\lambda$. 

In fact, it is possible to write down a differential equation for the pole flow. Since the locations of the poles fulfill $1+\lambda \Sigma_\nu(\tilde{p_l}) = 0$, by differentiating with respect to $\lambda$, we obtain $\Sigma_\nu(\tilde{p}_l) + \lambda \Sigma_\nu'(\tilde{p}_l) \dot{\tilde{p}}_l =0$ and hence
\begin{equation}
    \frac{\mathrm{d}\tilde{p}_l}{\mathrm{d}\lambda}  = -\frac{\Sigma_\nu(\tilde{p}_l)}{\lambda \Sigma_\nu'(\tilde{p}_l)}.
\end{equation}
This is reminiscent of the analysis in \cite{Sachs:2023eph} where the analogous feature was found for the anomalous dimension of double trace operators. This is consistent since these double trace operators correspond to bulk-to-boundary propagators in dS and thus become visible in our split representation, while they would be harder to identify in the Käll\'en-Lehmann representation \cite{Xianyu:2022jwk,Zhang:2025nzd}. 

\begin{figure}[t]
  \centering
  \includegraphics[width=0.95\linewidth]{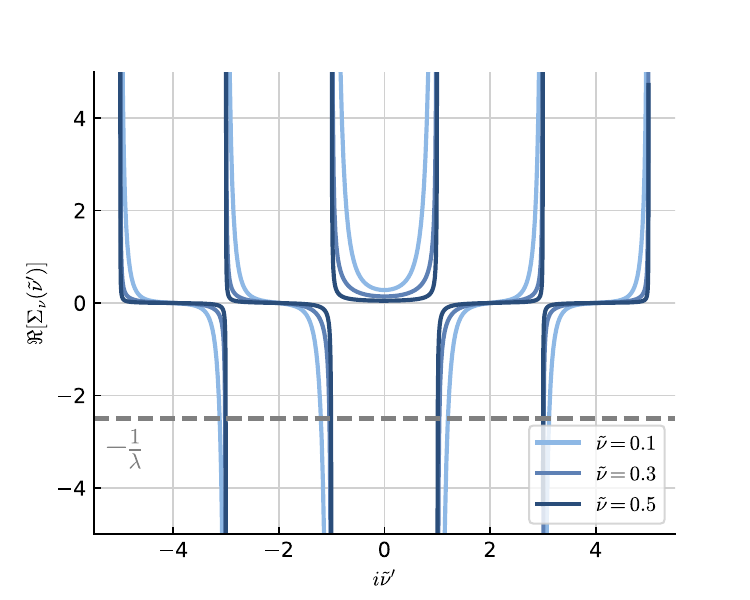}
  \caption{Pole structure of the 1-loop spectral function $\Sigma_\nu(\tilde{\nu}')$ along the imaginary axis. For simplicity, we plotted the spectral function in $d=2$, which has poles on the imaginary axis at $\pm i(2m+1),\,m\in\mathbb{N}$. This plot was also shown in \cite{DiPietro:2023inn}, however there the contributions from the time-ordered propagators were missing. The poles of the resummed spectral function are found at the intersection with the constant plane at $-1/\lambda$ (the sign is also different in our case compared to \cite{DiPietro:2023inn} where there seem to be some issues regarding the positivity of the spectral function, and consequently unitarity). There is one subtlety in this plot: We only show $\Re[\Sigma_\nu]$ and hence intersections in this plot are not necessarily the actual zeros in the complex plane, but the general idea should be clear.}
\end{figure}

Additionally, we can also make some statements about the residues of the shifted poles. For a pole $\tilde{p}_l^\lambda $ of $\tilde{\Sigma}_\nu^\lambda=(1+\lambda \Sigma_\nu)^{-1}$ (which is a zero of the denominator), the residue formula shows
\begin{equation}
    \tilde{r}_l^\lambda = \mathrm{Res}_{\tilde{p}_l^\lambda} \tilde{\Sigma}_\nu^\lambda = \frac{1}{\left(1+\lambda \Sigma_\nu\right)'(\tilde{p}_l^\lambda)} = \frac{1}{\lambda \Sigma_\nu'(\tilde{p}_l^\lambda)}.
\end{equation}
In the vicinity of $p_l$, we can again expand $\Sigma_\nu$ around $p_l$ and find
\begin{equation}
    \Sigma_\nu'(\tilde{p}_l^\lambda) = -\frac{r_l}{(\tilde{p}_l^\lambda-p_l)^2} + h_l'(\tilde{p}_l^\lambda).
\end{equation}
Using the pole equation $\tilde{p}_l^\lambda = p_l -\lambda r_l + \mathcal{O}(\lambda^2)$ we find
\begin{equation}
    \Sigma_\nu'(\tilde{p}_l^\lambda) = -\frac{1}{\lambda^2 r_l} +\mathcal{O}(1).
\end{equation}
Therefore, we arrive at
\begin{equation}
    \tilde{r}_l^\lambda = -\lambda r_l + \mathcal{O}(\lambda^2).
\end{equation}
Remember, that in our definition of the 1-loop bubble integral, we would still need to multiply with a factor $(-\lambda)$ to compute the actual diagram while for the re-summed case the spectral function has already absorbed all these factors. Therefore, to the first order, the residue of the resummed spectral function is actually the same as in the 1-loop case. 

\begin{figure}[t]
  \centering
  \begin{subfigure}{0.48\textwidth}
        \centering
        \includegraphics[width=\linewidth]{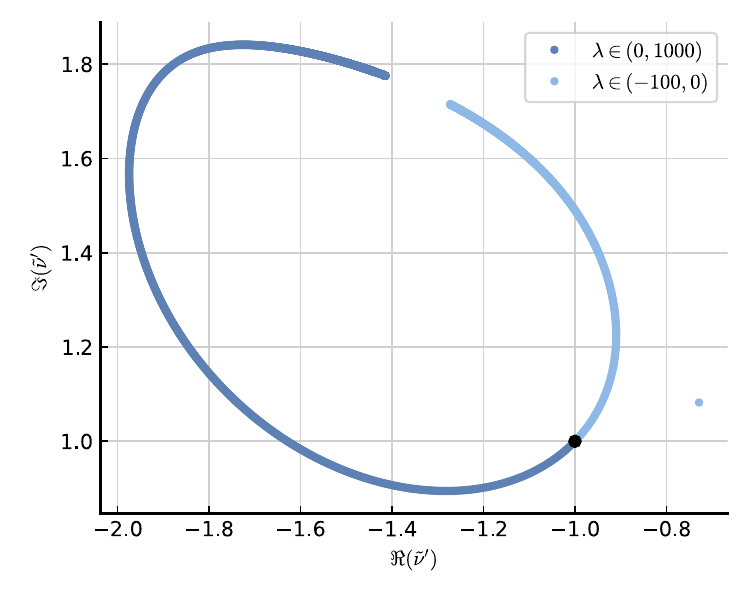}
    \end{subfigure}
    \hfill
    \begin{subfigure}{0.48\textwidth}
        \centering
        \includegraphics[width=\linewidth]{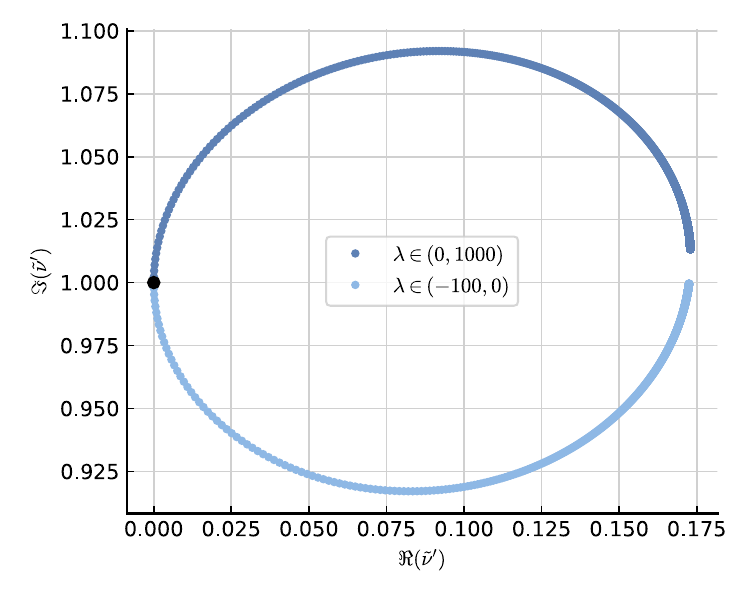}
    \end{subfigure}
  \caption{Representation of the pole flow for the spectral function in $d=2$. For a mass $\tilde{\nu}= 0.5$, we look at two poles of the 1-loop spectral function at $\tilde{\nu}' =-1+i$ (left) and $\tilde{\nu}'=i$ (right). 
  In these plots, we vary the coupling strength $\lambda\in(-100, 0)$ (light blue) and $\lambda\in(0,1000)$ (dark blue) and plot the poles of the re-summed spectral function $(1+\lambda \Sigma_\nu)^{-1}$ in steps of $\Delta \lambda =0.1$, creating this flow-like trajectory. The location of the poles was found numerically by locating the root of $1+\lambda \Sigma_\nu$, using the previous poles as an initial guess for the search of the next poles after increasing $\lambda$ by $\Delta\lambda$.}
  \label{fig:pole-flow}
\end{figure}

Let us come back to the initial problem of finding the signal part of the resummed correlator. The signal stems from the poles of the spectral function and as we have seen, these poles are related to the poles of the 1-loop bubble spectral function by an analytic flow in the complex plane. Recall that the poles of the 1-loop spectral function lie at $\pm i(\tfrac{d}{2} +2m \pm i \tilde{\nu}\pm i\tilde{\nu}),\,m\in\mathbb{N}$. 
As can be seen in Figure \ref{fig:pole-flow}, the shift of the poles has a non-zero real part. Hence the new poles $\tilde{p}_l^\lambda = p_l -\lambda r_l + \mathcal{O}(\lambda^2)$ can be interpreted (to first order) as a mass shift $\pm 2\nu \to \pm 2\nu - \lambda \Re[r_l]$.
However, since the different poles do not have equal residues, this is not a \textit{global} mass shift but a shift of every term in the sums that describe the non-local and local signal. Since the residues also stay at the same location as in the 1-loop case, to first order the resummation purely amounts to this mass shift and does not change any of the other factors in the series expansion of the signals. 
However, we can see much more going on in these plots for the $d=2$ spectral function (Figure \ref{fig:pole-flow}). Surprisingly, the pole flow closes on itself, meaning that the location of the poles is the same in the limits $\lambda \to -\infty$ and $\lambda\to +\infty$. Something similar was observed in \cite{Nowinski:2025cvw}, where the resummed correlator for conformally coupled scalars approached similar values in these limits. 
From the point of view of the spectral function, this behavior is totally expected. In the limits $\lambda \to \pm \infty$, the zeros of the function $1+\lambda \Sigma_\nu$ are simply the zeros of the spectral function itself and the sign of $\lambda$ does not matter. Therefore, also the residues should coincide in these two limits so that the correlator itself is the same (up to a global sign). 

Moreover, it is interesting that the trajectory of the pole flow for $\tilde{\nu}'=\pm(2m+1)i$ is following an elliptic orbit in the right half of the complex plane - breaking the expected $\mathbb{Z}_2$ symmetry of the spectral function. It is quite possible, that there is a second, symmetric, trajectory in the left half plane because initially these poles stem from the doubled $\mathrm{\Gamma}(\tfrac{d}{4}\pm\tfrac{i\tilde{\nu}'}{2})$ factors in the spectral function, so that in the resummed case they are actually acquiring a positive and a negative mass. 

Up to the usual pre-factors, the signals (in general) take the form
\begin{align}
    \mathcal{I}_{\text{np},NS}^\nu (\lambda;r_1,r_2) &\sim \sum\limits_{l\in\mathcal{L}}\tilde{r}_l^\lambda \mathcal{N}_{i\tilde{p}_l^\lambda} F_+^{i\tilde{p}_l^\lambda,\frac12}(r_1)F_+^{i\tilde{p}_l^\lambda,\frac12}(r_2) + \text{c.c.}, \\
    \mathcal{I}_{\text{np},LS}^\nu (\lambda;r_1,r_2) &\sim \sum\limits_{l\in\mathcal{L}}\tilde{r}_l^\lambda \mathcal{N}_{i\tilde{p}_l^\lambda} F_+^{i\tilde{p}_l^\lambda,\frac12}(r_1)F_-^{i\tilde{p}_l^\lambda,\frac12}(r_2) + \text{c.c.},
\end{align}
which amounts to oscillations with a shifted mass parameter in the squeezed limit. In appendix \ref{sec:shift-formula}, we also derive a formula for the full resummed result in terms of a generalized shift. For small $\lambda$ we can employ the approximate shift formulas we derived and find the leading order in $\lambda$:

\begin{align}
    \mathcal{I}_{\text{np},NS}^{\nu,\lambda} = \frac{\lambda\eta_*^4 \left[1+\sin\left(\frac{\pi}{2}(d+4\nu)\right)\right]}{64 \pi^{\frac{d+3}{2}}\sqrt{k_{12}k_{34}}k_1k_2k_3k_4} \sum\limits_{m=0}^\infty (-1)^m  F_+^{\frac{d}{2}+2m+2\nu+i \lambda r_l,\frac12}(r_1)F_+^{\frac{d}{2}+2m+2\nu+i \lambda r_l,\frac12}(r_2) \nonumber\\ \mathcal{N}_{\frac{d}{2}+2m+2\nu+i \lambda r_l}\mathrm{\Gamma}\left[\begin{matrix}\frac{d+1}{2}, \frac{d}{2}+m, -m-\nu, -m-2\nu, \frac{d}{2}+m+\nu,\frac{d}{2}+m+2\nu \\ d,m+1,\frac12 -m -\nu, \frac{d+1}{2}+m+\nu\end{matrix}\right] \nonumber \\
    +(\nu \leftrightarrow -\nu),
\end{align}
\begin{align}
    \mathcal{I}_{\text{np},LS}^{\nu,\lambda} = \frac{\lambda\eta_*^4 \left[1+\sin\left(\frac{\pi}{2}(d+4\nu)\right)\right]}{64 \pi^{\frac{d+3}{2}}\sqrt{k_{12}k_{34}}k_1k_2k_3k_4} \sum\limits_{m=0}^\infty (-1)^m  F_+^{\frac{d}{2}+2m+2\nu+i \lambda r_l,\frac12}(r_1)F_-^{\frac{d}{2}+2m+2\nu+i \lambda r_l,\frac12}(r_2) \nonumber\\ \mathcal{N}_{\frac{d}{2}+2m+2\nu+i \lambda r_l}\mathrm{\Gamma}\left[\begin{matrix}\frac{d+1}{2}, \frac{d}{2}+m, -m-\nu, -m-2\nu, \frac{d}{2}+m+\nu,\frac{d}{2}+m+2\nu \\ d,m+1,\frac12 -m -\nu, \frac{d+1}{2}+m+\nu\end{matrix}\right] \nonumber \\
    +(\nu \leftrightarrow -\nu).
\end{align}

So far, our discussion revolved around the case $\lambda >0$ which ensures positivity of the spectral function. When the coupling constant $\lambda $ is taken to be \textit{negative}, the structure of the resummed four-point function changes qualitatively. In this case, the factor $1+\lambda \Sigma_\nu$ in the denominator of the resummed spectral function can vanish or become negative in regions where the 1-loop spectral density is positive. Since $\Sigma_\nu$ itself is fixed by kinematics and corresponds to a sum over physical intermediate states, this behavior cannot be compensated by a change of variables or conventions. As a result, the effective spectral function of the resummed correlator is no longer guaranteed to be positive. 

From the point of view of unitarity, this signals a potential breakdown of the probabilistic interpretation. Positivity of the spectral function is required to ensure that discontinuities across physical cuts correspond to sums over positive-norm intermediate states, as dictated by the optical theorem. When $1+\lambda \Sigma_\nu <0$, the discontinuity of the resummed correlator acquires the wrong sign, indicating that the corresponding contribution cannot be interpreted as a physical correlation function. In particular, this behavior violates the positivity implied by cutting rules and suggests the appearance of negative-norm or ghost-like contributions in the effective description. 

Closely related to this issue is the appearance of poles in the resummed four-point function when $1+\lambda \Sigma_\nu=0$. For negative $\lambda$, such poles can occur not only as shifts of the 1-loop poles like in the $\lambda>0$ case, but also on the physical sheet and correspond to tachyonic or unstable bound states in the exchange channel. In flat space, this phenomenon is well known in large-$N$ scalar theories, where a negative quartic coupling leads to vacuum instability and the breakdown of the large-$N$ expansion due to the potential being unbounded from below. In the present de Sitter setting, the same mechanism manifests itself at the level of the four-point function as a loss of spectral positivity and the emergence of unphysical singularities.

It is important to emphasize that these pathologies are not artifacts of the resummation procedure. Rather, the large-$N$ resummation faithfully exposes the underlying inconsistency of the theory for $\lambda <0$. While perturbation theory at finite loop order may obscure these issues, the resummed result makes clear that a negative coupling generically conflicts with unitarity and stability, unless additional physics intervenes to regulate or reinterpret the spectrum.

However, the instability occurs only on a finite interval $(\lambda_{\min}, \lambda_{\max})\subset \mathbb{R}_-$ in which the physical sheet can generate zeros of the denominator. For any given $\nu$ in the principle series we have observed that the 1-loop spectral function $\Sigma_\nu$ is also bounded from above by some $\Sigma_\nu^{\max} = \sup_{\tilde{\nu}'\in\mathbb{R}} \Sigma_\nu(\tilde{\nu}')$. 
Hence, for $\lambda < -1/\Sigma_\nu^{\max}$, the re-summed spectral function is strictly negative on the real axis and similar arguments as in the $\lambda >0$ case show that the spectral function can only have poles (where the denominator changes sign) close to the poles of the 1-loop spectral function $\Sigma_\nu$. In particular, the respective shift of the poles somewhat mirrors the behavior in the $\lambda >0$ configuration. This fully explains the oscillatory behavior of the correlator on a finite negative interval of coupling constants that was observed in \cite{Nowinski:2025cvw}.

If we return to $d=3$, we recall that  positivity of the spectral function depends on the renormalization scheme, since we can always add or subtract any constant by adding a finite contact counterterm. However, adding such a term amounts to a redefinition of the coupling $\lambda$ in just such a way that $\frac{\lambda}{1+\lambda \Sigma_\nu}$ is invariant. This is the content of the Callan-Symanzik equation. The important property is therefore just that the spectral function is \emph{bounded from below}. If this bound is below zero, then there might be some additional poles for the resummed spectral function, but they should amount to contributions that can be removed by adding a cross diagram.

Since, unlike for $d=2$, there is not a closed expression for $\Sigma_\nu$ we are not able to repeat the numerical analysis  with the same rigor as for $d=2$ above. However, we do not expect a qualitative difference since the spectral function for $d=3$ can be reasonably well approximated by  the same class of functions as for $d=2$. We illustrate a numerical fit of $\Sigma_\nu$ in appendix \ref{sec:3d-spectral}.

\section{Conclusions}
In this paper we succeeded in resumming an infinite number of loops of a massive scalar field in de Sitter space. While the resummation of the corresponding flat space diagrams is standard text book material, this is a non-trivial problem in a de Sitter (or any time dependent) background due to energy non-conservation. What made this calculation possible is a combination of Källén-Lehmann representations of 1-loop graphs and split representations of Schwinger-Keldysh vertices which allowed us to factorize $N$-loop diagrams in the spectral representation. 

An additional complication in de Sitter space is the presence of several irreducible representations of the de Sitter isometry group that may contribute to the decomposition of unity. We showed that for the bubble chain considered here the principal series gives a complete partition of unity. This is in agreement with arguments out forward in \cite{Loparco:2023rug}, however, the proof involves novel functional identities for hypergeometric functions. Furthermore, the spectral representation on the Schwinger-Keldysh contour requires additional components compered to the decomposition of the Wightman functions given in \cite{Loparco:2023rug}. These have similar behavior at small momentum transfer yet are important to recover the correct UV-divergencies of loop integrals which we renormalized with a manifestly de Sitter invariant dimensional regularization and minimal subtraction. The existence of two components of the spectral function representing time-ordered and Wightman correlators of the contour also clarifies the 'convolution' product suggested in \cite{DiPietro:2023inn}.

While spectral representations of the 1-loop integral can equally well be carried out resorting to the K\"all\'en-Lehmann decomposition alone \cite{Zhang:2025nzd} the split representation is crucial to make the geometric series of the higher loop contribution manifest, which eventually allows to sum up these contributions. However, it turns out that the split representation has another important feature: It provides the frequency analysis of the cosmological collider signal directly at the level of the spectral function without having to actually perform the spectral integral. This poses a substantial simplification when predicting the cosmological signal of the model at hand. Furthermore, this shows that the frequencies of the signal can be obtained form the position space approach as well since this gives rise to the same spectral function (e.g. \cite{Sachs:2023eph}). On the other hand the latter has a dual representation in terms of conformal field theory, where poles correspond to exchanged single-trace operators and residues give operator product expansion coefficients. This then  allows for a unified description of comological correlators in terms of universality classes of conformal field theories.   

Finally, the spectral function of the resummed cosmological correlator is given by a simple geometric series that can be analyzed analytically when the 1-loop spectral function is known. In particular, this is the case in dS$_3$, where the spectral function is expressed in terms of digamma functions. In dS$_4$ the renormalized spectral function is more complicated, but we were still able to extract important non-perturbative information from it. In particular, it is possible to understand the non-perturbative dependence of the signals on the coupling constant $\lambda$, and to show that  $\lambda\to\infty$ is a regular point. On the other hand, at negative $\lambda$, we can identify pathological behaviors of the theory in de Sitter space and give a physical interpretation in terms of a unitary violating spectral function. 

Generalizations of our spectral approach to other forms of exchange diagrams (not necessarily 4-point correlators) are straightforward and possible by substituting the building blocks $\mathcal{I}_{\Yright}^\pm$ by different $n$-point functions (that do not need to be tree-level or de Sitter invariant). Moreover, when it comes to the classes of exchange diagrams that can be tackled with the help of the split representation, we are not at all restricted to tree-level computations or bubble chains. If the spectral functions of sunset or melon-type diagrams in de Sitter were known, we could immediately re-use our computations by substituting these spectral functions in the respective integrals. Related discussions can also be found in \cite{Lee:2025kgs}. 

\acknowledgments

We would like to thank Giovanni Cabass, Maximilian Haensch, Johannes Henn, Matthias Nowinski, Prashanth Raman, Denis Werth and  Weichen Xiao  for helpfull discussions. This work is supported in parts by the Excellence Cluster Origins of the DFG under Germany's Excellence Strategy EXC-2094 390783311 as well as EXC 2094/2: ORIGINS 2.

\appendix

\section{Useful formulae and computations}
\subsection{Special functions}
For the derivation of the orthogonality relation of Hankel functions in appendix \ref{sec:App:Hankel} we utilize the following Mellin-Barnes representations of Hankel functions:
\begin{align}\label{eq:HankelMB}
    H_{\nu}^{(1)} (z) &= \frac{1}{\pi}\int_{-i\infty}^{+i\infty}\frac{\mathrm{d}s}{2\pi i} \left(\frac{z}{2}\right)^{-2s} e^{+i\frac{\pi}{2}(2s-\nu-1)}\mathrm{\Gamma}(s+\tfrac{\nu}{2}) \mathrm{\Gamma}(s-\tfrac{\nu}{2}), \nonumber\\
    H_{\nu}^{(2)} (z) &= \frac{1}{\pi}\int_{-i\infty}^{+i\infty}\frac{\mathrm{d}s}{2\pi i} \left(\frac{z}{2}\right)^{-2s} e^{-i\frac{\pi}{2}(2s-\nu-1)}\mathrm{\Gamma}(s+\tfrac{\nu}{2}) \mathrm{\Gamma}(s-\tfrac{\nu}{2}).
\end{align}
For the evaluation of the Mellin-Barnes integrals we also use the following time integral:
\begin{equation}\label{eq:basicTimeInt}
    \int_{-\infty}^0\mathrm{d}\eta \,(-\eta)^{p-1}e^{ik\eta} = \frac{\mathrm{\Gamma}(p)}{i^pk^p}.
\end{equation}
Throughout the main text, the shorthand notation
\begin{equation}
    \mathrm{\Gamma}\left[\begin{matrix}a_1, a_2, \dots,a_n\\ b_1,b_2,\dots,b_n \end{matrix}\right] \equiv \frac{\mathrm{\Gamma}(a_1)\mathrm{\Gamma}(a_2)\dots\mathrm{\Gamma}(a_n)}{\mathrm{\Gamma}(b_1)\mathrm{\Gamma}(b_2)\dots\mathrm{\Gamma}(b_n)}
\end{equation}
for products of $\mathrm{\Gamma}$-functions is used. Furthermore, the results of massive exchange diagrams involve hypergeometric functions
\begin{equation}
    {}_pF_q\left.\left[\begin{matrix}a_1, \dots,a_p\\ b_1,\dots,b_q  \end{matrix}\right\vert z\right] \equiv\sum\limits_{n=0}^\infty \frac{(a_1)_n\dots(a_p)_n}{(b_1)_n\dots(b_q)_n}\frac{z^n}{n!},
\end{equation}
where $(a)_n$ is the Pochhammer symbol
\begin{equation}
    (a)_n \equiv\frac{\mathrm{\Gamma}(a+n)}{\mathrm{\Gamma}(a)}.
\end{equation}
To evaluate the spectral integrals we also introduced a regularized form of these hypergeometric functions,
\begin{equation}
    {}_p\tilde{\mathcal{F}}_q\left.\left[\begin{matrix}a_1, \dots,a_p\\ b_1,\dots,b_q  \end{matrix}\right\vert z\right] \equiv \frac{1}{\mathrm{\Gamma}[b_1,\dots,b_q]}{}_pF_q\left.\left[\begin{matrix}a_1, \dots,a_p\\ b_1,\dots,b_q  \end{matrix}\right\vert z\right]
\end{equation}
which is an entire function of all of its parameters (away from the singular points $z=1$ or $z=\infty$).

\subsection{Shift formula for pole resummation}\label{sec:shift-formula}
We would like to evaluate integrals of the form
\begin{equation}
    I_N =\int_{-\infty}^{+\infty}\frac{\mathrm{d}\tilde{\nu}'}{2\pi i} f(\tilde{\nu}') \Sigma_\nu(\tilde{\nu}')^N
\end{equation}
and resum them in the generating functional 
\begin{equation}
    \mathcal{G}(\lambda) = \sum\limits_{N=0}^\infty \lambda^N I_N = \int_{-\infty}^{+\infty}\frac{\mathrm{d}\tilde{\nu}'}{2\pi i} \frac{f(\tilde{\nu}')}{1-\lambda \Sigma_\nu(\tilde{\nu}')}.
\end{equation}
To do so, we apply the residue theorem since we know the pole structure of the integrand and the integrand has a suitable fall-off behavior at infinity. The function $f$ has simple poles and we can use our standard formula (\ref{eq:basicResidueFormula}) to pick up the respective residues. The more challenging part is the contribution from the poles of $\Sigma_\nu$. \\
We know that $\Sigma_\nu$ has a set of simple poles $p_l,\,l\in\mathcal{L}$ where $\mathcal{L}$ is some index set. In the case of the bubble spectral function we have
\begin{equation}
    \mathcal{L}= \left\{\pm i(\tfrac{d}{2}+2m \pm\nu\pm\nu),\, m\in\mathbb{N}\right\}.
\end{equation}
Locally, we can decompose $\Sigma_\nu$ as 
\begin{equation}
    \Sigma_\nu(\tilde{\nu}') = \frac{r_l}{\tilde{\nu}'-p_l} + h_l(\tilde{\nu}'),
\end{equation}
where $r_l$ is the residue of $\Sigma_\nu$ at $p_l$ and $h_l$ is an analytic function. Instead of solving the resummed integral, we will solve all integrals separately and resum at the end. To do so, we first expand
\begin{equation}
    \Sigma_\nu(\tilde{\nu}')^N = \sum\limits_{k=0}^N \binom{N}{k} \left(\frac{r_l}{\tilde{\nu}'-p_l}\right)^k h_l(\tilde{\nu}')^{N-k}
\end{equation}
using the local expansion of $\Sigma_\nu$. The integral $I_N$ then has a set of $k$-th order poles for $k\in\{1,\dots,N\}$ at $p_l$, so we can rewrite
\begin{equation}
    \text{Res}_{p_l}\left[f(\tilde{\nu}')\Sigma_\nu(\tilde{\nu}')^N\right] = \sum\limits_{k=1}^N \binom{N}{k} r_l^k \,\text{Res}_{p_l}\left[\frac{f(\tilde{\nu}')h_l(\tilde{\nu}')^{N-k}}{(\tilde{\nu}'-p_l)^k}\right].
\end{equation}
For a $k$-th order pole, the residue is given by the formula
\begin{equation}
    \text{Res}_{p_l}\left[\frac{g(\tilde{\nu}')}{(\tilde{\nu}'-p_l)^k}\right] = \frac{1}{(k-1)!}\frac{\mathrm{d}^{k-1}}{\mathrm{d}\tilde{\nu}'^{k-1}}g(\tilde{\nu}')\Big\vert_{\tilde{\nu}'=p_l}.
\end{equation}
Hence, we find
\begin{equation}
    \text{Res}_{p_l} \left[f(\tilde{\nu}')\Sigma_\nu(\tilde{\nu}')^N\right] = \sum\limits_{k=1}^N\binom{N}{k} \frac{r_l^k}{(k-1)!}\frac{\mathrm{d}^{k-1}}{\mathrm{d}\tilde{\nu}'^{k-1}}\left[f(\tilde{\nu}')h_l(\tilde{\nu}')^{N-k}\right]\Big\vert_{\tilde{\nu}'=p_l}.
\end{equation}
Next, if we want to evaluate the generating functional $\mathcal{G}(\lambda)$, we must consider the contributions from all integrals $I_N$ to the pole at $p_l$. We denote these contributions as $\mathcal{G}_l$ so that $\mathcal{G} =\mathcal{G}_f +\sum_{l\in\mathcal{L}} \mathcal{G}_l$ ($\mathcal{G}_f$ being the contribution from the poles of $f$). The $N=0$ term does not contribute because it does not contain $\Sigma_\nu$. We obtain:
\begin{align}
    \mathcal{G}_l(\lambda) &= \sum\limits_{N=1}^\infty \lambda^N \,\text{Res}_{p_l}\left[f \Sigma_\nu^N\right] \\
    &= \sum\limits_{N=1}^\infty \sum\limits_{k=1}^N\binom{N}{k} \frac{\lambda^N r_l^k}{(k-1)!}\frac{\mathrm{d}^{k-1}}{\mathrm{d}\tilde{\nu}'^{k-1}}\left[f(\tilde{\nu}')h_l(\tilde{\nu}')^{N-k}\right]\Big\vert_{\tilde{\nu}'=p_l}.
\end{align}
By setting $m\equiv N-k \geq 0$ so that $N=m+k$ we change the order of summation:
\begin{equation}
    \mathcal{G}_l = \sum\limits_{k=1}^\infty \frac{r_l^k}{(k-1)!} \sum\limits_{m=0}^\infty \lambda^{m+k} \binom{m+k}{k} \frac{\mathrm{d}^{k-1}}{\mathrm{d}\tilde{\nu}'^{k-1}}\left.\left[f(\tilde{\nu}')h_l(\tilde{\nu}')^{m}\right]\right\vert_{\tilde{\nu}'=p_l}.
\end{equation}
Using the standard binomial identity,
\begin{equation}
    \sum\limits_{m=0}^\infty \binom{m+k}{k} x^m = \frac{1}{(1-x)^{k+1}},
\end{equation}
we can evaluate the second sum exactly and simplify to
\begin{equation}
    \mathcal{G}_l(\lambda) = \sum\limits_{k=1}^\infty \frac{(\lambda r_l)^k}{(k-1)!} \frac{\mathrm{d}^{k-1}}{\mathrm{d}\tilde{\nu}'^{k-1}} \left.\left[\frac{f(\tilde{\nu}')}{(1-\lambda h_l(\tilde{\nu}'))^{k+1}}\right]\right\vert_{\tilde{\nu}'=p_l}.
\end{equation}
This is \textit{almost} a translation operator acting on some function. Remember, that for an analytic function we have
\begin{equation}
    \sum\limits_{k=0}^\infty \frac{a^k}{k!} \frac{\mathrm{d}^k}{\mathrm{d}x^k} g(x) = e^{a\partial_x} g(x) = g(x+a).
\end{equation}
There remains an obstruction by the $k$-dependence of the denominator of the function, hence the expression we found is a form of a \emph{generalized shift}.

\subsection{Orthogonality relation for Hankel functions}\label{sec:App:Hankel}
In the following we will explicitly show the orthogonality relation from (\ref{eq:orthoHankel}). We want to solve the integral
\begin{equation}
    A = \int_{-\infty}^0\frac{\mathrm{d}\eta}{(-\eta)^{1+\epsilon}} H_{\nu}^{(1)}(-k\eta)H_{\nu'}^{(1)}(-k\eta).
\end{equation}
We introduced the parameter $\epsilon$ for regularization purposes. We can make use of a Mellin-Barnes representation of the Hankel function given in (\ref{eq:HankelMB}) and rewrite the integral as
\begin{align}
    A=\frac{1}{\pi^2}\int_{-i\infty}^{+i\infty}\frac{\mathrm{d}s_1}{2\pi i}\frac{\mathrm{d}s_2}{2\pi i}\left(\frac{k}{2}\right)^{-2s_1-2s_2}e^{i\frac{\pi}{2}(2s_1+2s_2-\nu-\nu'-2)}\mathrm{\Gamma}(s_1+\tfrac{\nu'}{2})\mathrm{\Gamma}(s_1-\tfrac{\nu'}{2}) \nonumber\\
    \mathrm{\Gamma}(s_2+\tfrac{\nu}{2})\mathrm{\Gamma}(s_2-\tfrac{\nu}{2})  \int_{-\infty}^0\mathrm{d}\eta\,(-\eta)^{-1-\epsilon-2s_1-2s_2}
\end{align}
Evaluating the arising time integral with contributions from both Hankel functions
\begin{equation}
    \int_{-\infty}^0\mathrm{d}\eta\,(-\eta)^{-1-\epsilon-2s_1-2s_2} = i\pi \delta(s_1+s_2+\tfrac{\epsilon}{2})
\end{equation}
leads to a single Mellin-Barnes integral
\begin{equation}
    A = \frac{1}{2\pi^2}\left(\frac{k}{2}\right)^\epsilon e^{-i\frac{\pi}{2}(\nu+\nu'+\epsilon+2)}\int_{-i\infty}^{+i\infty}\frac{\mathrm{d}s_1}{2\pi i} \mathrm{\Gamma}(s_1+\tfrac{\nu'}{2})\mathrm{\Gamma}(s_1-\tfrac{\nu'}{2})\mathrm{\Gamma}(-s_1-\tfrac{\epsilon}{2}-\tfrac{\nu}{2})\mathrm{\Gamma}(-s_1-\tfrac{\epsilon}{2}+\tfrac{\nu}{2})
\end{equation}
which can be solved with the help of the residue theorem. The integrand has poles stemming from the $\mathrm{\Gamma}$-functions. We can close the Mellin contour in the left half plane and therefore only pick up the left poles stemming from the first two $\mathrm{\Gamma}$-factors:
\begin{align}
    A = \frac{1}{2\pi^2} \left(\frac{k}{2}\right)^\epsilon e^{-i\frac{\pi}{2}(\nu+\nu'+\epsilon+2)}\left\{\sum\limits_{n=0}^\infty \frac{(-1)^n}{n!}\mathrm{\Gamma}(-n+\nu')\mathrm{\Gamma}(n-\tfrac{\epsilon}{2}-\tfrac{\nu'}{2}-\tfrac{\nu}{2})\mathrm{\Gamma}(n-\tfrac{\epsilon}{2}-\tfrac{\nu'}{2}+\tfrac{\nu}{2})\right.  \nonumber\\ 
    \left.+\sum\limits_{n=0}^\infty \frac{(-1)^n}{n!}\mathrm{\Gamma}(-n+\nu')\mathrm{\Gamma}(n-\tfrac{\epsilon}{2}+\tfrac{\nu'}{2}-\tfrac{\nu}{2})\mathrm{\Gamma}(n-\tfrac{\epsilon}{2}+\tfrac{\nu'}{2}+\tfrac{\nu}{2})\right\}
\end{align}
Using the reflection formula of the $\mathrm{\Gamma}$-function, $\mathrm{\Gamma}(-n+z)=(-1)^{n-1}\mathrm{\Gamma}(-z)\mathrm{\Gamma}(1+z)/\mathrm{\Gamma}(n+1-z)$, as well as Gauss' summation theorem for hypergeometric functions from which we deduce
\begin{equation}
    \sum\limits_{n=0}^\infty \frac{1}{n!}\frac{\mathrm{\Gamma}(a+n)\mathrm{\Gamma}(b+n)}{\mathrm{\Gamma}(c+n)} = \frac{\mathrm{\Gamma}(a)\mathrm{\Gamma}(b)\mathrm{\Gamma}(c-a-b)}{\mathrm{\Gamma}(c-a)\mathrm{\Gamma}(c-b)},
\end{equation}
we find
\begin{equation}
    A= \frac{1}{2\pi^2}e^{-i\frac{\pi}{2}(\nu+\nu'+\epsilon)}\left(\frac{k}{2}\right)^\epsilon \mathrm{\Gamma}(1+\epsilon)\mathrm{\Gamma}(\nu')\mathrm{\Gamma}(1-\nu') \mathcal{E}_\epsilon(\nu,\nu')
\end{equation}
where we introduced the function
\begin{equation}
    \mathcal{E}_\epsilon(\nu,\nu')=\frac{\mathrm{\Gamma}(\frac{-\epsilon-\nu'-\nu}{2})\mathrm{\Gamma}(\frac{-\epsilon-\nu'+\nu}{2})}{\mathrm{\Gamma}(1+\frac{\epsilon-\nu'+\nu}{2})\mathrm{\Gamma}(1+\frac{\epsilon-\nu'-\nu}{2})} - \frac{\mathrm{\Gamma}(\frac{-\epsilon+\nu'-\nu}{2})\mathrm{\Gamma}(\frac{-\epsilon+\nu'+\nu}{2})}{\mathrm{\Gamma}(1+\frac{\epsilon+\nu'+\nu}{2})\mathrm{\Gamma}(1+\frac{\epsilon+\nu'-\nu}{2})}.
\end{equation}
For $\nu'\neq\pm\nu$ we can simply take the $\epsilon\to0$ limit in which the two terms are actually the same and cancel exactly so that $\mathcal{E}_\epsilon \equiv 0$. However, when $\nu'=\pm\nu$, naively taking the limit $\epsilon\to 0$ leads to divergent $\mathrm{\Gamma}(0)$ factors in the numerators. Therefore, we will expand the $\mathrm{\Gamma}$-functions in the parameters $x\equiv \nu'+\nu$, $y\equiv \nu'-\nu$ and $\epsilon$. By expanding the functions in the denominator, we can extract the leading divergence. First, assume $y\to 0$ and $x \neq 0$. Then:
\begin{align}
    \mathcal{E}_\epsilon(x,y) &= \frac{\mathrm{\Gamma}(\tfrac{-\epsilon-x}{2})\mathrm{\Gamma}(\tfrac{-\epsilon-y}{2})}{\frac{\epsilon-x}{2}\frac{\epsilon-y}{2}\mathrm{\Gamma}(\tfrac{\epsilon-x}{2})\mathrm{\Gamma}(\tfrac{\epsilon-y}{2})}-\frac{\mathrm{\Gamma}(\tfrac{-\epsilon+x}{2})\mathrm{\Gamma}(\tfrac{-\epsilon+y}{2})}{\frac{\epsilon+x}{2}\frac{\epsilon+y}{2}\mathrm{\Gamma}(\tfrac{\epsilon+x}{2})\mathrm{\Gamma}(\tfrac{\epsilon+y}{2})} \\
    &\sim \frac{4}{(\epsilon-x)(\epsilon-y)}-\frac{4}{(\epsilon+x)(\epsilon+y)} = -\frac{8\epsilon}{x(\epsilon^2-y^2)} \to -\frac{4\pi}{\nu} \delta(\tilde{\nu}-\tilde{\nu}')
\end{align}
Here we used the fact that $\nu=i\tilde{\nu}$ and $\nu'=i\tilde{\nu}'$ are purely imaginary for fields in the principal series of de Sitter and hence we end up with the usual Poisson kernel for a $\delta$-function. A similar analysis can be performed for the $x\to 0$ and $y\neq 0$ case.
In total, in the limit $\epsilon\to 0$ we pick up the $\delta$-functions from these two contributions
\begin{equation}
    \lim\limits_{\epsilon\to 0} \mathcal{E}_\epsilon(\nu,\nu') = -\frac{4\pi}{\nu}\left[\delta(\tilde{\nu}-\tilde{\nu}')+\delta(\tilde{\nu}+\tilde{\nu}')\right]
\end{equation}
and the final answer is thus given by (\ref{eq:orthoHankel}).

\subsection{Three-point function}\label{sec:buildingBlocks}

In this section we explicitly compute the 3-point contact diagram with two conformally coupled external legs and one massive scalar external leg which is frequently used in the main text to build up 4-point exchange diagrams of various types. We define:
\begin{align}
    \mathcal{I}_{\Yright}^+(r_1;\nu) &= \int_{-\infty}^0\frac{\mathrm{d}\eta}{(-\eta)^4} G_+(k_1;\eta)G_+(k_2;\eta) u_\nu^*(k_s;\eta) \nonumber \\
    &= \frac{\sqrt{\pi}\eta_*^2}{8k_1k_2} e^{-i\frac{\pi}{2}(\nu^*+\frac12)} \int_{-\infty}^0\frac{\mathrm{d}\eta}{(-\eta)^{1/2}} e^{ik_{12}\eta} H_{\nu^*}^{(2)}(-k_s\eta).
\end{align}
Be aware that we do not use the full bulk-to-boundary propagator for the massive field. Instead, we strip off the factors related to the evaluation of the boundary vertex in some late-time limit and work directly with the mode function $u_\nu^*(k_s;\eta)$ of the massive field.
For fields in the principal series we have $\nu^*=-\nu$ and we can use the Mellin-Barnes representation of the Hankel function (\ref{eq:HankelMB}) as well as the time integral (\ref{eq:basicTimeInt}) to obtain
\begin{equation}
    \mathcal{I}_{\Yright}^+(r_1;\nu) = \frac{\eta_*^2}{8\sqrt{\pi k_{12}}k_1k_2}  \int_{-i\infty}^{+i\infty} \frac{\mathrm{d}s}{2\pi i}\left(\frac{k_s}{2k_{12}}\right)^{-2s} \mathrm{\Gamma}(s+\tfrac{\nu}{2})\mathrm{\Gamma}(s-\tfrac{\nu}{2})\mathrm{\Gamma}(\tfrac12-2s).
\end{equation}
Again, we can close the integration contour in the left half-plane and pick up only the residues at the poles from the first two $\mathrm{\Gamma}$-functions: 
\begin{equation}
    \mathcal{I}_{\Yright}^+(r_1;\nu) = \frac{\eta_*^2}{8\sqrt{\pi k_{12}}k_1k_2}\left\{\sum\limits_{n=0}^\infty \frac{(-1)^n}{n!}\mathrm{\Gamma}(-n-\nu)\mathrm{\Gamma}(\tfrac12+\nu+2n)\left(\frac{r_1}{2}\right)^{2n+\nu} + (\nu\leftrightarrow -\nu)\right\}.
\end{equation}
Using the reflection and Legendre duplication formula for the $\mathrm{\Gamma}$-function we can recast the result in terms of a hypergeometric function. Ultimately, we end up with
\begin{equation}
    \mathcal{I}_{\Yright}^+(r_1;\nu) = \frac{\eta_*^2}{8\sqrt{\pi k_{12}}k_1k_2}\left\{F_+^{\nu,\frac12}(r_1) + F_-^{\nu,\frac12}(r_1)\right\},
\end{equation}
where the functions $F_\pm^{\nu,p}(r)$ were defined in (\ref{eq:defF}). 

\subsection{Spectral function in $d=3$}\label{sec:3d-spectral}

\begin{figure}[t]
  \centering
  \includegraphics[width=0.5\linewidth]{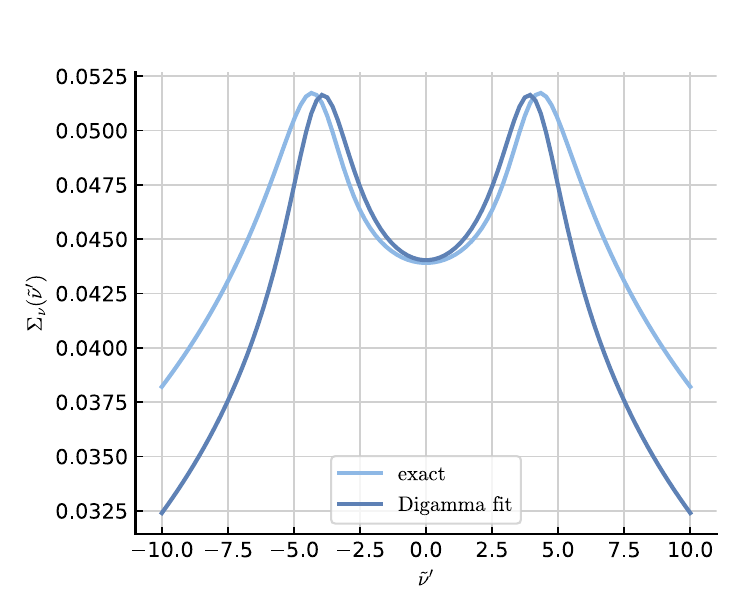}
  \caption{Comparison of the spectral function $\Sigma_\nu(\tilde{\nu}')$ in $d=3$ to a sum of digamma functions. We chose $\tilde{\nu}=2.0$ in this plot.}
  \label{fig:3d-fit}
\end{figure}

Since there exists a closed-form expression for the spectral function in $d=2$ in terms of digamma functions, it is a natural question whether the series expansion $P_\nu(\tilde{\nu}')$ also has a closed-form expression in $d=3$. It is not clear whether such a closed form exists but, interestingly, we were able to fit the function
\begin{equation}
    \sigma_\nu(\tilde{\nu}') = C_1\sum\limits_{a,b=\pm}\psi\left(\tfrac{d}{2}+a\,i \tilde{\nu}+b\tfrac{i\tilde{\nu}'}{2}\right) + C_0
\end{equation}
to $\Sigma_\nu(\tilde{\nu}')$ qualitatively. As can be seen in Figure \ref{fig:3d-fit}, the behavior of the two functions is very similar with almost identical maxima. The fall-off behavior seems slightly different, so there might be some $\tilde{\nu}'$-dependent envelope function required. 

\bibliographystyle{JHEP}
\bibliography{biblio.bib}






\end{document}